\documentclass[twocolumn]{aastex63}

\usepackage{natbib}
\usepackage{graphicx}
\usepackage{amsmath} 			
\usepackage{amssymb}			

\usepackage{mathtools}

\usepackage{calligra}
\DeclareMathAlphabet{\mathcalligra}{T1}{calligra}{m}{n}
\DeclareFontShape{T1}{calligra}{m}{n}{<->s*[2.5]callig15}{}
\newcommand{\scriptr}{\mathcalligra{r}\,}

\usepackage{listings}
\usepackage{bm}
\usepackage{subfigure}

\newcommand{\change}[1]{{\color{black} #1}}

\graphicspath{{Figures/}}

\newcommand{\ORNLphys}{Physics Division, Oak Ridge National Laboratory, Oak Ridge, TN 37831-6354, USA}
\newcommand{\genasis}{\textsc{GenASiS}}

\submitjournal{ApJS}

\shorttitle{GenASiS: Self-gravitating Baryonic Matter}
\shortauthors{Cardall et al.}

\begin{document}

\title{\textsc{GenASiS}: \textit{Gen}eral \textit{A}strophysical \textit{Si}mulation \textit{S}ystem. II. Self-gravitating Baryonic Matter\footnote{This manuscript has been authored by UT-Battelle, LLC, under contract DE-AC05-00OR22725 with the US Department of Energy (DOE). The US government retains and the publisher, by accepting the article for publication, acknowledges that the US government retains a nonexclusive, paid-up, irrevocable, worldwide license to publish or reproduce the published form of this manuscript, or allow others to do so, for US government purposes. DOE will provide public access to these results of federally sponsored research in accordance with the DOE Public Access Plan (\texttt{http://energy.gov/downloads/doe-public-access-plan}).}}

\correspondingauthor{Christian Y. Cardall}
\email{cardallcy@ornl.gov}

\author[0000-0002-0086-105X]{Christian Y. Cardall}
\affiliation{\ORNLphys}

\author[0000-0003-0395-8532]{Reuben D. Budiardja}
\affiliation{National Center for Computational Sciences, Oak Ridge National Laboratory, Oak Ridge, TN 37831-6008, USA}

\author[0000-0003-4459-2557]{R. Daniel Murphy}
\affiliation{Department of Physics and Astronomy, University of Tennessee, Knoxville, TN 37996-1200, USA}

\author[0000-0003-1251-9507]{Eirik Endeve}
\affiliation{Computer Science and Mathematics Division, Oak Ridge National Laboratory, Oak Ridge, TN 37831-6221, USA}
\affiliation{Department of Physics and Astronomy, University of Tennessee, Knoxville, TN 37996-1200, USA}



\begin{abstract}
\textsc{GenASiS} ({\em Gen}eral {\em A}strophysical {\em Si}mulation {\em S}ystem) is a code being developed initially and primarily, though not exclusively, for the simulation of core-collapse supernovae on the world's leading capability supercomputers.
This paper---the second in a series---documents capabilities for Newtonian self-gravitating fluid dynamics, including tabulated microphysical equations of state treating nuclei and nuclear matter (`baryonic matter').
Computation of the gravitational potential of a spheroid, and simulation of the gravitational collapse of dust and of an ideal fluid, provide tests of self-gravitation against known solutions. 
In multidimensional computations of the adiabatic collapse, bounce, and explosion of spherically symmetric pre-supernova progenitors---which we propose become a standard benchmark for code comparisons---we find that the explosions are prompt and remain spherically symmetric (as expected), with an average shock expansion speed and total kinetic energy that are inversely correlated with the progenitor mass at the onset of collapse and the compactness parameter.
\end{abstract}



\section{Introduction}
\label{sec:Introduction}

Astrophysical events generally---and core-collapse supernovae in particular---are multiscale and multiphysics phenomena. 
See for instance a recent review by \citet{Janka2025Long-Term-Multi} for an overview of efforts to understand core-collapse supernovae via large-scale simulations.
The physics that must be addressed in order to treat the collapse, bounce, and explosion of a pre-supernova progenitor star includes the nuclear composition and fluid dynamics of matter comprising baryons, charged leptons, and photons; neutrino radiation transport in regimes ranging from tightly-coupled equilibrium with matter to free streaming; and self-gravity. Magnetic fields may also be relevant at some level, probably dominantly so in the case of hypernovae, the especially energetic jet-like supernovae associated with massive and rapidly rotating progenitors. 

\genasis\ ({\em Gen}eral {\em A}strophysical {\em Si}mulation {\em S}ystem) is a code under development that is aimed at the simulation of core-collapse supernovae and, potentially, other multiphysics problems.
Its more fundamental layers have been publicly released \citep{Budiardja2022GENASISBasics:-,Cardall2023GENASIS-MATHEMA}.
Earlier versions of \genasis\ were used to study turbulent magnetic field amplification \citep{Endeve2010Generation-of-M,Endeve2012Turbulent-Magne,Endeve2013Turbulence-and-}
and the stochasticity of convection-dominated vs. stationary-accretion-shock-instability-dominated explosions \citep{Cardall2015Stochasticity-a} in highly simplified parametrized models of the region between the nascent neutron star and the shock in the post-bounce supernova environment.
Paper~I in this series of methods papers described a centrally refined mesh suitable for collapse problems and basic fluid dynamics capabilities and tests \citep{Cardall2014GenASiS:-Genera}. 

The purpose of this work, Paper~II in this series, is to present and exercise a Poisson solver and an updated fluid dynamics solver in \genasis, both of which make efficient use of hardware accelerators (e.g. GPUs).
For present purposes the centrally refined mesh presented in Paper~I is set aside in favor of a single-level spherical coordinate mesh, with coarsening near the coordinate singularities at the origin and polar axis to avoid crippling time step restrictions.
The multipole Poisson solver draws inspiration both from those implemented in the FLASH code\footnote{\url{https://flash.rochester.edu/site/flashcode/user\_support/flash\_ug\_devel.pdf}}
 \citep{Couch:2013} and the approach of \citet{Muller:1995}.
\change{(See also \citet{Wongwathanarat2010An-axis-free-ov} on the use of overlapping spherical coordinate patches to avoid problems near the polar axis in self-gravitating flows, an improvement we will consider in future work.)}
In the work presented here \change{the Poisson solver} is used only for Newtonian self-gravity.
Full general relativity would be ideal for core-collapse supernova simulations, but an efficient Poisson solver is foundational not only to Newtonian gravity but to other widely-used approximations to general relativity, including the substitution of a relativistic monopole in an otherwise Newtonian multipole expansion \citep{Marek2006Exploring-the-r} and the imposition of the conformal flatness condition on a relativistic metric \citep{Isenberg2008Waveless-Approx,Wilson1996Relativistic-nu,Flanagan1999Possible-Explan,Cordero-Carrion2009Improved-constr}.
The fluid dynamics solver now includes parabolic reconstruction and the use of tabulated microphysical equations of state treating nuclei with a representative heavy nucleus and a phase transition to nuclear matter (along with charged leptons and photons; `baryonic matter').
After the mesh and solvers are discussed in \S\ref{sec:Solvers}, several test problems are presented \S\ref{sec:Tests} before concluding remarks are given in \S\ref{sec:Conclusion}.
\change{An Appendix follows up on an improvement suggested in the main text.}

%

%

\section{Solvers}
\label{sec:Solvers}

In this section we describe the mesh and solvers used for the computations presented in this paper.
All simulations are performed in spherical coordinates, with a coarsening strategy designed to avoid severe Courant--Friedrichs--Lewy (CFL) time step restrictions near the coordinate singularities at the  origin and the polar axis.
We use a multipole expansion to solve the Poisson equation for the Newtonian gravitational potential.
To compute the fluid dynamics of baryonic matter, we use the finite-volume method in order to handle shocks \citep{Cardall2014GenASiS:-Genera,Cardall2023GENASIS-MATHEMA} and tabulated microphysical equations of state treating nuclei and nuclear matter.
These are the solvers needed for the adiabatic gravitational collapse simulations presented here; simulations with neutrino radiation hydrodynamics will be reported elsewhere.

We have written the Poisson solver, finite-volume solver, and equation of state interpolation to take significant advantage of hardware accelerators when available. 
This is accomplished by using OpenMP \texttt{target} directives to offload computational kernels to the accelerators.
The general techniques we use to manage data movement and data mapping for offloading these kernels are discussed in \cite{Budiardja:2019}.

\subsection{Mesh}
\label{sec:Mesh}

\begin{figure*}[t]
\centering
\includegraphics[height=4in]{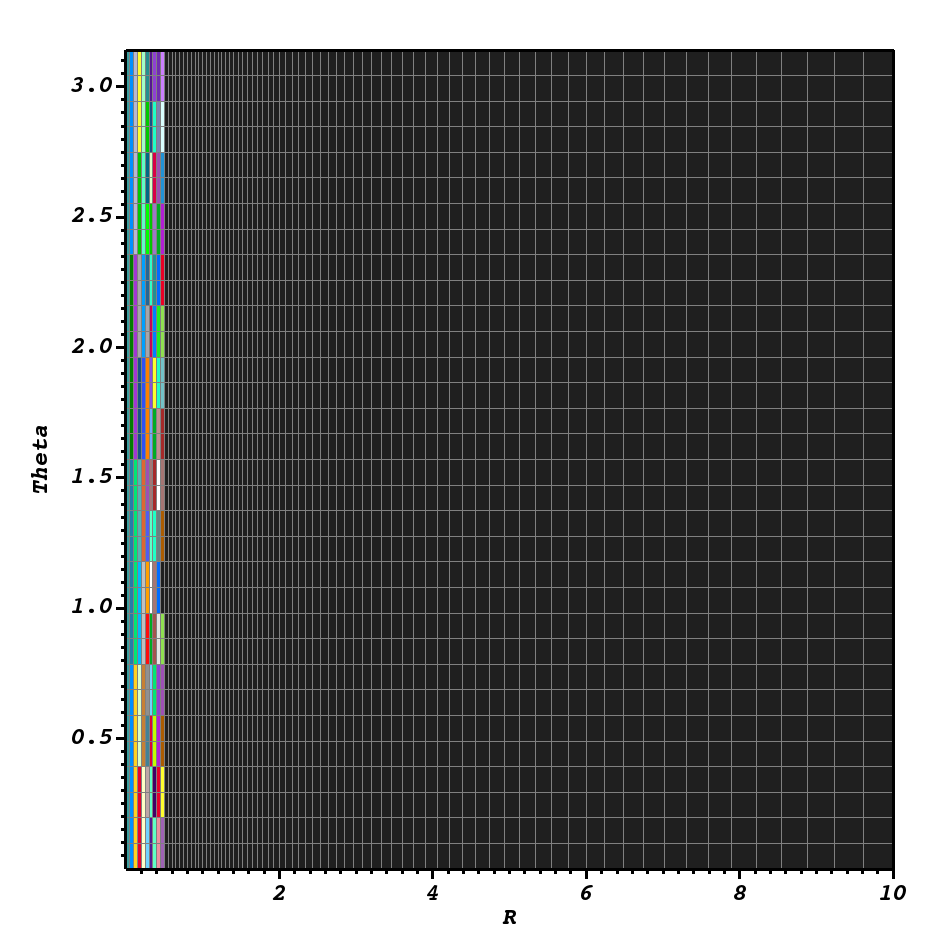}
\includegraphics[height=4in]{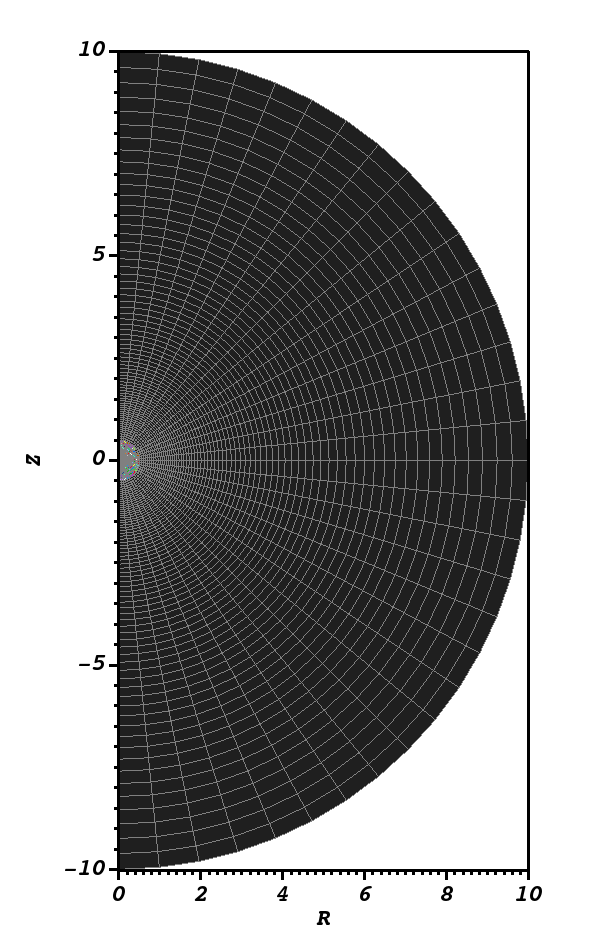}
\includegraphics[height=4in]{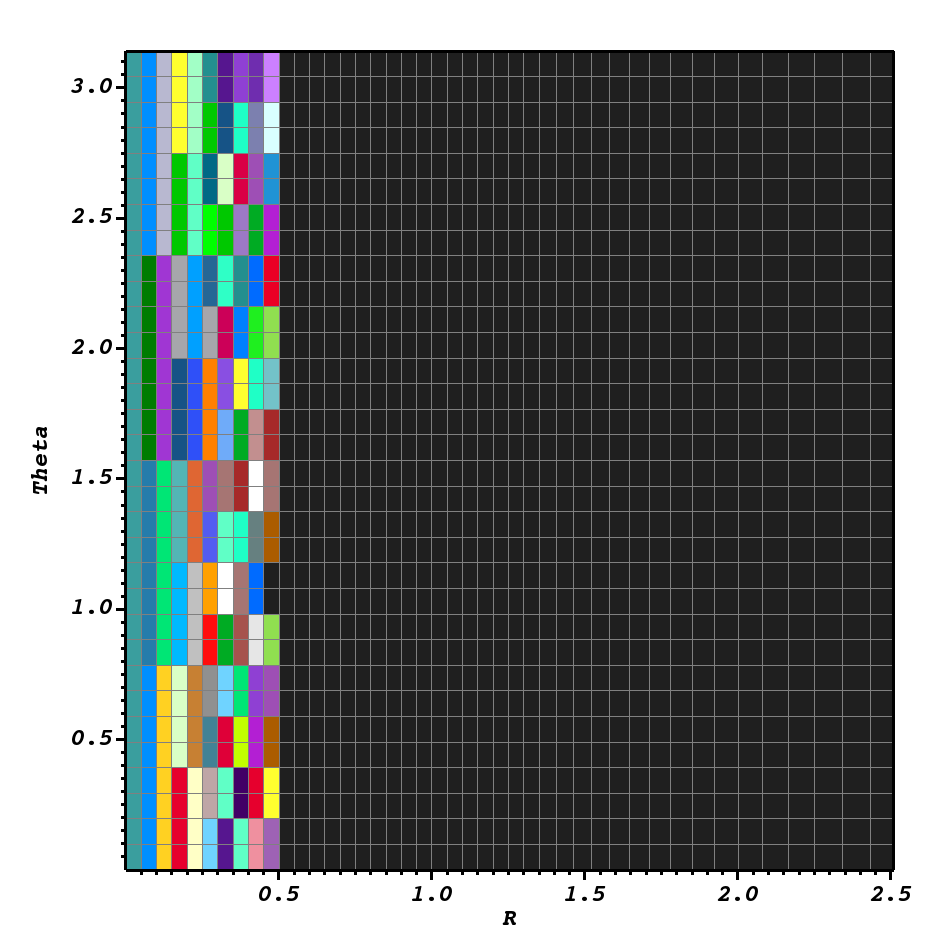}
\includegraphics[height=4in]{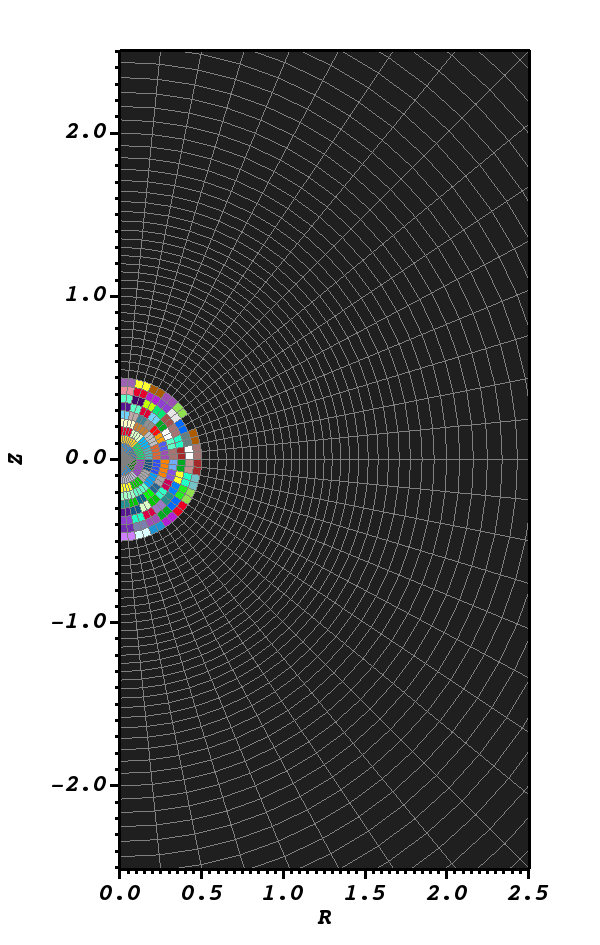}
\caption{An example 2D spherical coordinate mesh, in coordinate space (left) and physical space (right), showing the full mesh extent (top) and a region closer to $r = 0$ (bottom).
For $r > r_\mathrm{core} = 1.25$ the radial cell width $\Delta r \propto r$, yielding a constant polar/radial cell aspect ratio $r \, \Delta \theta / \Delta r$ (here $\approx 2.45$). 
For $r < r_\mathrm{core}$ the cell radial width $\Delta r_\mathrm{min}$ is uniform and the polar/radial aspect ratio rapidly decreases with decreasing $r$.
Approaching the origin, neighboring cells at a given radius with $r \, \Delta \theta < \Delta r  _\mathrm{core}$ are grouped into polar-angle `coarsening blocks' (randomly colored) consisting of $2, 4, 8, \dots$ cells as needed until the block width exceeds $\Delta r  _\mathrm{core}$.
Averaging over these blocks to suppress small-wavelength perturbations allows explicit time steps to be limited only by $\Delta r  _\mathrm{core}$.
}
\label{Fig:CoarseningBlocks_2D}
\end{figure*}

\begin{figure*}[t]
\centering
\includegraphics[height=2.5in]{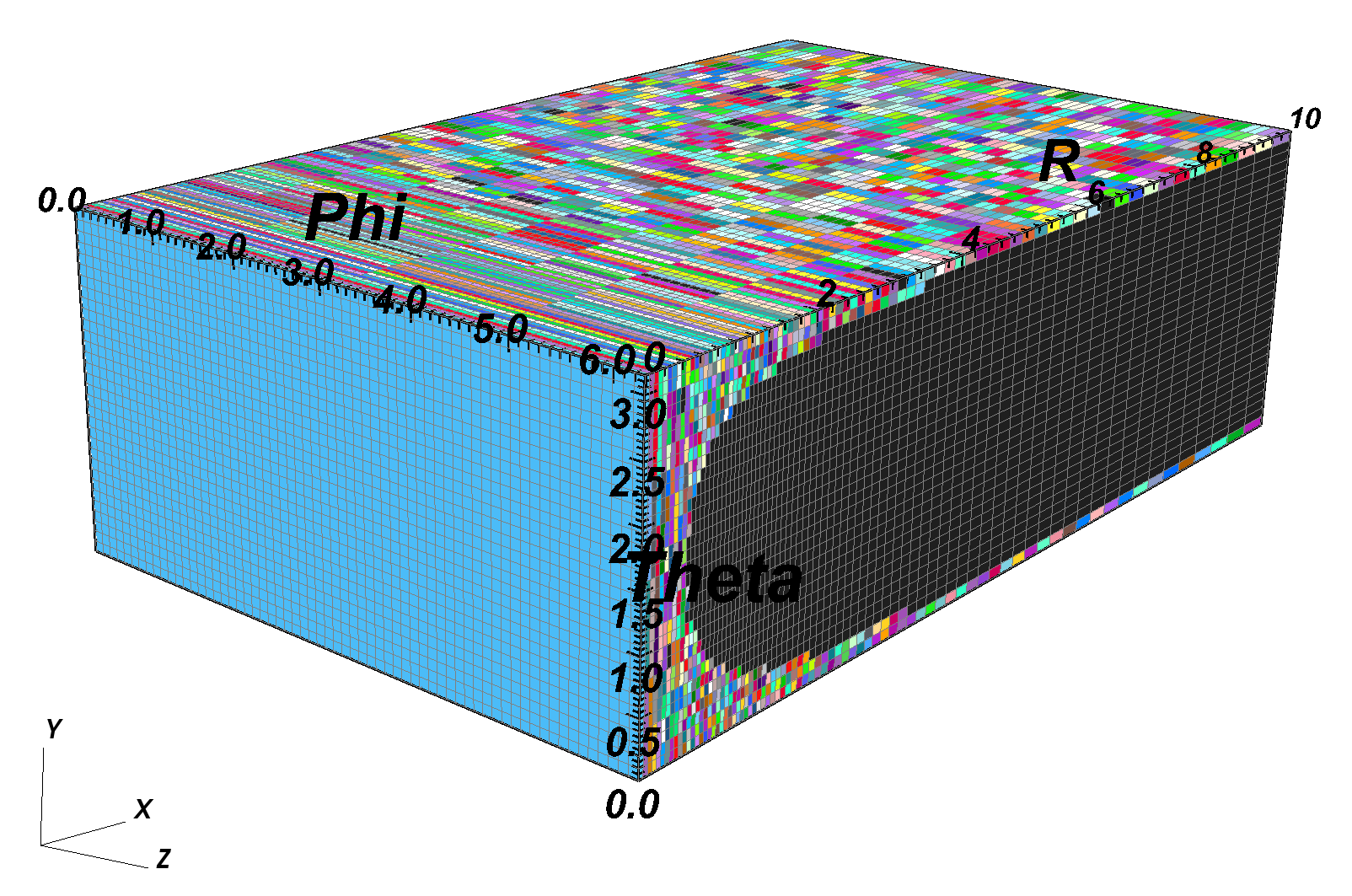}
\includegraphics[height=2.5in]{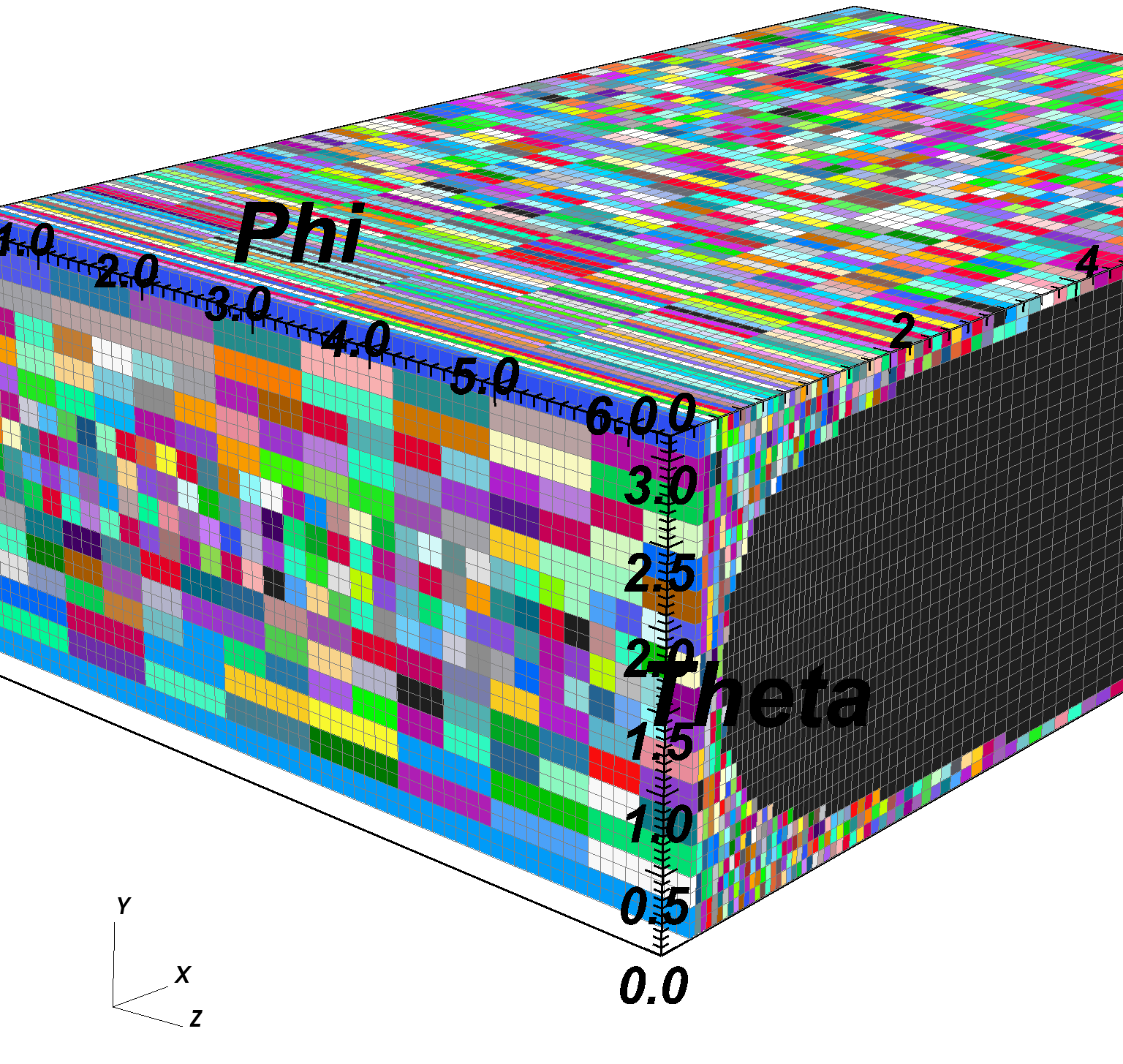}
\includegraphics[height=3.5in]{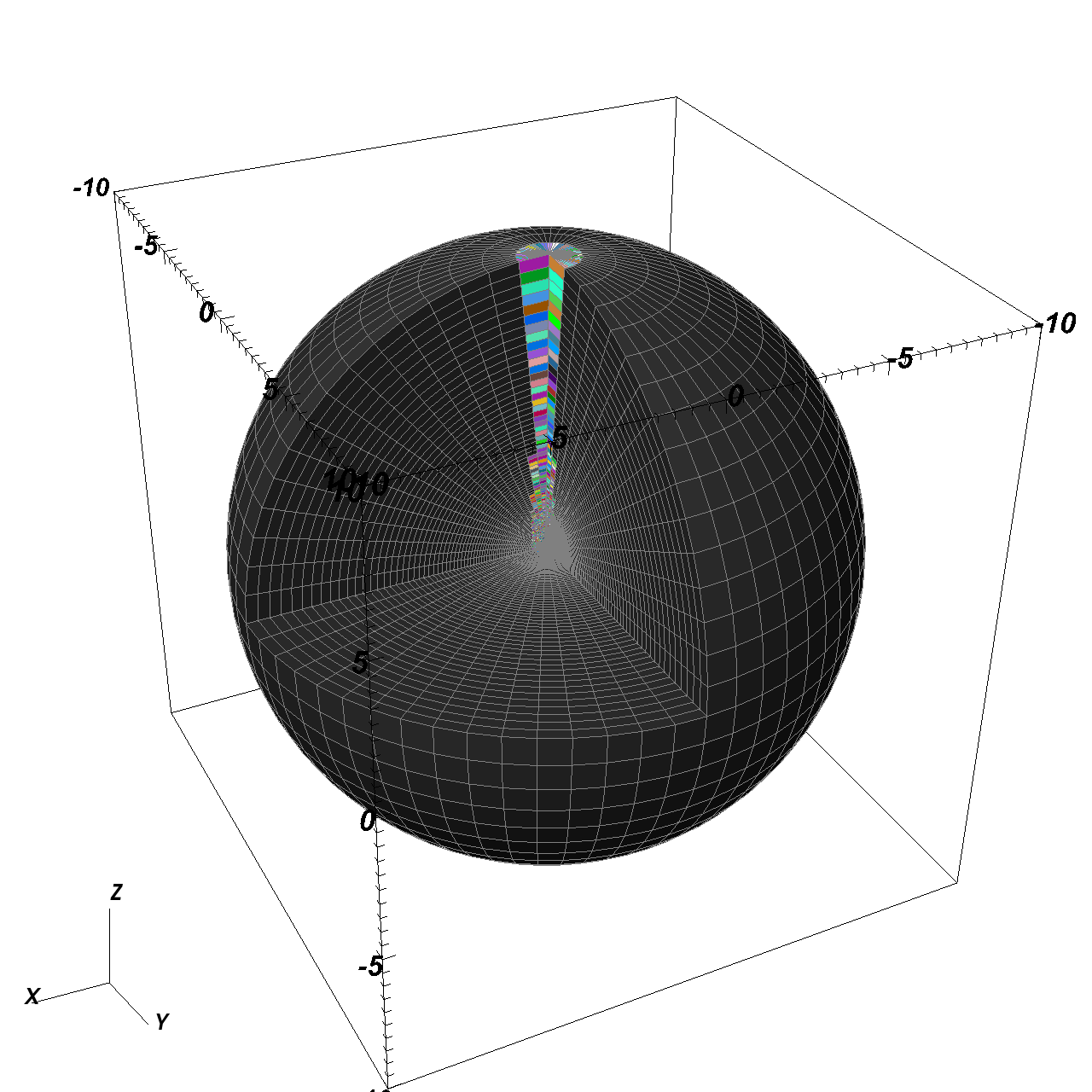}
\includegraphics[height=3.5in]{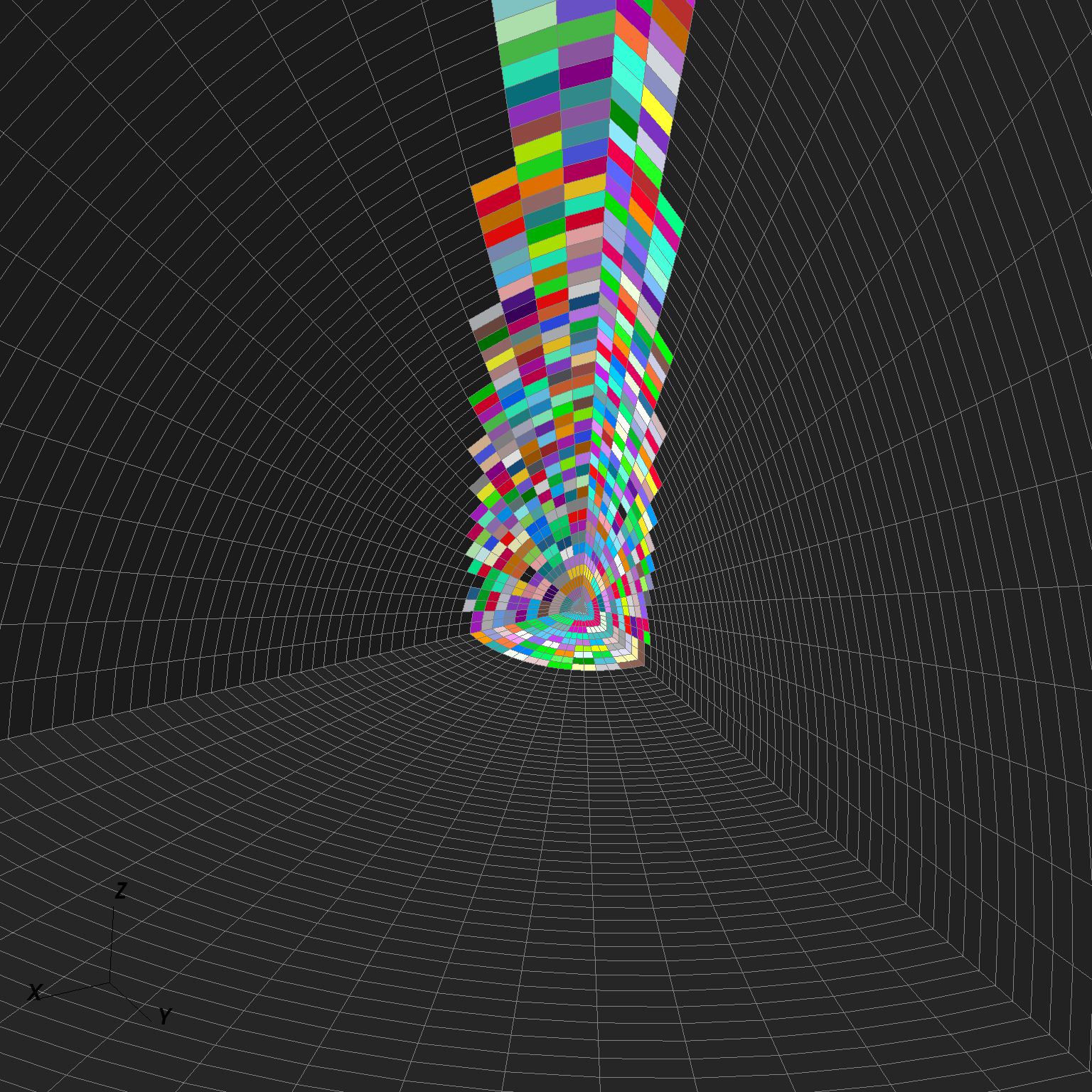}
\caption{An example 3D spherical coordinate mesh, in coordinate space (top) and physical space (bottom), showing the full mesh extent (left) and a region closer to $r = 0$ (right), with the $r = 0.25$ plane exposed in coordinate space (upper right).
The coarsening blocks (randomly colored) now appear along the polar axis as well as near the origin and are now two-dimensional, with the block size in each angular dimension determined by comparing $r \, \sin \theta \, \Delta \phi$ and $r \, \Delta \theta$ with $\Delta r  _\mathrm{core}$. 
}
\label{Fig:CoarseningBlocks_3D}
\end{figure*}

\begin{figure*}[t]
\centering
\includegraphics[height=3in]{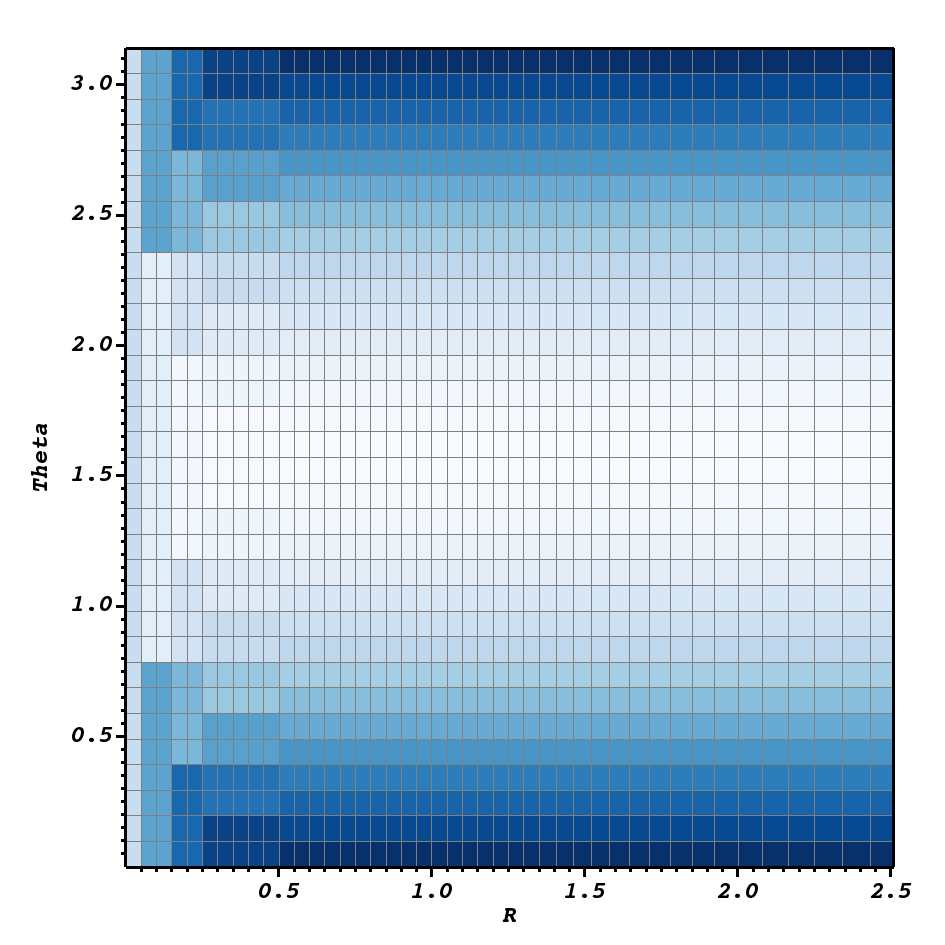}
\hspace{0.2in}
\includegraphics[height=3in]{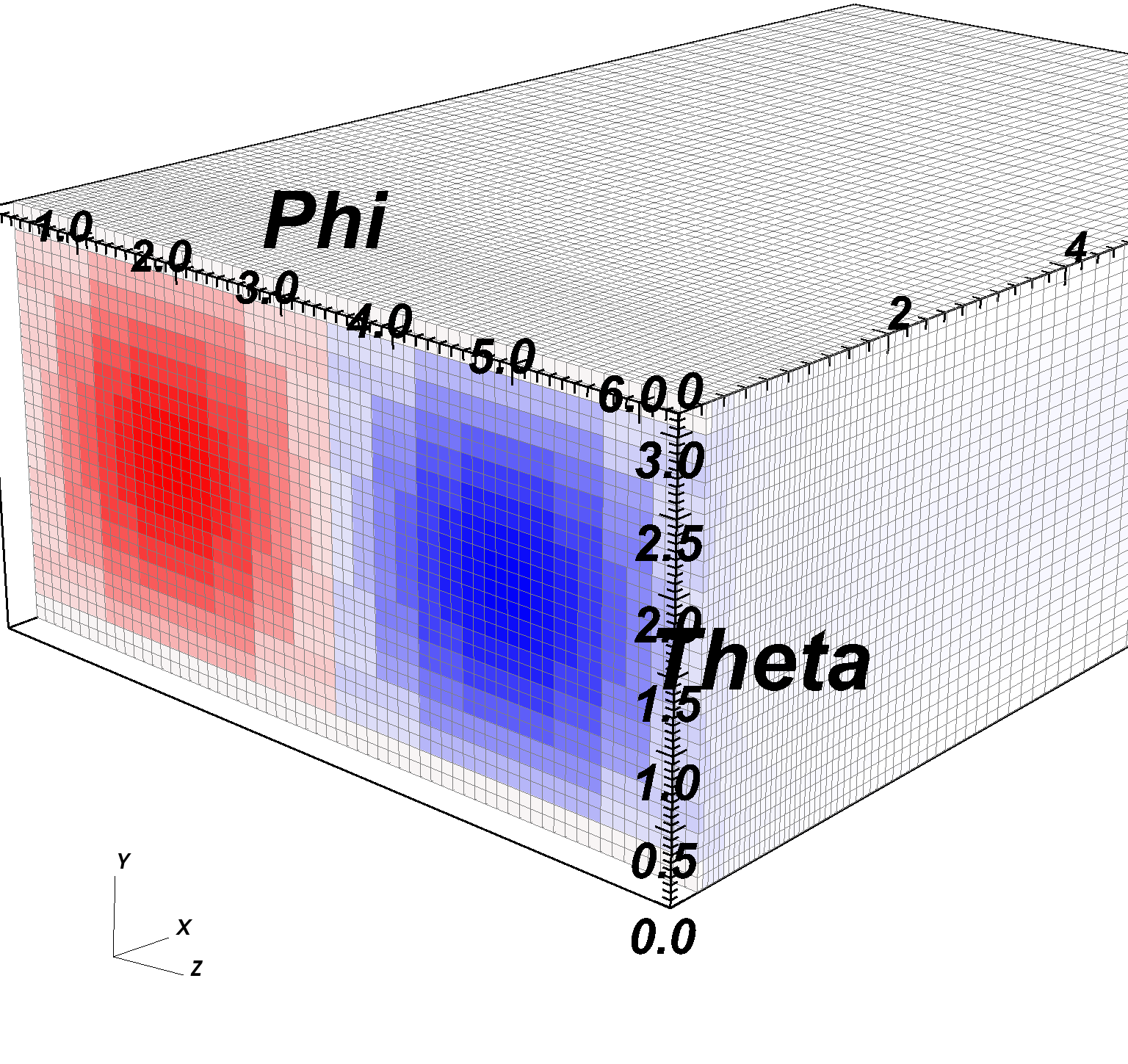}
\caption{An example field, sinusoidal in polar (2D, 3D) and azimuthal (3D) angles and displayed in coordinate space, has been coarsened by averaging over the blocks displayed in Figures~\ref{Fig:CoarseningBlocks_2D} and \ref{Fig:CoarseningBlocks_3D}. 
}
\label{Fig:CoarsenedField}
\end{figure*}

In the tests and examples in this paper, three-dimensional (3D) position space is regarded as a manifold described by an atlas consisting of a single chart with spherical coordinates $r$ (radial distance from the origin), $\theta$ (polar angle), and $\phi$ (azimuthal angle).
For numerical computation this chart is discretized into a mesh, a fixed structured grid of finite volumes referred to as `cells'.
Spherical coordinates are used regardless of whether spherical symmetry (effectively 1D), axial symmetry (effectively 2D), or no symmetry (full 3D) is assumed.  
This discretization differs from Paper I, in which a level-by-level approach to cell-by-cell mesh refinement of a single coordinate chart is described. 
Paper I includes examples of a Sedov blast wave evolved on a fixed centrally refined multi-level mesh of a type suitable for gravitational collapse, using spherical coordinates in 1D, cylindrical coordinates in 2D, and rectangular coordinates in 3D.
While our interest in this type of refinable mesh continues, the work described here is restricted to a single-level mesh using spherical coordinates regardless of the symmetry assumed in, and the corresponding effective dimensionality of, a particular computation.

A spherical coordinate mesh is naturally adapted to centrally condensed systems generally, and can be particularly useful for core-collapse supernovae. 
Without the complications of mesh refinement it can cover the large dynamic range in radius incident to the gravitational collapse of a massive stellar core and the subsequent propagation to large radius of the shock wave produced by core bounce.
In the work presented here the polar angle dimension is divided into $N_\theta$ cells of equal coordinate width $\Delta \theta = \pi / N_\theta$, while the azimuthal angle dimension is divided into $N_\phi = 2 N_\theta$ cells of equal coordinate width $\Delta \phi = 2 \pi / N_\phi = \Delta \theta$.
In terms of radius, the computational domain is conceptually divided by a fixed radius $r = r_\mathrm{core}$ into an inner region intended to resolve the `core' of the system and an outer region that provides economical coverage to large radius. 
The radial extent of the core region is covered by $N_\mathrm{core}$ radial cells of uniform radial cell width $\Delta r_\mathrm{min} = r_\mathrm{core} / N_\mathrm{core}$.
Outside the core $\Delta r \propto r$, yielding a constant polar/radial cell aspect ratio $r_\mathrm{core} \, \Delta \theta / \Delta r_\mathrm{min}$.
Setting $N_\theta / N_\mathrm{core} = 32 / 25$ (corresponding to 128 cells in $\theta$ for every 100 cells in $r$ covering the core) yields $r_\mathrm{core} \, \Delta \theta / \Delta r_\mathrm{min} = \pi \, N_\mathrm{core} / N_\theta \approx 2.45$; this deviation from unity prioritizes computational effort on the resolution of radial variations.
A final parameter is the ratio $R_\mathrm{radial} = N_r / N_\theta$, which can be tuned to yield an outer boundary $r = r_\mathrm{out}$ close to some desired target value, and which determines the total number $N_r$ of radial cells.  
In resolution studies we fix the core radius $r_\mathrm{core}$, the aspect ratio parameter $N_\theta / N_\mathrm{core}$, and the radial ratio $R_\mathrm{radial}$, thereby determining $N_\mathrm{core}$ and $N_r$ based on the single varying parameter $N_\theta$.
Example 2D and 3D meshes of limited size and low resolution are shown for illustrative purposes in Figures~\ref{Fig:CoarseningBlocks_2D} and \ref{Fig:CoarseningBlocks_3D} respectively.

Strictly speaking a single spherical coordinate chart does not constitute a mathematically proper atlas because of the coordinate singularities at the origin and along the polar axis, and these singularities also cause numerical issues even in a discrete context.
Of particular concern is that the stability of explicit schemes requires that time steps be smaller than cell widths divided by the signal speed (e.g. the sound speed of a fluid). 
While $\Delta r_\mathrm{min}$ is uniform for $r < r_\mathrm{core}$, the polar cell width $r \, \Delta \theta$ tends to zero as $r \rightarrow 0$, and the azimuthal cell width $r \, \sin \theta \, \Delta \phi$ tends to zero as $\theta \rightarrow 0$ and $\theta \rightarrow \pi$ for all $r$.
This difficulty has been dealt with in a number of ways; see for instance \citet{Wongwathanarat2010An-axis-free-ov,Asaithambi2017A-note-on-a-con,Skinner2019FORNAX:-A-Flexi,Muller2020Hydrodynamics-o,Ji2023Ameliorating-th}.
Here we ameliorate this with a coarsening scheme that allows $\Delta r_\mathrm{min}$ to be the minimum width governing explicit time steps: when angular cell widths fall below $\Delta r_\mathrm{min}$ they are grouped into angular blocks at each radius, as illustrated in Figures~\ref{Fig:CoarseningBlocks_2D} and \ref{Fig:CoarseningBlocks_3D}, over which averages are performed in order to suppress small-wavelength perturbations.
For illustrative purposes the coarsening of fields that are sinusoidal in angle are shown in Figure~\ref{Fig:CoarsenedField}.
In order to avoid communication between MPI tasks in these coarsening operations, the domain is decomposed into full-solid-angle radial shells so that each MPI task contains all the angular data for its share of radial cells.
We also mitigate cost by distributing the coarsening blocks over CPU cores or GPU threads for averaging.
Nevertheless, in 3D adiabatic collapse runs utilizing GPUs,
coarsening operations take up to $10 \%$ of the solver time. 
While apparently not 
negligible in cost, this is a small price to pay for significantly increased time steps.
Additional details about the application of coarsening to fluid dynamics are given in \S\ref{sec:FluidDynamics}.

\subsection{Poisson Solver}
\label{sec:Poisson}

The Poisson equation for the potential $\Phi(\mathbf{x})$ generated by a source $S(\mathbf{x})$ is
\begin{equation}
\nabla^2\Phi = S,
\label{eq:PoissonEquation}
\end{equation}
and for a finite isolated source the Green function solution
\begin{equation}
\Phi(\mathbf{x})=-\frac{1}{4\pi}\int_{\mathbb{R}^3} \mathrm{d}\mathbf{x'}\,\frac{S(\mathbf{x'})}{|\mathbf{x}-\mathbf{x'}|}
\label{eq:IntPoisson}
\end{equation}
vanishes at spatial infinity. 
Direct numerical integration would be an $N^2$ proposition for $N = N_r N_\theta N_\phi$ total spatial cells in a discretized domain in spherical coordinates $( r, \theta, \phi) = (r, \mathbf{\Omega})$: the solution for \textit{each} spatial cell would require a sum over \textit{all} cells.
The cost can be reduced by introducing the multipole expansion
\begin{equation}
\frac{1}{|\mathbf{x}-\mathbf{x'}|}=\sum_{\ell=0}^\infty \sum_{m=-\ell}^\ell\frac{4\pi}{2\ell+1}\frac{r^\ell_<}{r^{\ell+1}_>}Y_{\ell m}(\mathbf{\Omega}) \, Y^*_{\ell m}(\mathbf{\Omega}'),
\label{eq:Multipole}
\end{equation}
where $r_<$ ($r_>$) is the smaller (larger) of $|\mathbf{x}|$ and $|\mathbf{x'}|$, and $Y_{\ell m}(\mathbf{\Omega})$ are the spherical harmonics.
At first blush, it may seem that truncation at $\ell = L$ results in an operation count proportional to $(L+1)^2 N_r^2 N_\theta N_\phi$, already significantly less than $N^2$ when $(L+1)^2 \ll N_\theta N_\phi$.
However, as will be seen below, it turns out that the moments of $S$ at different radii are not independent, but can be obtained in the course of outward and inward radial integrations, further reducing the cost by an additional factor of $N_r$ to be $\propto (L+1)^2 N_r N_\theta N_\phi$ \citep{Muller:1995}. 
And for quasi-spherical source distributions, results of desired accuracy can be obtained with a  modest number of multipole terms ($L \approx 10-20$).

\change{The basic outline of our approach is very much like that of \citet{Muller:1995}, but we follow the FLASH code\footnote{\url{https://flash.rochester.edu/site/flashcode/user\_support/flash\_ug\_devel.pdf}} in expressing the sum over the $m$-values of spherical harmonics for a given $\ell$ in terms of real angular functions.}
The spherical harmonics, defined in terms of the associated Legendre polynomials $P_\ell^m(x)$ as
\begin{equation}
Y_{\ell m}(\theta,\phi) = \sqrt{ \frac{ 2\ell + 1 }{ 4\pi } \frac{ (\ell - m)! }{ (\ell + m)! } } \; P_\ell^m(\cos\theta) \; \mathrm{e}^{\mathrm{i} m \phi},
\end{equation}
satisfy
\begin{equation}
Y_{\ell,-m}(\mathbf{\Omega}) = (-1)^m \; Y_{\ell m}^*(\mathbf{\Omega}). 
\end{equation}
 This allows the sum over $m$ in Equation~(\ref{eq:Multipole}) to be expressed as
 \begin{eqnarray}
 & &\sum_{m=-\ell}^\ell Y_{\ell m}(\mathbf{\Omega}) \, Y^*_{\ell m}(\mathbf{\Omega}') \nonumber \\
 & &\hphantom{ \sum_{m=-\ell}^\ell }  = \sum_{m=0}^\ell \sum_{a \in \{c, s\}}  \Delta_m \, A_{\ell m}^a (\mathbf{\Omega}) \, A_{\ell m}^a (\mathbf{\Omega}'),
 \end{eqnarray}
 in which
 \begin{equation}
 \Delta_m = 2 - \delta_{m0}
 \label{eq:Delta}
 \end{equation}
 is a coefficient function, and
\begin{eqnarray}
A_{\ell m}^c (\theta,\phi) &=& \tilde{P}_\ell^m (\cos\theta) \, \cos(m\phi), 
\label{eq:AngularKernel_c} \\
A_{\ell m}^s (\theta,\phi) &=& \tilde{P}_\ell^m (\cos\theta) \, \sin(m\phi)
\label{eq:AngularKernel_s}
\end{eqnarray}
are `cosine' and `sine' angular kernels indexed by $a = c, s$ respectively.
The renormalized associated Legendre polynomials
\begin{equation}
 \tilde{P}_\ell^m (x) =   \sqrt{ \frac{ 2\ell + 1 }{ 4\pi } \frac{ (\ell - m)! }{ (\ell + m)! } } \; P_\ell^m (x)  
\end{equation}
are more suitable for numerical recursion, being less prone to errors from nearly cancelling terms \citep{Press2007Numerical-Recip}.
For a given $m$, recursion in $\ell$ begins with the values
\begin{eqnarray}
\tilde{P}_m^m (x) &=& (-1)^m \sqrt{ \frac{ 2m + 1 }{ 4\pi  (2m)! } } \; (2m - 1)!!  \nonumber \\
& & \times (1-x^2)^{m/2}, \\
\tilde{P}_{m+1}^m (x) &=& x \, \sqrt{ 2m + 3 } \; \tilde{P}_m^m (x)
\end{eqnarray}
and continues with
\begin{eqnarray}
\tilde{P}_\ell^m (x) &=& \sqrt{ \frac{ 4 \ell^2 - 1 }{ \ell^2 - m^2 } } \left[ x \, \tilde{P}_{\ell-1}^m (x) \vphantom{ \sqrt{ \frac{ (\ell - 1)^2 - m^2 }{ 4(\ell - 1)^2 - 1 } } } \right. \nonumber \\
& &\left. - \sqrt{ \frac{ (\ell - 1)^2 - m^2 }{ 4(\ell - 1)^2 - 1 } } \tilde{P}_{\ell-2}^m (x) \right].
\end{eqnarray}
For modest $L$ and with their purely angular dependence, the kernels $A_{\ell m}^c (\mathbf{\Omega})$ and $A_{\ell m}^s (\mathbf{\Omega})$ (averaged over each angular cell) impose a sufficiently small storage burden that they can be computed once and for all at the beginning of a simulation.

Having defined real angular kernel functions, angular moments of the source $S(r,\mathbf{\Omega})$ in each radial shell can be defined.
For each $0 \le \ell \le L$, and $0 \le m \le \ell$, and $a \in \{c, s\}$, these are 
\begin{equation}
\mathcal{A}_{\ell m}^a(r') = \int_{S^2} \mathrm{d}\mathbf{\Omega}' \;  A_{\ell m}^a (\mathbf{\Omega}') \; S(r',\mathbf{\Omega}').
\label{eq:AngularMomentsContinuum}
\end{equation}
In a finite volume discretization this can be approximated as
\begin{equation}
\left( \mathcal{A}_{\ell m}^a \right)_{\bar\imath} 
  = \sum_{\bar\jmath \bar k} \left( \Delta\mathbf{\Omega} \right)_{\bar\jmath \bar k} \,
  \left( A_{\ell m}^a \right)_{\bar\jmath \bar k} \, 
  S_{\bar\imath \bar\jmath \bar k},
\label{eq:AngularMomentsDiscrete} 
\end{equation}
where $\bar\imath,\bar\jmath,\bar k$ respectively index cell-averaged values in the $r,\theta,\phi$ dimensions.
The cell-averaged angular kernel functions $\left( A_{\ell m}^a \right)_{\bar\jmath \bar k}$ are computed and stored at the beginning of a run.
The averaging is performed numerically, via Romberg integration.

Computation of the angular moments via Equation~(\ref{eq:AngularMomentsDiscrete}) with hardware accelerators (GPUs) using an OpenMP \texttt{target} directive requires special attention.
The computation is a reduction of an effectively dimension-6 entity ($a,\ell,m,\bar\imath,\bar\jmath,\bar k$)  to a dimension-4 entity ($a,\ell,m,\bar\imath $) via summation over two dimensions, the position space angular bins ($\bar\jmath,\bar k$). 
In practice we reorganize this dimensionality by collapsing the indices ($a,\ell,m$) to a single `angular moment index' \texttt{iAM}; this converts Equation~(\ref{eq:AngularMomentsDiscrete}) to reduction of an effectively dimension-4 entity ($\mathtt{iAM},\bar\imath,\bar\jmath,\bar k$) to a dimension-2 entity ($\mathtt{iAM},\bar\imath$).
Because of domain decomposition---partition of position space among MPI processes---an MPI reduction is one aspect of the parallelized computation of Equation~(\ref{eq:AngularMomentsDiscrete}).
The moment sums local to each MPI process, reduced to a dimension-2 array \texttt{MyAM}, can be parallelized using OpenMP. 
The most straightforward approach, which works for threading over CPU cores and is shown in Listing~\ref{lst:AngularMoments_CPU}, involves an OpenMP reduction on the array \texttt{MyAM} (line 19).
However, OpenMP reduction of the array \texttt{MyAM} requires the system to spontaneously allocate a large number of private arrays to perform the reduction, potentially costly in terms of memory available on the GPU. 
We avoid this by exploiting two levels of parallelism available on the GPU via OpenMP, as shown in Listing~\ref{lst:AngularMoments_GPU}: the outer two loops indexing \texttt{MyAM} are distributed over teams of threads, allowing the OpenMP reduction in the inner two loops to proceed on an intermediate scalar \texttt{MyAME}.

\begin{lstlisting}[float,frame=tb,numbers=left,numbersep=5pt,xleftmargin=10pt,label=lst:AngularMoments_CPU,caption={Computing angular moments on the CPU.}]
integer(KDI) :: &
  iR, &   !-- iRadius
  iT, &   !-- iTheta
  iP, &   !-- iPhi
  iAM, &  !-- iAngularMoment
  nAM     !-- nAngularMoments
integer(KDI), dimension(3) :: &
  nC  !-- nCells
real(KDR), dimension(:,:) :: &
  dSA, &  !-- dSolidAngle
  MyAM    !-- MyAngularMoment
real(KDR), dimension(:,:,:) :: &
  A, &  !-- Angular kernel functions
  S     !-- Source

!-- MyM already initialized to 0.0\_KDR

!\$OMP parallel do collapse ( 4 ) \&
!\$OMP reduction ( + : MyAM )
do iAM = 1, nAM
 do iP = 1, nC(3)
  do iT = 1, nC(2) 
   do iR = 1, nC(1)
      
     MyAM(iR,iAM) = &
       MyAM(iR,iAM) &
       + dSA(iT,iP) * A(iT,iP,iAM) * S(iR,iT,iP)

   end do !-- iR
  end do !-- iT
 end do !-- iP
end do !-- iAM
!\$OMP  end parallel do
\end{lstlisting}

\begin{lstlisting}[float,frame=tb,numbers=left,numbersep=5pt,xleftmargin=10pt,label=lst:AngularMoments_GPU,caption={Computing angular moments on the GPU.}]
!-- Declarations in Listing~\ref{lst:AngularMoments_CPU} omitted
real(KDR) :: &
  MyAME  !-- MyAngularMomentElement

!\$OMP target teams distribute collapse ( 2 ) \&
!\$OMP private ( MyAME )
do iAM = 1, nAM
 do iR = 1, nC(1)
 
   MyME = 0.0_KDR
  
  !\$OMP parallel do collapse ( 2 ) \&
  !\$OMP reduction ( + : MyAME ) \&
  do iP = 1, nC(3) 
   do iT = 1, nC(2)
      
     MyAME = &
       MyAME &
       + dSA(iT,iP) * A(iT,iP,iAM) * S(iR,iT,iP)

   end do !-- iT
  end do !-- iP
  !\$OMP end parallel do

  MyAM(iR,iAM) = MyAME
  
 end do !-- iR
end do !-- iAM
!\$OMP end target teams distribute

\end{lstlisting}

The angular moments must be integrated into radial moments, which then can be assembled into the final solution.
In terms of the angular moments $\mathcal{A}_{\ell m}^a(r')$ defined in Equation~(\ref{eq:AngularMomentsContinuum}), the angular kernel functions $A_{\ell m}^a (\mathbf{\Omega})$ in Equations~(\ref{eq:AngularKernel_c})-(\ref{eq:AngularKernel_s}), and $\Delta_m$ in Equation~(\ref{eq:Delta}), the potential as given in Equations~(\ref{eq:IntPoisson}) and (\ref{eq:Multipole}) reads 
\begin{eqnarray}
\Phi(r, \mathbf{\Omega}) &=& - \sum_{\ell=0}^\infty  \sum_{m=0}^\ell \sum_{a \in \{c, s\}}  \frac{\Delta_m}{2\ell + 1} A_{\ell m}^a (\mathbf{\Omega}) \nonumber \\
& & \times \int_0^\infty dr' \, r'^2 \, \frac{r_<^\ell}{r_>^{\ell+1}} \, \mathcal{A}_{\ell m}^a(r').
\end{eqnarray}
The radial integral is broken into two pieces by defining the regular $(\mathcal{R})$ and irregular $(\mathcal{I})$ radial kernel functions
\begin{eqnarray}
R_\ell^\mathcal{R}(r) &=& r^\ell, \label{eq:RadialRegular} \\
R_\ell^\mathcal{I}(r) &=& \frac{1}{r^{\ell+1}}  \label{eq:RadialIrregular}
\end{eqnarray}
and the corresponding radial moments
\begin{eqnarray}
\mathcal{M}_{\ell m}^{\mathcal{R} a}(r) &=& \int_0^r  dr' \, r'^2 \, R_\ell^\mathcal{R}(r')  \, \mathcal{A}_{\ell m}^a(r'), \\
\mathcal{M}_{\ell m}^{\mathcal{I} a}(r) &=& \int_r^\infty  dr' \, r'^2 \, R_\ell^\mathcal{I}(r')  \, \mathcal{A}_{\ell m}^a(r'),
\end{eqnarray}
in terms of which
\begin{eqnarray}
\Phi(r, \mathbf{\Omega}) &=& - \sum_{\ell=0}^\infty  \sum_{m=0}^\ell \sum_{a \in \{c, s\}}  \sum_{\scriptr \in \{\mathcal{R},\mathcal{I} \}}  \frac{\Delta_m}{2\ell + 1} \nonumber \\
& & \times  A_{\ell m}^a (\mathbf{\Omega}) \, R_\ell^\scriptr (r) \, \mathcal{M}_{\ell m}^{(\neg\scriptr) a}(r), \label{eq:Solution}
\end{eqnarray}
in which the toggled values $\neg \scriptr$ of the regular/irregular indices are $\neg \mathcal{R} = \mathcal{I}$ and $\neg \mathcal{I} = \mathcal{R}$.
On the discretized radial grid---consisting of $N_r$ radial shells (indexed as $1 \le \bar\imath \le N_r$) with $N_r + 1$ shell edges (indexed as  $1 \le i \le N_r + 1$)---the radial moments are initially computed on the shell edges.
The regular moments are computed by outward integration,
\begin{eqnarray}
\left( \mathcal{M}_{\ell m}^{\mathcal{R} a} \right)_{1} &=& 0,  \\
\left( \mathcal{M}_{\ell m}^{\mathcal{R} a} \right)_{i+1} &=& \left( \mathcal{M}_{\ell m}^{\mathcal{R} a} \right)_{i} + \left( \Delta\left( \frac{r^3}{3} \right) \right)_{\bar\imath} \left( R_\ell^\mathcal{R}\right)_{\bar\imath} \left( \mathcal{A}_{\ell m}^a \right)_{\bar\imath}, \nonumber
\end{eqnarray}
while the irregular moments are computed by inward integration,
\begin{eqnarray}
\left( \mathcal{M}_{\ell m}^{\mathcal{I} a} \right)_{N_r + 1} &=& 0,  \\
\left( \mathcal{M}_{\ell m}^{\mathcal{I} a} \right)_{i} &=& \left( \mathcal{M}_{\ell m}^{\mathcal{I} a} \right)_{i+1} + \left( \Delta\left( \frac{r^3}{3} \right) \right)_{\bar\imath} \left( R_\ell^\mathcal{I}\right)_{\bar\imath} \left( \mathcal{A}_{\ell m}^a \right)_{\bar\imath}. \nonumber
\end{eqnarray}
The cell-averaged radial kernel functions $\left( R_\ell^\mathcal{R}\right)_{\bar\imath}$ and $\left( R_\ell^\mathcal{I}\right)_{\bar\imath}$ are computed and stored at the beginning of a run.
The averaging is performed analytically based on Equations~(\ref{eq:RadialRegular}) and (\ref{eq:RadialIrregular}).
Cell-averaged values $\left( \mathcal{M}_{\ell m}^{\scriptr a} \right)_{\bar\imath}$ are obtained from the surrounding edge values $\left( \mathcal{M}_{\ell m}^{\scriptr a} \right)_{i}$ and $\left( \mathcal{M}_{\ell m}^{\scriptr a} \right)_{i+1}$ by linear interpolation to the cell center, and are used in the cell-averaged discretization of Equation~(\ref{eq:Solution}) to obtain solution values $\Phi_{\bar\imath \bar\jmath \bar k}$.
In our experience this procedure seems sufficient to avoid the pathology discussed by \citet{Couch:2013}.

\subsection{Fluid Dynamics of Baryonic Matter}
\label{sec:FluidDynamics}

Fluid dynamics is governed by hyperbolic balance equations of the form
\begin{equation}
  \frac{\partial\mathcal{U}}{\partial t}+\bm{\nabla} \cdot {\mathcal{F}(\mathcal{U})}
  =\mathcal{S}(\mathcal{U}), \label{eq:BalanceEquation}
\end{equation}
where $\mathcal{U}$ is a vector of `conserved' or `balanced' fields and $\mathcal{F}(\mathcal{U})$ and $\mathcal{S}(\mathcal{U})$ are the fluxes and sources. 
As described in Paper I, our treatment of fluid dynamics is based on explicit Runge-Kutta time integration of ordinary differential equations obtained from finite-volume discretization (method of lines), along with solution of a Riemann problem at each cell interface.
Changes relative to Paper I include our reconstruction method, time step determination, measures taken to ameliorate coordinate singularities, and implementation of a tabulated equation of state for baryonic matter.

`Reconstruction' refers to the process of obtaining field values at cell interfaces---needed to compute fluxes between cells---from the cell-averaged values solved for in the finite-volume approach.
We now use a reconstruction method similar to that described by \cite{Skinner2019FORNAX:-A-Flexi}.
Consider a field $f$ with cell-averaged values $f_{\bar\imath \bar\jmath \bar k}$, and focus on a particular dimension for which cell-averaged values are indexed by $\bar{q} \in \{\bar\imath, \bar\jmath, \bar k\}$.
The results presented in this paper are obtained with parabolic reconstruction: within each cell $\bar{q}$, the field $f$ is represented by a parabola 
\begin{equation}
f(u) = a + b \, u + c \, u^2.
\label{eq:Parabola}
\end{equation}
Here $u$ is the (generally curvilinear) coordinate in the given dimension.
The constants $a, b, c$ are determined from the cell-averaged value $\langle f \rangle_0 = f_{\bar{q}}$ and the values $\langle f \rangle_- = f_{\bar{q} - 1}$ and $\langle f \rangle_+ = f_{\bar{q}+1}$ of its nearest neighbors, subject to the requirement---important for stability---that the parabola be monotonic within the cell.
Once the final monotonized parabola is obtained, the reconstructed face values $f_{\leftarrow 0} = f ( u_q )$ and $f_{0 \rightarrow} = f ( u_{q+1} )$ are given by Equation~(\ref{eq:Parabola}); here $u_q$ and $u_{q+1}$ are the coordinate values on the inner and outer faces respectively of cell $\bar q$.
The reconstructed face values $f_{- \rightarrow}$ (determined from the reconstruction in the left neighbor) and $f_{\leftarrow 0}$ are elements of the `left' and `right' states used by the Riemann solver to determine fluxes at interface $q$, the outer face of cell $\bar q - 1$ and the inner face of cell $\bar q$.
Similarly, the reconstructed face values $f_{0 \rightarrow}$ and $f_{\leftarrow +}$ are elements of the `left' and `right' states used by the Riemann solver to determine fluxes at interface $q+1$, the outer face of cell $\bar q$ and the inner face of cell $\bar q + 1$.
 
The monotonicity requirement results in a three-step reconstruction procedure.
The first step is to check if $\langle f \rangle_0$ is an extremum relative to $\langle f \rangle_-$ and $\langle f \rangle_+$, in which case constant reconstruction is employed: $a = \langle f \rangle_0$ and $b = c = 0$.
Otherwise, in each of the following two steps a parabola is constructed and then adjusted---`flattened', in fact, similar in spirit to the  `slope limiting' of linear reconstruction---if required by monotonicity.
As will be seen below, the cell-averaged relation
\begin{equation}
\langle f \rangle_0 = a + b \, \langle u \rangle_0 + c \, \langle u^2 \rangle_0
\end{equation}
of cell $\bar{q}$ is enforced throughout.
For a fixed mesh the coordinate averages $\langle u \rangle$ and $\langle u^2 \rangle$ for all cells can be computed and stored once and for all upon initialization.

The second step is to construct a parabola spanning the cell and its neighbhors from the solution of the system
\begin{equation}
\begin{bmatrix}
1 & \langle u \rangle_- & \langle u^2 \rangle_- \\
1 & \langle u \rangle_0 & \langle u^2 \rangle_0 \\
1 & \langle u \rangle_+ & \langle u^2 \rangle_+ \\
\end{bmatrix}
\begin{bmatrix}
a \\ b \\ c 
\end{bmatrix}
= 
\begin{bmatrix}
\langle f \rangle_- \\ \langle f \rangle_0 \\ \langle f \rangle_+
\end{bmatrix}.
\label{eq:FirstParabola}
\end{equation}
The resulting parabola yields candidate face values $f_{\leftarrow 0}$ and $f_{0 \rightarrow}$, but an adjustment is required if the parabola's extremum at $u_E = - b \, / \, 2 \, c$ falls between the inner edge $u_{\leftarrow -}$ of the left neighbor and the outer edge $u_{+ \rightarrow}$ of the right neighbor.
Let $u_0 = ( u_{\leftarrow 0} + u_{0 \rightarrow}) / 2$. 
If $u_{\leftarrow -} < u_E < u_0$, a revised value of $f_{\leftarrow 0}$ is obtained by imposing a zero slope at $u_{\leftarrow -}$, in accord with the solution of the system
\begin{equation}
\begin{bmatrix}
0 & 1 & 2 \, u_{\leftarrow -} \\
1 & \langle u \rangle_- & \langle u^2 \rangle_- \\
1 & \langle u \rangle_0 & \langle u^2 \rangle_0 \\
\end{bmatrix}
\begin{bmatrix}
a \\ b \\ c 
\end{bmatrix}
= 
\begin{bmatrix}
0 \\ \langle f \rangle_- \\ \langle f \rangle_0
\end{bmatrix}.
\end{equation}
If instead $u_0 < u_E < u_{+ \rightarrow}$, a revised value of $f_{0 \rightarrow}$ is obtained by imposing a zero slope at $u_{+ \rightarrow}$, in accord with the solution of the system
\begin{equation}
\begin{bmatrix}
1 & \langle u \rangle_0 & \langle u^2 \rangle_0 \\
1 & \langle u \rangle_+ & \langle u^2 \rangle_+ \\
0 & 1 & 2 \, u_{+ \rightarrow} \\
\end{bmatrix}
\begin{bmatrix}
a \\ b \\ c 
\end{bmatrix}
= 
\begin{bmatrix}
\langle f \rangle_0 \\ \langle f \rangle_+ \\ 0
\end{bmatrix}.
\end{equation}
This completes the second step. 

A third step is necessary if $f_{\leftarrow 0}$ or $f_{0 \rightarrow}$ has been revised in order to enforce monotonicity in the second step, because in this case the new parabola determined by 
\begin{equation}
\begin{bmatrix}
1 & u_{\leftarrow 0} & u_{\leftarrow 0}^2  \\
1 & \langle u \rangle_0 & \langle u^2 \rangle_0 \\
1 & u_{0 \rightarrow} & u_{0 \rightarrow}^2  \\
\end{bmatrix}
\begin{bmatrix}
a \\ b \\ c 
\end{bmatrix}
= 
\begin{bmatrix}
f_{\leftarrow 0} \\ \langle f \rangle_0 \\ f_{0 \rightarrow}
\end{bmatrix}.
\label{eq:SecondParabola}
\end{equation}
is no longer the same as the first parabola determined by Equation~(\ref{eq:FirstParabola}).
In particular a correction to this new parabola is required if its extremum lies within the cell $\bar{q}$, that is, if $u_{\leftarrow 0} < u_E < u_{0 \rightarrow}$.
If $u_{\leftarrow 0} < u_E < u_0$ (extremum in the left half of the cell), it would have been $f_{\leftarrow 0}$ that was revised in the second step. 
At this point the verbal description in \cite{Skinner2019FORNAX:-A-Flexi} suggests that $f_{\leftarrow 0}$ be revised again with a zero slope condition at $u_{\leftarrow 0}$, presumably keeping $f_{0 \rightarrow}$ fixed; but we find that this tends to undo the flattening achieved by the revision of $f_{\leftarrow 0}$ in the second step, and can lead to instability.
Therefore we keep $f_{\leftarrow 0}$ fixed while imposing the zero slope condition at $u_{\leftarrow 0}$, and achieve additional flattening by revising $f_{0 \rightarrow}$ according to the solution of the system 
\begin{equation}
\begin{bmatrix}
0 & 1 & 2 \, u_{\leftarrow 0} \\
1 & u_{\leftarrow 0} & u_{\leftarrow 0}^2 \\
1 & \langle u \rangle_0 & \langle u^2 \rangle_0 \\
\end{bmatrix}
\begin{bmatrix}
a \\ b \\ c 
\end{bmatrix}
= 
\begin{bmatrix}
0 \\ f_{\leftarrow 0} \\ \langle f \rangle_0
\end{bmatrix}.
\end{equation}
If instead $u_0 < u_E < u_{\leftarrow 0}$ (extremum in the right half of the cell), we keep $f_{0 \rightarrow}$ fixed while imposing the zero slope condition at $u_{0 \rightarrow}$, and achieve additional flattening by revising $f_{\leftarrow 0}$ according to the solution of the system 
\begin{equation}
\begin{bmatrix}
1 & \langle u \rangle_0 & \langle u^2 \rangle_0 \\
1 & u_{0 \rightarrow} & u_{0 \rightarrow}^2 \\
0 & 1 & 2 \, u_{0 \rightarrow} \\
\end{bmatrix}
\begin{bmatrix}
a \\ b \\ c 
\end{bmatrix}
= 
\begin{bmatrix}
\langle f \rangle_0 \\ f_{\leftarrow 0} \\ 0
\end{bmatrix}.
\end{equation}
This completes the third and final reconstruction step.

Turning from reconstruction to the Courant-Friedrichs-Lewy (CFL) condition for stable explicit evolution of the fluid, we note small changes relative to Paper I.
A CFL-restricted time step can be expressed
\begin{equation}
\Delta t = C \, \Delta t_\mathrm{CFL}.
\label{eq:CourantFactor}
\end{equation}
In Paper I we determined $\Delta t_\mathrm{CFL}$ by finding a minimum over the dimensions $i$ in each cell and then taking the minimum over all cells:
\begin{equation}
\Delta t_\mathrm{CFL} 
	= \min_\mathrm{cells} \left( \min_i 
		\left( \frac{\Delta u_i}{ \max( | \lambda_i^- |,  | \lambda_i^+ | ) } \right) \right),
\end{equation}
where $\Delta u_i$ is a cell coordinate width and $| \lambda_i^- |,  | \lambda_i^+ | $ are the characteristic coordinate speeds in dimension $i$.
In this approach the Courant factor $C$ depends on the number of dimensions $d$ according to $C \lesssim 1/ d$.
We now instead take
\begin{equation}
{\Delta t_\mathrm{CFL}}^{-1} 
	= \max_\mathrm{cells} \left( \sum_{i=1}^d  
		\frac{ \max( | \lambda_i^- |,  | \lambda_i^+ | ) }{c_i \, \Delta u_i} \right),
\end{equation}
so that the `$1/d$' is in effect built into $\Delta t_\mathrm{CFL}$ itself.
In this way the user-tuned parameter $C \lesssim 1$ is now more straightforwardly uniform across different runtime dimensionalities of a given problem.

This new expression for $\Delta t_\mathrm{CFL}$ also makes available coarsening factors $c_i$ to reduce the impact on the time step of cells near coordinate singularities that are gathered into `coarsening blocks' as described in Section~\ref{sec:Mesh} ($c_i$ being equal to the number of cells in dimension $i$ of the coarsening block to which a cell belongs).
In each Runge-Kutta substep the updates of balanced density variables are volume-averaged over these coarsening blocks: for each cell $i$ in a particular coarsening block,
\begin{equation}
K_{i \, \in \, \text{block}} 
	= \frac{ \sum_{j \, \in \, \text{block}} V_j \, K_j }
		{ \sum_{j \, \in \, \text{block}} V_j },
\end{equation}
where the $K_j$ are the Runge-Kutta updates of balanced densities and the $V_j$ are the cell volumes.
This preserves the conservative nature of our finite-volume solver.

In another concession to the presence of coordinate singularities in our spherical coordinate mesh, we include the option of zeroing out the lateral (polar and azimuthal) momenta near the origin and the polar axis.
We did not find this necessary for our test problems with `dust' (a fluid with vanishing pressure), but otherwise we typically zero out lateral momenta in the first polar cell next to the axis and within radius $0.1 \, r_\mathrm{core}$ (see Section~\ref{sec:Mesh} for the meaning of $r_\mathrm{core}$).

Another change in our treatment of fluid dynamics relative to Paper I is the capability to evolve baryonic matter (nuclei via a representative heavy nucleus and bulk nuclear matter, along with electrons/positrons and photons).
The simulations of adiabatic stellar collapse, bounce, and explosion presented in Section~\ref{sec:AdiabaticExplosion} utilize infrastructure for tabulated equations of state based on publicly available code;\footnote{\texttt{https://stellarcollapse.org}} see also \citet{OConnor2010A-new-open-sour}.
These routines involve multidimensional interpolation and root-finding to solve for temperature from internal energy. 
While the publicly available routines accept input corresponding to a single cell, we have modified them to process a batch of cells in parallel with massive numbers of GPU threads.
The result is over 50X speedup for equation of state calls when GPUs are used.

\section{Tests}
\label{sec:Tests}

In this section we present solutions to five test problems computed with \genasis: the gravitational potential of homogeneous spheroids; 
the gravitational collapse of a homogeneous sphere and spheroid of `dust', that is, a pressureless fluid;  
the self-similar gravitational collapse of a polytropic fluid; 
and the adiabatic collapse, bounce, and explosion of numerically evolved pre-supernova progenitor stars. 
Computations were performed on Frontier at the Oak Ridge Leadership Computing Facility (OLCF) at Oak Ridge National Laboratory.

All of these tests utilize a spherical coordinate mesh in up to three position space dimensions as described in \S\ref{sec:Mesh}.
The first three tests are taken to be dimensionless, and in this case the standard mesh parameters are $N_\theta = 256$ for the homogeneous spheroids and $N_\theta = 192$ for the dust collapse problems,  
$r_\mathrm{core}=0.25$ (corresponding to $\Delta r_\mathrm{min} = 1.25 \times 10^{-3}$ or $1.67 \times 10^{-3}$), and $R_\mathrm{radial}$ = 3.68, yielding $r_\mathrm{out} \approx 1.012 \times 10^1$ or $r_\mathrm{out} \approx 1.006 \times 10^1$. 
The last two tests are computed with physical units, and in this case the standard mesh parameters are $N_\theta = 192$ and   
$r_\mathrm{core}=16~\mathrm{km}$ (corresponding to $\Delta r_\mathrm{min} \approx 1.067 \times 10^{-1}~\mathrm{km}$), and $R_\mathrm{radial}$ = 5.9, yielding $r_\mathrm{out} \approx 1.091\times 10^4~\mathrm{km}$.
\change{The choices of $r_\mathrm{core}$ and $N_\theta$ (through its connection to $N_\mathrm{core} = 25/32 \, N_\theta$) respectively reflect expectations regarding the typical size of a collapsed core in a given problem and the number of cells desired to cover it.
Given those parameters, a value of $R_\mathrm{radial} = N_r / N_\theta$ is determined that yields a suitable value for the radius of the outer boundary.}
For the problems with physical units, mesh coarsening near coordinate singularities as discussed in \S\ref{sec:Solvers} increases the time steps by a factor of
\begin{equation}
\begin{aligned}
\frac{\Delta r_\mathrm{min} }{ r_\mathrm{min} \, \Delta \theta} 
	&\approx 1.222 \times 10^2 \quad\quad\text{in 2D}, \\
\frac{\Delta r_\mathrm{min} }{ r_\mathrm{min} \, \sin\theta_\mathrm{min} \, \Delta \phi }
	&\approx 1.494 \times 10^4 \quad\quad\text{in 3D}
\end{aligned}
\end{equation}
relative to what the time steps would be without coarsening.
For the Courant factor in Equation~(\ref{eq:CourantFactor}) we use $C = 0.7$.

All of these tests involve the Newtonian gravitational potential $\Phi$, for which the source in Equation~(\ref{eq:PoissonEquation}) is
\begin{equation}
S = 4 \pi G \, \rho,
\quad\quad
\rho = m_b  N_b
\end{equation}
 where $G$ is the gravitational constant and $\rho$ is the mass density, given in terms of the average baryon mass $m_b$ and baryon number density $N_b$. 
When physical units are employed the atomic mass unit \change{$m_u$} has been used for $m_b$, while in the dimensionless problems the values $m_b = 1$ (and also $G = 1$) have been utilized.
The Poisson solver uses maximum multipole degree $L=12$ unless otherwise noted.
In setting the initial conditions for problems involving homogeneous spheroids we use a subgrid in cells at the spheroid surface to compute the cell volume fraction inside the spheroid, and reduce the density accordingly in order to numerically smooth the discontinuous edge of the density distribution. 

\begin{figure*}[!t]
\centering
\includegraphics[width=0.49\textwidth]{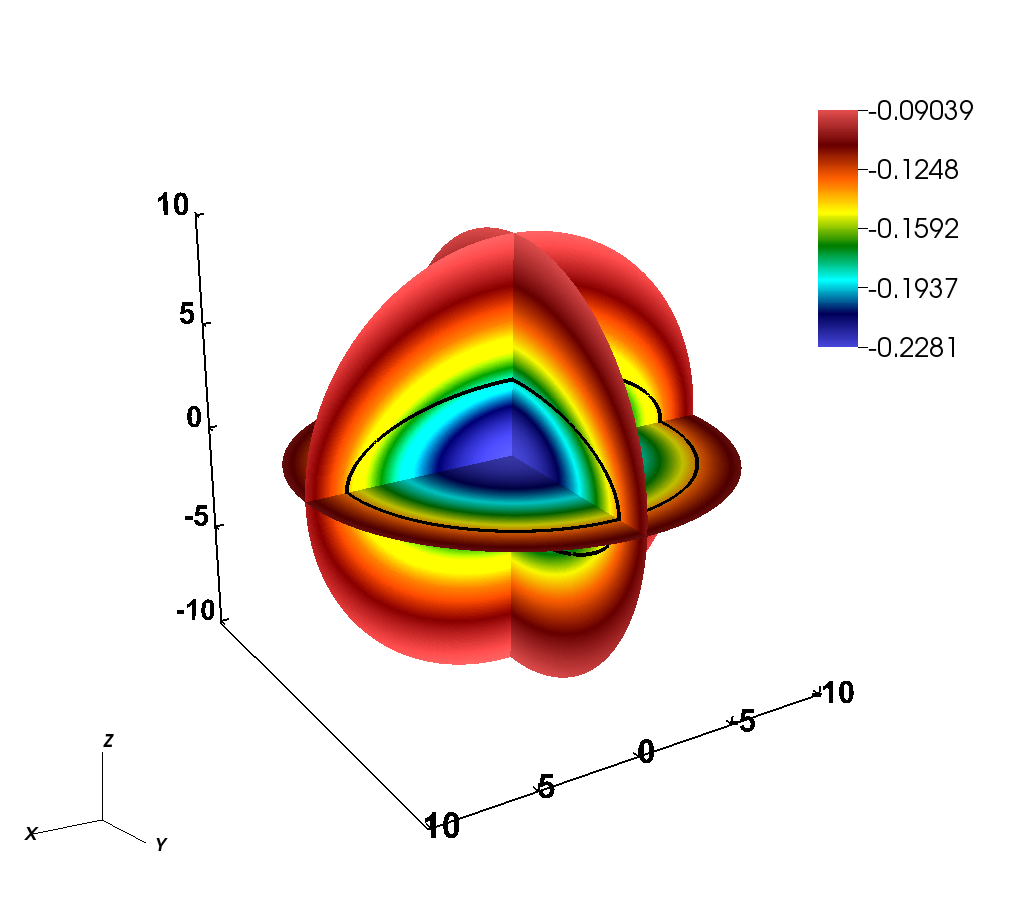}
\includegraphics[width=0.49\textwidth]{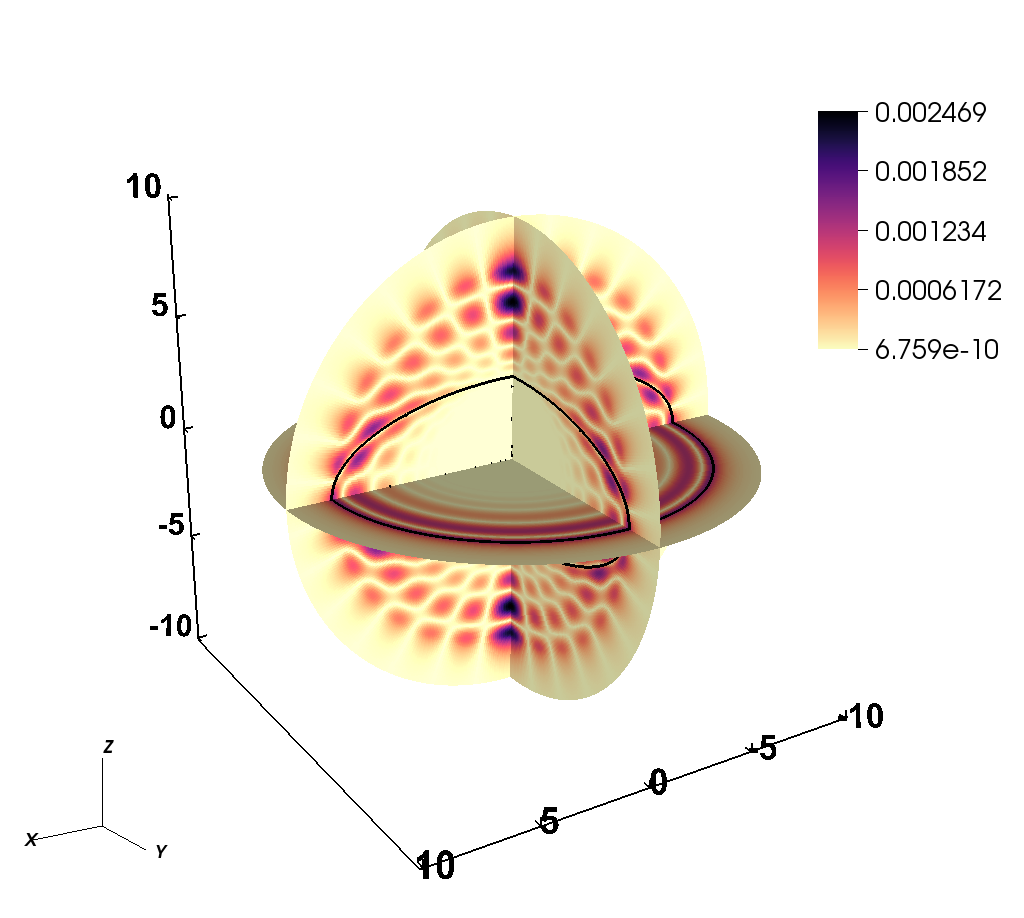}
\caption{Three-slice of the computed gravitational potential (left) and its relative error (right) for a three-dimensional oblate spheroid with $e = 0.9$. For both plots, the shape of the spheroid is outlined in black.}
\label{Fig:OblateSpheroid_Solution_Error}
\end{figure*}

\begin{figure*}
\centering
\includegraphics[width=0.49\textwidth]{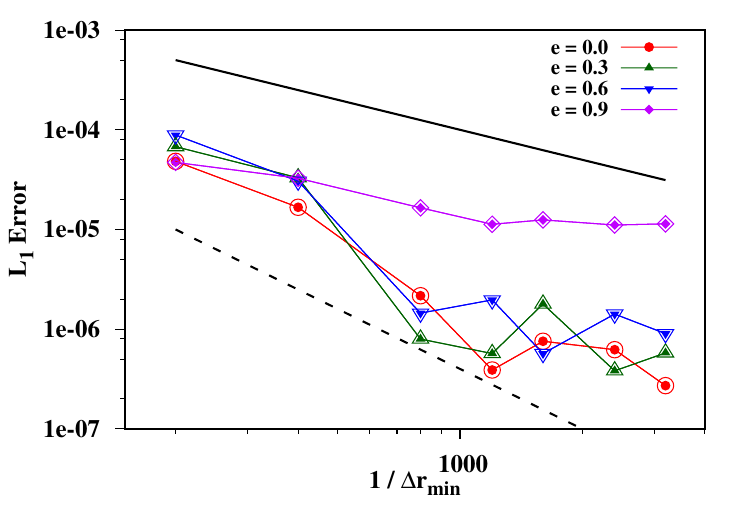}
\includegraphics[width=0.49\textwidth]{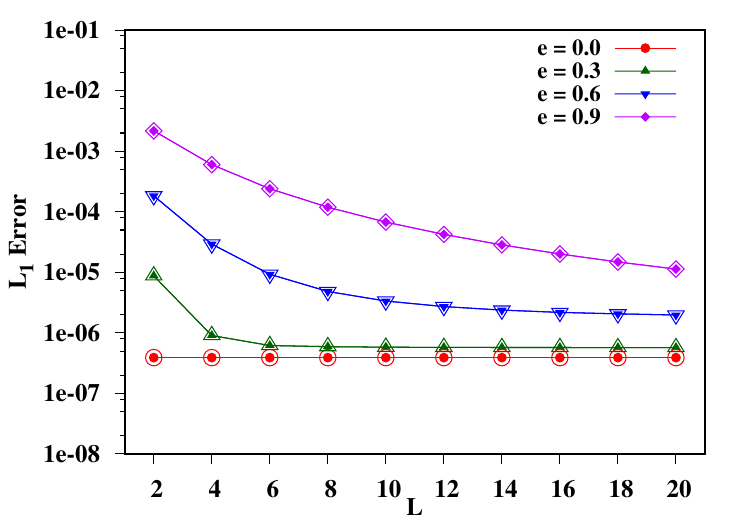}
\caption{$L_1$ error of the potential as a function of resolution (left) computed with $L=20$ and number of multipole terms $L$ (right) computed with $N_\theta = 384$ for spheroids with different eccentricities. The solid and dashed black lines are references for first- and second-order convergence respectively. }
\label{Fig:Spheroids_Convergence}
\end{figure*}

In the gravitational collapse problems the balanced variables, fluxes, and sources in Equation~(\ref{eq:BalanceEquation}) are
\begin{equation}
\mathcal{U} = \begin{bmatrix} N_b \\ \mathbf{S} \\ E \\ N_e \end{bmatrix},
\
\mathcal{F} = \begin{bmatrix} N_b \, \mathbf{v}
			\\ \mathbf{S} \otimes \mathbf{v} + p \, \mathbf{1} 
			\\ (E + p) \, \mathbf{v} 
			\\ N_e \, \mathbf{v} 
		\end{bmatrix}, 
\ 
\mathcal{S} = \begin{bmatrix} 0 
			\\ -\rho \, \bm{\nabla}\Phi 
			\\ -\rho \, \mathbf{v} \cdot \bm{\nabla}\Phi 
			\\ 0
		\end{bmatrix}.
\label{eq:BalancedVariablesGravitational}
\end{equation} 
The balanced (`lab frame') variables include the baryon number density $N_b$, the momentum 3-covector $\mathbf{S}$, the internal+kinetic energy $E$, and (for the collapse of pre-supernova progenitors) the electron number density $N_e$.
These are given in terms of the primitive (`comoving frame') variables---the baryon number density $n_b$, the velocity 3-vector $\mathbf{v}$, the internal energy density $e$, and (for the collapse of pre-supernova progenitors) the electron fraction $Y_e$---according to 
\begin{equation}
\begin{aligned}
N_b &= n_b, \\
\mathbf{S} &= m_b n_b \, \underline{\mathbf{v}}, \\
E &= e + \frac{1}{2} m_b n_b \, \mathbf{v}^2, \\
N_e &= n_b \, Y_e.
\end{aligned}
\end{equation}
The three-metric $\bm{\gamma}$ associated with spherical coordinates is used to obtain from the 3-vector $\mathbf{v}$ the 3-covector $\underline{\mathbf{v}} = \bm{\gamma}\cdot \mathbf{v} = \bm{\gamma}(\mathbf{v}, . )$ appearing in the expression for $\mathbf{S}$, and the squared velocity $\mathbf{v}^2 = \underline{\mathbf{v}} \cdot \mathbf{v} = \bm{\gamma}(\mathbf{v}, \mathbf{v})$ appearing in the expression for $E$.
The pressure $p$ is given in terms of the density $n_b$ (and, for pre-supernova progenitors, the temperature $T$ and electron fraction $Y_e$) by the equation of state.
It is the primitive variables that are reconstructed by the procedure described in \S\ref{sec:FluidDynamics}.
The components of the gradient $\bm{\nabla} \Phi$ in the source terms are given by the simplest possible centered difference, i.e. that computed from the two nearest neighbors.
Note also that, unless one were to evolve rectangular rather than curvilinear momentum components, the use of spherical coordinates introduces well-known geometric source terms from the divergence of the momentum flux, a rank-2 tensor; see for instance \citet{Cardall2021Fluid-Dynamics-}.
(Evolution of rectangular momentum components is not implemented in the tests presented here, but our discretization of the geometric source terms matches that of the divergence so as to minimize numerical artifacts near coordinate singularities.)

Except for the adiabatic collapse, bounce, and explosion of pre-supernova progenitor stars, we compare the solutions computed with \genasis\ to reference analytic or semi-analytic solutions.
In particular we display plots of $L_1$ errors, typically as a function of spatial resolution, given in terms of sums over position space cells by
\begin{equation}
L_1 ( f ) = \frac{\sum_{i j k} \left| f_{i j k} - f^0_{i j k} \right|}{\sum_{i j k}\left|f^0_{i j k}\right|}
\label{eq:L1error}
\end{equation}
for a computed variable $f$ compared with the reference value $f^0$.

\subsection{Potential of Homogeneous Spheroids}
\label{sec:HomogeneousSpheroid}


The gravitational potential of a spheroid with a homogeneous mass distribution is given by an analytic expression \citep{Stone1992ZEUS-2D:-A-Radi,BinneyTremaine:2ndEd} that can be compared against the potential computed by our Poisson solver.
We exercise our solver's multidimensional capability by computing the potential for spheroids of varying eccentricity.

The surface of an oblate spheroid is given by 
\begin{equation}
\frac{r_\perp^2}{{a_1}^2} + \frac{z^2}{{a_3}^2} = 1 
\end{equation}
with $r_\perp^2 = x^2 + y^2$ and $a_1 > a_3$, so that $a_1$ and $a_3$ are the lengths of the semi-major and semi-minor axes respectively.
The interior and exterior of the spheroid correspond to the expression on the left being less than or greater than 1 respectively.
In this particular problem we set $G = 1$.
Then the potential of such a spheroid with constant density $\rho$ is of the form
\begin{equation}
\Phi ( r_\perp, z ) =
	\left\{ 
	\begin{aligned}
		- \pi \, \rho \, ( I {a_1}^2 - A_1 r_\perp^2 - A_3 z ^ 2 ) \\
		- \pi \, \rho \, \frac{{a_1}^2 a_3}{{a'_1}^2 a'_3} \, 
			( I' {a'_1}^2 - A'_1 r_\perp^2 - A'_3 z ^ 2 )
	\end{aligned}
	\right.
\label{eqn:SpheroidPotential}
\end{equation}
for the interior (upper expression) and exterior (lower expression).
For the interior, the quantities $I$, $A_1$, and $A_3$ are determined by $a_1$ and $a_3$ according to
\begin{align}
I &= 2\frac{\sqrt{1-e^2}}{e}\mathrm{sin}^{-1}e, \\
A_1 &=  \frac{\sqrt{1-e^2}}{e^2} \left[ \frac{\mathrm{sin}^{-1}e}{e}-\sqrt{1-e^2}\right], \\
A_3 &=  2\frac{\sqrt{1-e^2}}{e^2} \left[ \frac{1}{\sqrt{1-e^2}} - \frac{\mathrm{sin}^{-1}e}{e}\right],
\end{align}
where 
\begin{equation}
e=\sqrt{1-\frac{{a_3}^2}{{a_1}^2}}
\end{equation}
is the eccentricity. 
For the exterior, the quantities $I'$, $A'_1$, and $A'_3$ are given by the same expressions with $a_1$ and $a_3$ replaced by $a'_1$ and $a'_3$, where the latter are given by
\begin{equation}
{a'_i}^2 = {a_i}^2 + \lambda ( r_\perp, z )
\end{equation}
with $i = 1$ or $i = 3$, and $\lambda ( r_\perp, z )$ is the positive solution of the quadratic equation
\begin{equation}
\frac{r_\perp^2}{{a_1}^2 + \lambda} + \frac{z^2}{{a_3}^2 + \lambda} = 1.
\end{equation}
A sphere is the special case $R = a_1 = a_3$ with potential
\begin{equation}
\Phi(r) = \left\{
	\begin{aligned}
		-2\pi \rho \left( R^2-\frac{1}{3}r^2\right)  & &\text{if } r\le R, \\
 		-\frac{4\pi \rho R^3}{3r}  & & \text{if } r>R.
	\end{aligned}
	\right.
\end{equation}

\begin{figure}
\centering
\includegraphics[width=0.46\textwidth]{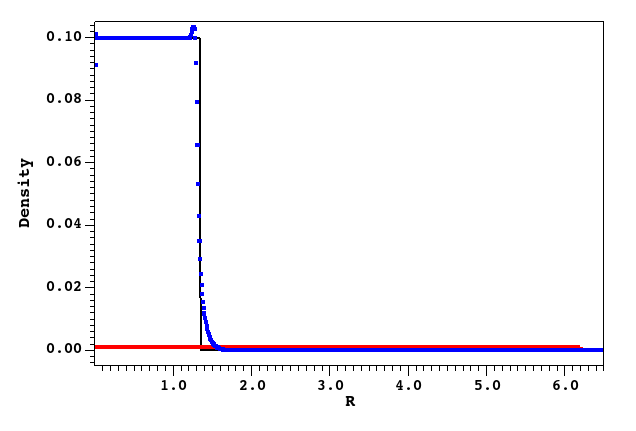}
\caption{Density profile of the spherical dust collapse problem with Newtonian gravity and $N_\theta=192$, showing the initial (red points) and final ($t \approx 0.9544~t_\infty$, blue points and black analytic solution) states. Computed results for 1D, 2D, and 3D are visually indistinguishable. 
}
\label{Fig:OppenheimerSnyder_Profile}
\end{figure}

\begin{figure}
\centering
\includegraphics[width=0.49\textwidth]{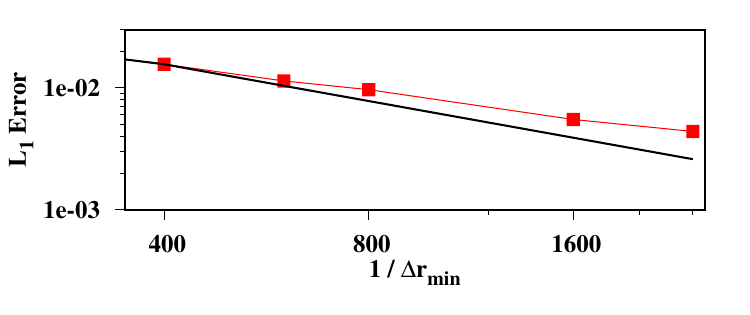}
\caption{$L_1$ error of density as a function of resolution for the spherical dust collapse problem in 1D. The solid black line is a reference for first-order convergence, expected for a discontinuous problem. 2D and 3D $L_1$ errors, not plotted here, have the same values as 1D for the resolutions we have tested (up to $1/ \Delta r_\mathrm{min} = [1600, 800]$ for [2D, 3D]).
}
\label{Fig:OppenheimerSnyder_Convergence}
\end{figure}


Figure \ref{Fig:OblateSpheroid_Solution_Error} depicts three-slices of the potential computed with \genasis\ and its relative error for a strongly oblate spheroid of eccentricity $e = 0.9$ and constant density $\rho=10^{-3}$ (and therefore $S = 4 \pi \rho$ in Equation~(\ref{eq:PoissonEquation}) for $G=1$) and total mass $M = 1$; this corresponds to semi-major axis $a_1 \approx 8.182$ and semi-minor axis $a_3 \approx 3.566$. 
The relative error for each cell is computed using Equation~(\ref{eq:L1error}) but without the summation in both the numerator and denominator.
\change{This error in the numerical solution exhibits a regular pattern with polar-angle dependence. This pattern arises because the multipole expansion of the potential is truncated at a finite maximum degree $ L  $; the dominant residual therefore carries the angular structure of the next higher-order terms (primarily associated with $  P_{L+1}(\cos\theta)  $ and $  P_{L+2}(\cos\theta)  $).
In this case with $L = 12$ and symmetry between the upper and lower halves, the pattern corresponds to $P_{14}(\cos\theta)$.}

\begin{figure*}
\centering
\includegraphics[width=0.49\textwidth]{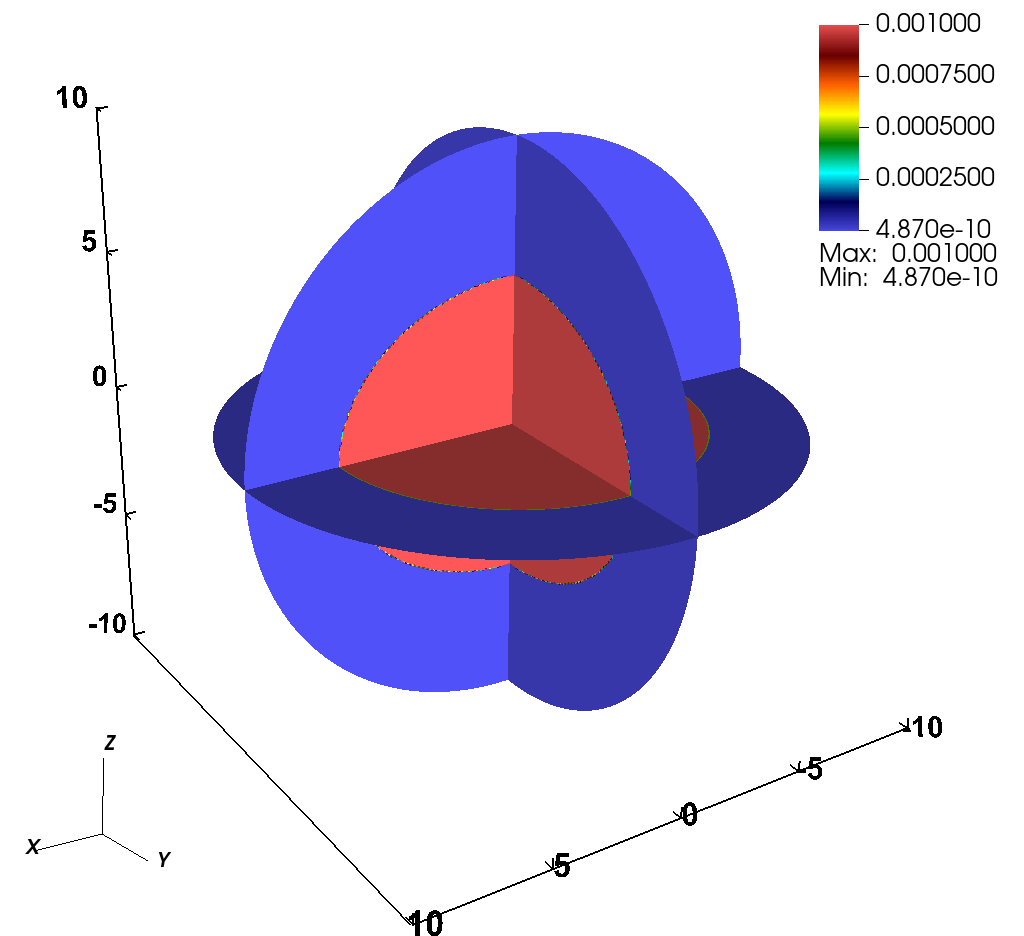}
\includegraphics[width=0.49\textwidth]{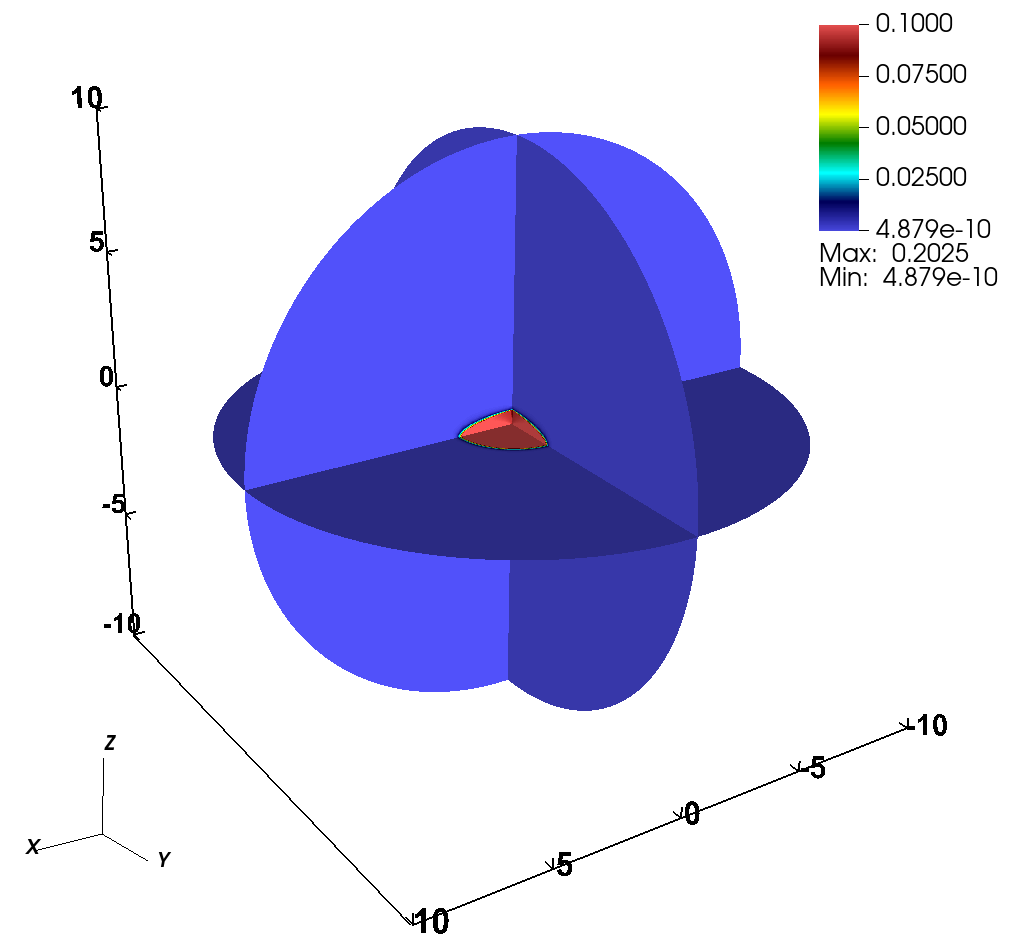}
\caption{Three-slice of initial (left) and final (right) density of the three-dimensional collapse of a dust spheroid with $N_{\theta} = 192$.}
\label{Fig:LinMestelShu_Density}
\end{figure*}

\begin{figure}
\centering
\includegraphics[width=0.5\textwidth]{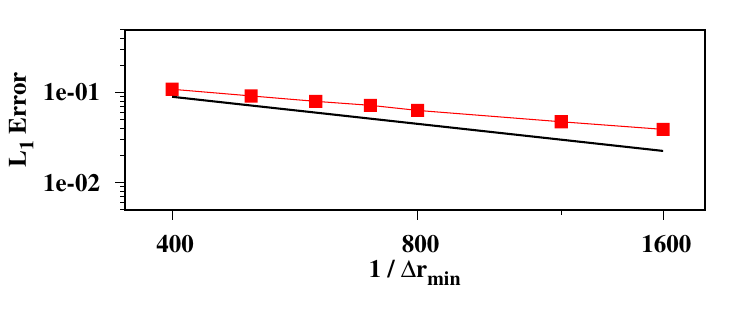}
\caption{$L_1$ error of density as a function of resolution for the collapse of a dust spheroid in 2D. The solid black line is a reference for first-order convergence, expected for a discontinuous problem. 3D $L_1$ errors, not plotted here, have the same values as 2D for the resolutions we have tested (up to $1/ \Delta r_\mathrm{min} = 800$).
}
\label{Fig:LinMestelShu_Convergence}
\end{figure}

To measure the convergence of the Poisson solver, in the left panel of Figure \ref{Fig:Spheroids_Convergence} we plot the $L_1$ error of the potential as a function of resolution for $L=20$ for homogeneous spheroids with $e = 0.0, 0.3, 0.6, 0.9$ while keeping the density $\rho = 10^{-3}$ and mass $M=1$ fixed.
The solid and dashed black lines are references for first- and second-order convergence respectively.
We observe that increasing resolution only initially improves the solution before plateauing, after which increased $L$ may be needed for increased resolution to be effective.
The right panel plots the $L_1$ error as function of $L$ for $N_\theta = 384$ (corresponding to $1/ \Delta r_\mathrm{min} = 1200$).
Increasing $L$ improves the solutions of the oblate spheroids before plateauing, after which increased resolution may be needed for increased $L$ to be effective.

\subsection{Collapse of a Pressureless Fluid Sphere}
\label{sec:OppenheimerSnyder}

Collapsing under its own Newtonian self-gravity, a homogeneous sphere of dust (pressureless fluid) that begins at rest at $t=0$ reaches infinite density at finite collapse time $t = t_\infty$.
During collapse the sphere maintains a spatially uniform but increasing density $\rho(t)$, while the radius $R(t)$ decreases so as to maintain constant mass $M = (4\pi / 3) \, \rho(t) \, R(t)^3$.  
Abusing notation, $R$ and $t$ are related parametrically through the evolution variable $\eta$:
\begin{equation}
\begin{aligned}
R(\eta) &= \frac{R_0}{2} \left( 1 + \cos\eta \right), \\[5pt]
t(\eta) &= \frac{t_\infty }{\pi} \left( \eta + \sin\eta \right).
\end{aligned}
\end{equation}
With $R_0 = R(0)$ and $\rho_0 = \rho(0)$ denoting the initial radius and density, and setting $G = 1 = c$ for this problem, the collapse time is
\begin{equation}
t_\infty = \frac{\pi}{2} \, R_0 \left( \frac{ R_0 }{2 M} \right)^{1/2} 
	= \frac{\pi}{2} \left( \frac{ 3 }{8 \pi \rho_0} \right)^{1/2}.
\end{equation}
We choose $M=1$ and $\rho_0 = 10^{-3}$, corresponding to $R_0 \approx 6.204$ and $t_\infty \approx 17.16$.
Figure~\ref{Fig:OppenheimerSnyder_Profile} shows the density profile computed with \genasis\ for $\rho_f = 10^2 \rho_0 = 10^{-1}$, corresponding to $R_f \approx 1.336 \approx 0.2154~R_0$ at $t_f \approx 16.38 \approx 0.9544~t_\infty$. 
Convergence of the final configuration with increasing resolution is displayed in Figure~\ref{Fig:OppenheimerSnyder_Convergence}.

\subsection{Collapse of a Pressureless Fluid Spheroid}
\label{sec:LinMestelShu}

A homogeneous oblate spheroid of dust also maintains a uniform density as it collapses under its own Newtonian self-gravity, remaining a spheroid but with increasing eccentricity \citep{Lin1965The-Gravitation}.
As with spherical dust collapse (a limiting case with vanishing eccentricity), we choose $M=1$ and $\rho_0 = 10^{-3}$, but initial eccentricity $e_0 = e(0) = 0.6$, which corresponds to initial semi-major and semi-minor axes $a_1(0) \approx 6.683$ and $a_3(0) \approx 5.346$.
Also as with spherical dust collapse, we set $G = 1 = c$ and use \genasis\ to evolve to $\rho_f = 10^2 \rho_0 = 10^{-1}$, which corresponds to eccentricity $e(t_f) \approx 0.9683$, with $a_1(t_f) \approx 0.3175 \, a_1(0) \approx 2.122$ and $a_3(t_f) \approx 9.918\times 10^{-2} \, a_3(0) \approx 0.5302$ at $t_f \approx 16.23$.
These values are determined by a reference solution generated by a pair of ordinary differential equations.
The axis ratios $X(t)$ and $Z(t)$ defined by $a_1(t) = X(t) \, a_1(0)$ and $a_3(t) = Z(t) \, a_3(0)$, subject to initial conditions $X = Z = 1$ and $\mathrm{d}X / \mathrm{d}t = \mathrm{d}Z / \mathrm{d}t = 0$, are governed by 
\begin{equation}
\begin{aligned}
\frac{\mathrm{d}^2 \! X}{\mathrm{d} t^2} 
	&= - \frac{\rho_0 \, A(e)}{X Z}, \\[5pt]
\frac{\mathrm{d}^2 \! Z}{\mathrm{d} t^2} 
	&= - \frac{\rho_0 \, C(e)}{X^2} \\[5pt]
\end{aligned}
\end{equation}
with
\begin{equation}
\begin{aligned}
A(e) &= \frac{ 2 \pi \sqrt{1 - e^2} }{e^2} \left( \frac{\sin^{-1}e}{e} - \sqrt{1 - e^2} \right), \\[5pt]
C(e) &= \frac{ 4 \pi \sqrt{1 - e^2} }{e^2} \left( \frac{1}{\sqrt{1 - e^2}} - \frac{\sin^{-1}e}{e}  \right) \\[5pt]
\end{aligned}
\end{equation}
and
\begin{equation}
e = e(t) = \sqrt{ 1 -  \frac{Z^2}{X^2} ( 1  -  {e_0}^2 ) }.
\end{equation}
Note the typographic error in Equation~(21) of \cite{Lin1965The-Gravitation}.
The uniform density evolves according to $\rho(t) = \rho_0 / X^2  Z$.
Figure~\ref{Fig:LinMestelShu_Density} shows three-slices of the initial and final density computed with \genasis. 
Convergence of the final configuration with increasing resolution is displayed in Figure~\ref{Fig:LinMestelShu_Convergence}.

\subsection{Collapse of a Polytropic Fluid Sphere}
\label{sec:YahilLattimer}

\begin{figure}
  \centering
  \includegraphics[width=0.5\textwidth]{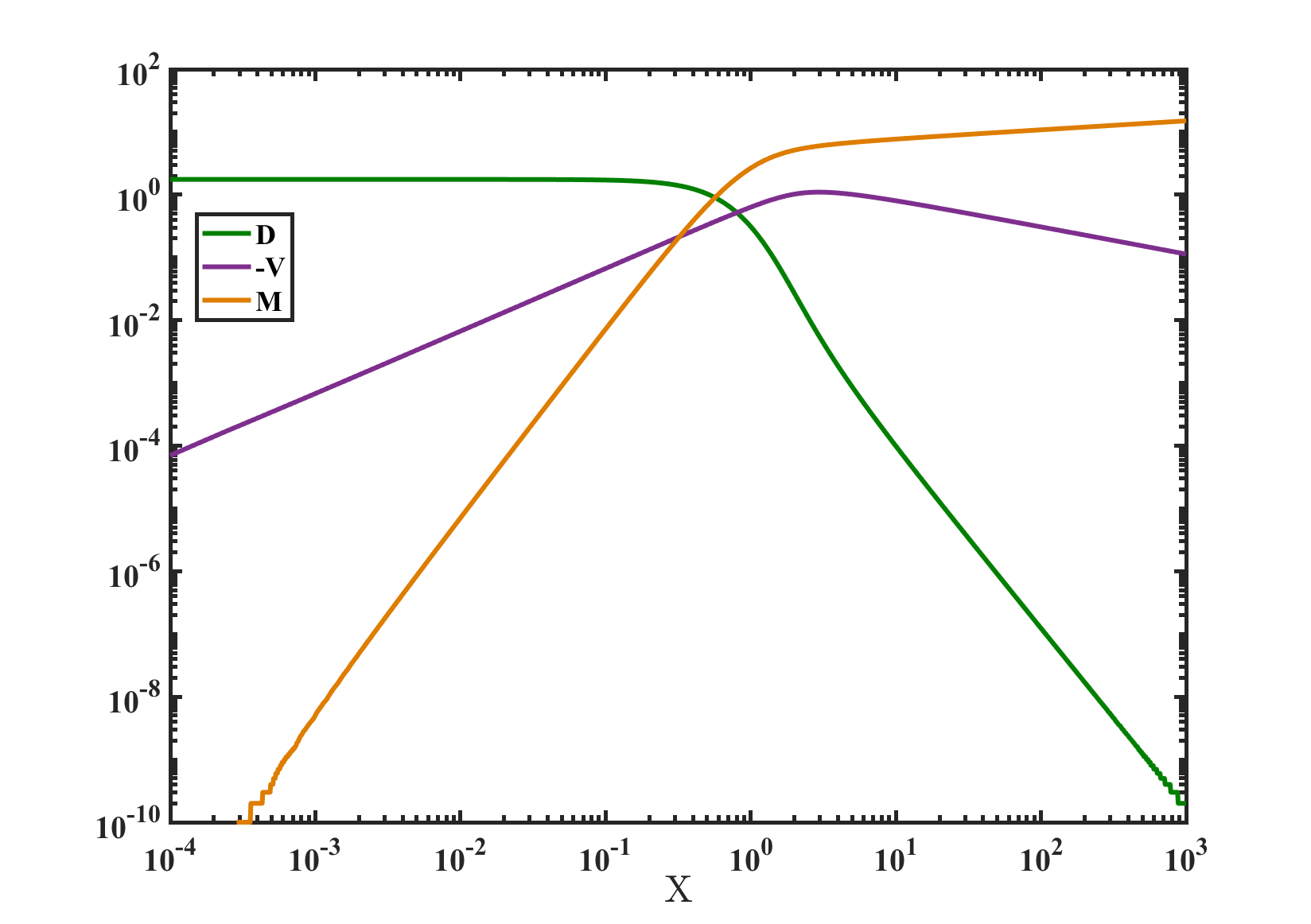}
  \caption{Plot of dimensionless variables $D$, $-V$, and $M$ versus $X$, obtained by numerically integrating Equations~\eqref{eq:odeYL:M}-\eqref{eq:YL:Constraint} with $D(0)=1.75$ and $\gamma=1.30$.} 
  \label{Fig:YL_analytic}
\end{figure}

\begin{figure*}[!t]
\centering
\includegraphics[width=0.49\textwidth]{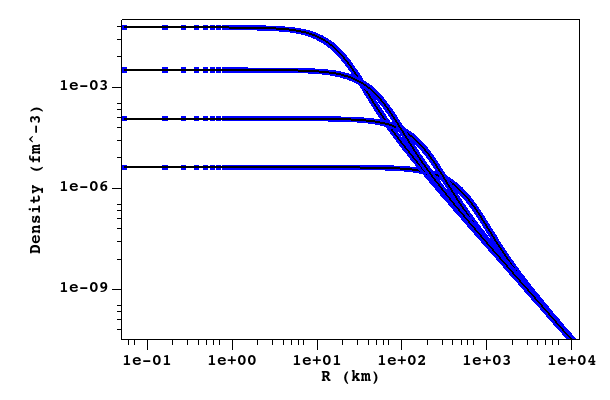}
\includegraphics[width=0.49\textwidth]{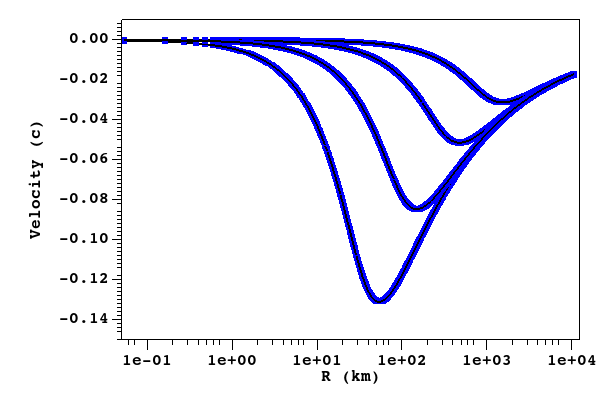}
\caption{
Number density and velocity profiles of the 1D collapse of a polytropic sphere with $N_\theta=192$ and $\Delta r_\mathrm{min} \approx 1.067 \times 10^{-1}~\mathrm{km}$), showing the computed (blue points) and reference (black line) solutions at $t \approx [0, 4.961\times10^{-2}, 5.900\times10^{-2}, 6.069\times10^{-2}]~\rm{s}$. Computed results for 1D, 2D, and 3D are visually indistinguishable. 
}
\label{Fig:YahilLattimer_Profile}
\end{figure*}

\begin{figure}
\centering
\includegraphics[width=0.5\textwidth]{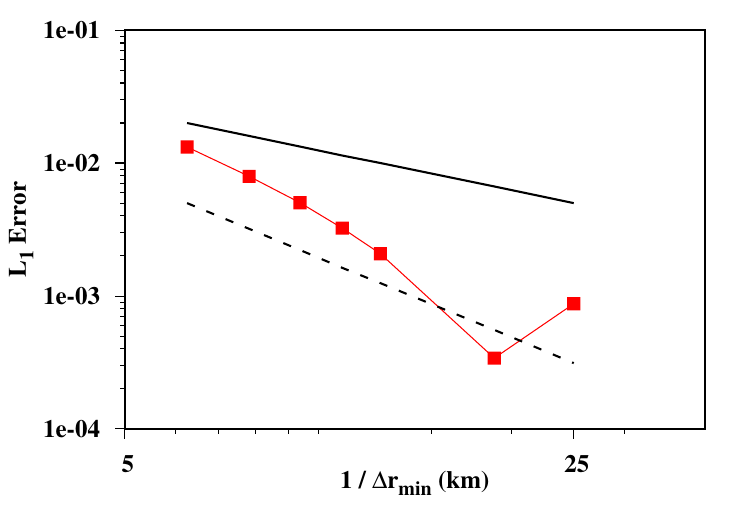}
\caption{
$L_1$ error of density as a function of resolution for the collapse of a polytropic sphere in 1D. The solid and dashed black lines are references for first- and second-order convergence, respectively. 2D and 3D $L_1$ errors, not plotted here, are indistinguishable from 1D for the resolutions we have tested (up to $1/ \Delta r_\mathrm{min} = [25.0, 12.5]~\rm{km}$ for [2D, 3D]).
}
\label{Fig:YahilLattimer_Convergence}
\end{figure}


\citet{YahilLattimer:1982} investigated the self-similar collapse of polytropic stars, i.e. with equation of state
\begin{equation}
  p = \kappa \rho^\gamma,
  \label{eq:polytrope}
\end{equation}
where $\kappa$ is the polytropic constant and $\gamma$ is the adiabatic index.  
The structure of spherically symmetric self-similar collapse is well understood \citep{Yahil:1983,YahilLattimer:1982}.  
The stellar core separates into a subsonic homologous inner core ($v\propto r$) and a supersonically infalling outer core, with velocity limited by the free fall speed ($|v| < \sqrt{2GM / r}$).  

Self-similar solutions exist because only two dimensional parameters exist in the problem---the polytropic constant $\kappa$ and the gravitational constant $G$.
As a result, the only dimensionless combination of radius and time (the similarity variable) is
\begin{equation}
  X = \kappa^{-1/2} \, G^{(\gamma - 1)/2} \, r \, t_\infty^{\gamma - 2},
  \label{eq:similarityVariableYL}
\end{equation}
where $t_\infty$ is the time to `catastrophe', when the density becomes infinite.  
Solutions exist for $6/5 \le \gamma\le 4/3$ \citep{Yahil:1983,YahilLattimer:1982}. 

For a self-similar solution, all hydrodynamic variables are functions of $X$ alone and its structure remains the same for all times.  
Following \citet{YahilLattimer:1982}, dimensionless variables depending on $X$ can be related to the corresponding dimensional variables through scale factors defined in terms of powers of $\kappa$, $G$, and $t_\infty$.  
Specifically, 
\begin{align}
  \rho &= G^{-1} \, t_\infty^{-2} \, D ( X ) \label{YL:rho}, \\
  v &= \kappa^{1/2}\, G^{(1 - \gamma)/2} \, t_\infty^{1 - \gamma} \, V ( X ) \label{YL:v}, \\
  m & = \kappa^{3/2} \, G^{(1-3\gamma)/2} \, t_\infty^{4-3\gamma} \, M (X) \label{YL:m},
\end{align}
where $\rho$, $v$, and $m$ are the dimensional mass density, fluid velocity, and enclosed mass, and $D$, $V$, and $M$ are the corresponding dimensionless variables.  
Using the relations in Equation~\eqref{eq:similarityVariableYL} and Equations~\eqref{YL:rho}-\eqref{YL:m}, the Euler equations can be recast as a system of ordinary differential equations for the dimensionless variables:
\begin{align}
  \frac{\mathrm{d}M}{\mathrm{d}X} &= 4 \pi X ^ 2 D, \label{eq:odeYL:M} \\
  \frac{\mathrm{d}D}{\mathrm{d}X} &= \frac{D\,\big[\,F_{1}-U\,F_{2}\,\big]}{\big[\,\gamma\,D^{\gamma-1}-U^{2}\,\big]}, \label{eq:odeYL:D}
\end{align}
where
\begin{align}
  F_{1} &= (4-3\,\gamma)-2\,U/X, \\
  F_{2} &= (\gamma-1)\,(2-\gamma)\,X+(3-2\,\gamma)\,U - M / X^{2},
\end{align}
and where $U$, the dimensionless fluid velocity with respect to the homologous frame, is related to the dimensionless fluid velocity by
\begin{equation}
  V = (\gamma - 2) X + U.
  \label{eq:YL:U}
\end{equation}
The algebraic constraint
\begin{equation}
  4\pi X^2 \, D \, U = (4-3\gamma)\,M.  
  \label{eq:YL:Constraint}
\end{equation}
relates the dimensionless density and velocity variables to the dimensionless enclosed mass.

\begin{figure*}[!t]
\centering
\includegraphics[width=0.49\textwidth]{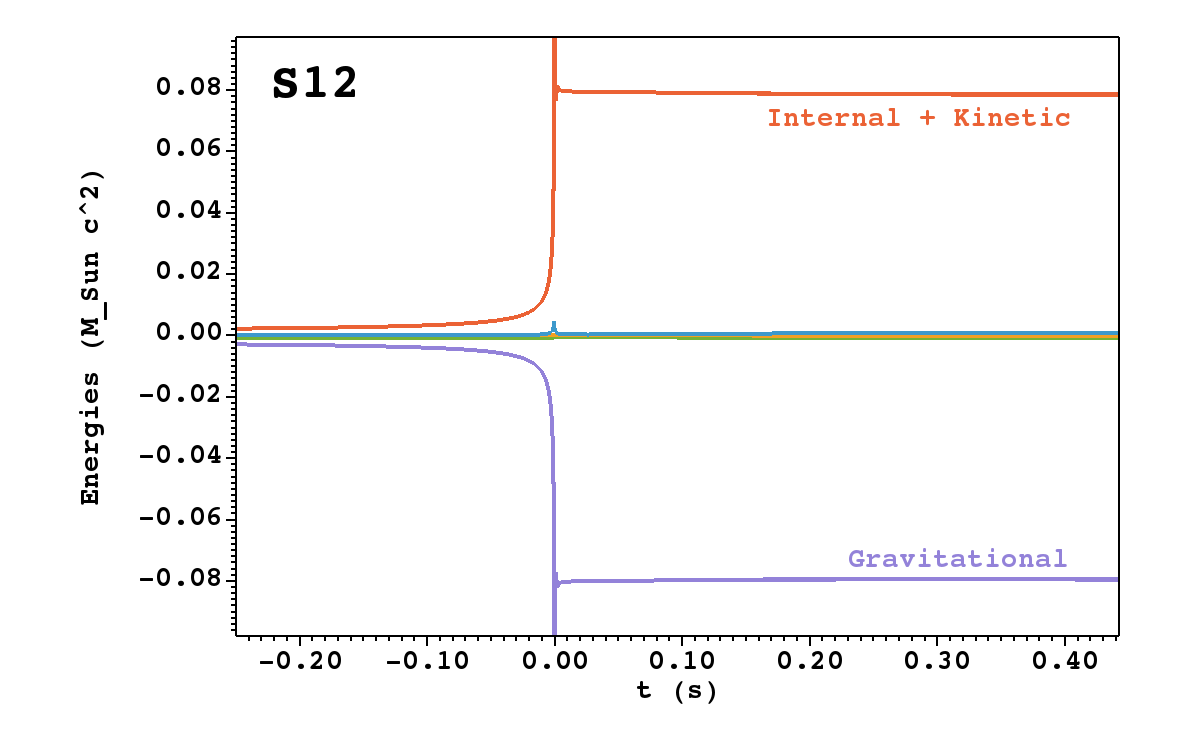}
\includegraphics[width=0.49\textwidth]{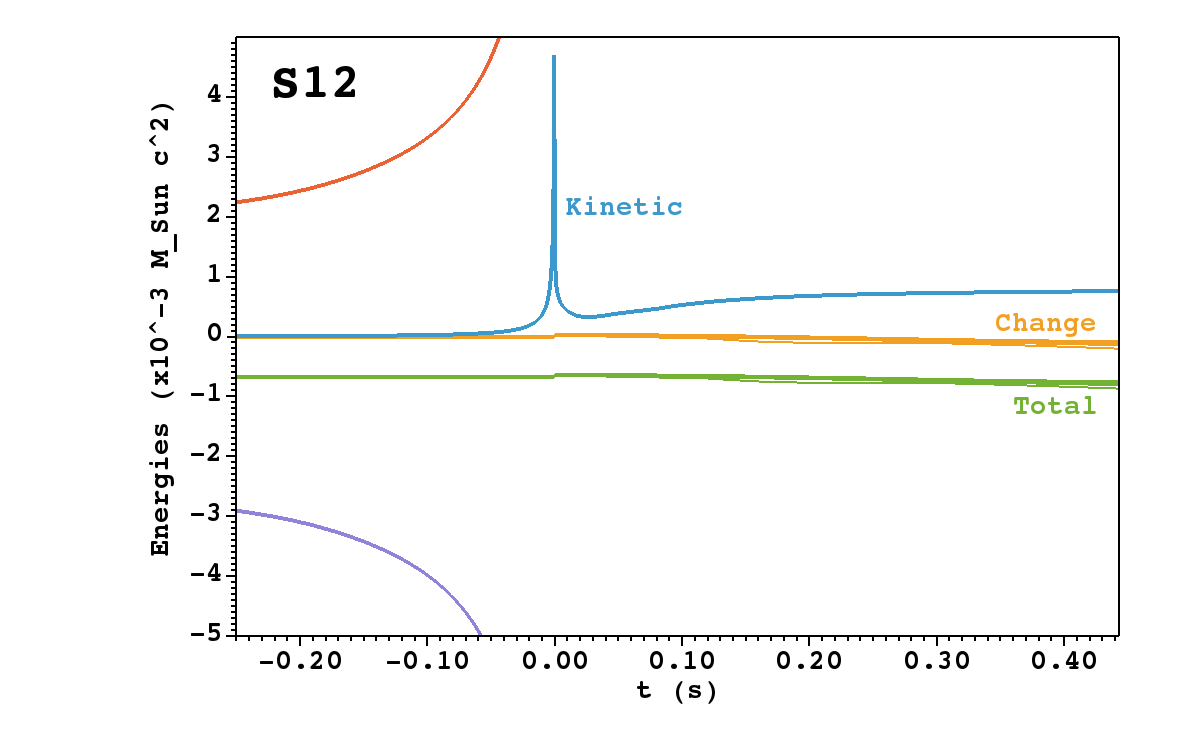}
\caption{Global energy measures (internal+kinetic, gravitational, kinetic, total, and change in the total) for adiabatic model S12 as a function of time for three different resolutions ($N_\theta = 128, 192, 256$ and $\Delta r_\mathrm{min} \approx (1.600, 1.067, 0.8000) \times 10^{-1}~\mathrm{km}$, denoted by curves of increasing thickness). The kinetic energy, total energy, and change in total energy, barely discernible in the left panel on the scale of the internal and gravitational energies, are more clearly distinguished in the right panel.}
\label{Fig:WH_A_Energies_S12}
\end{figure*}

\begin{figure*}
\centering
\includegraphics[width=0.49\textwidth]{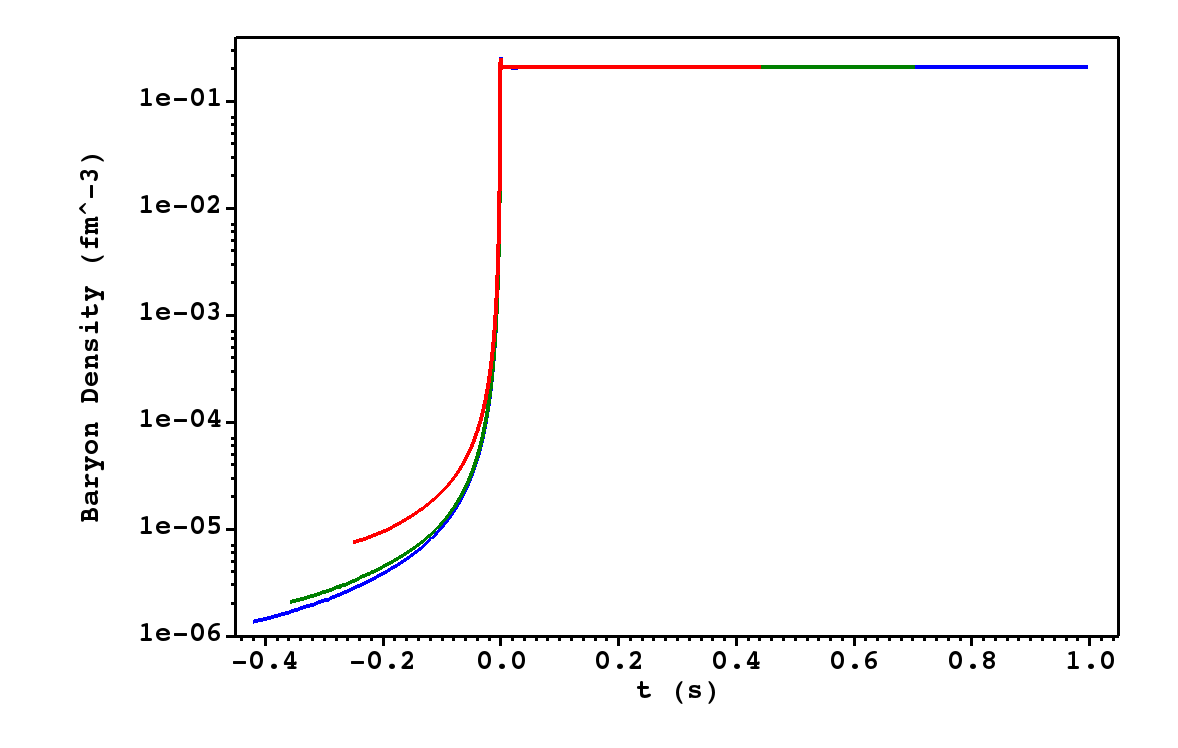}
\includegraphics[width=0.49\textwidth]{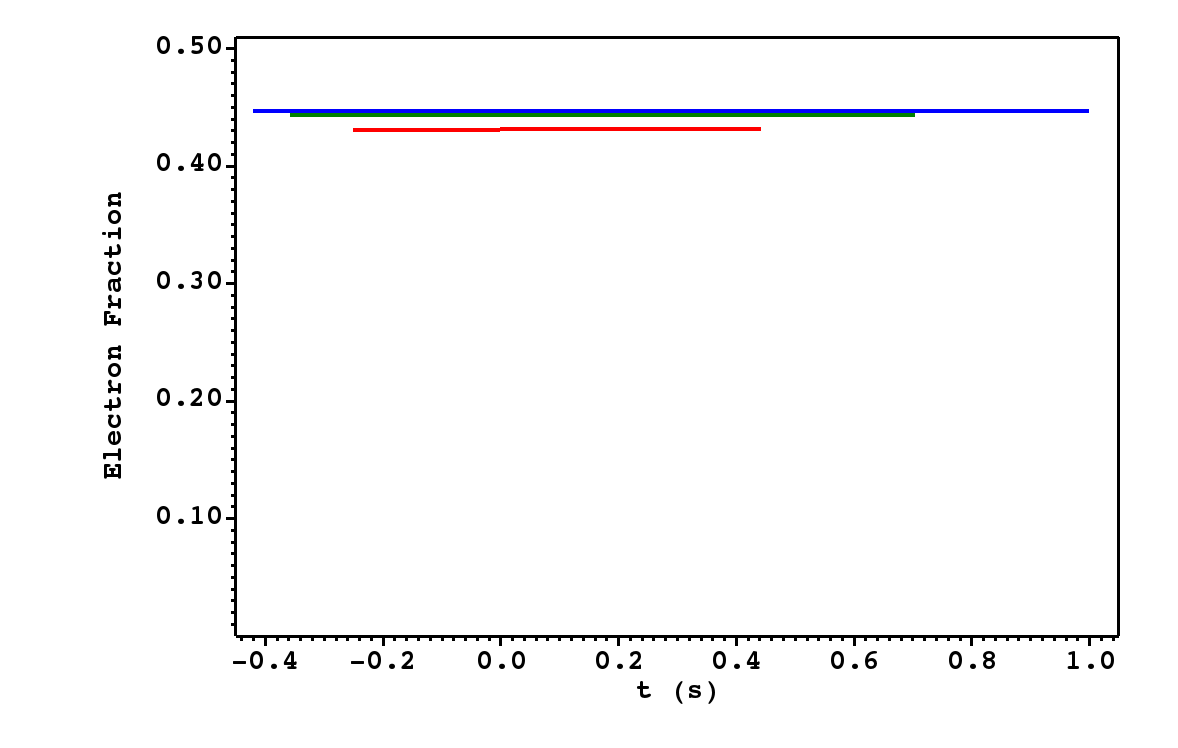}
\includegraphics[width=0.49\textwidth]{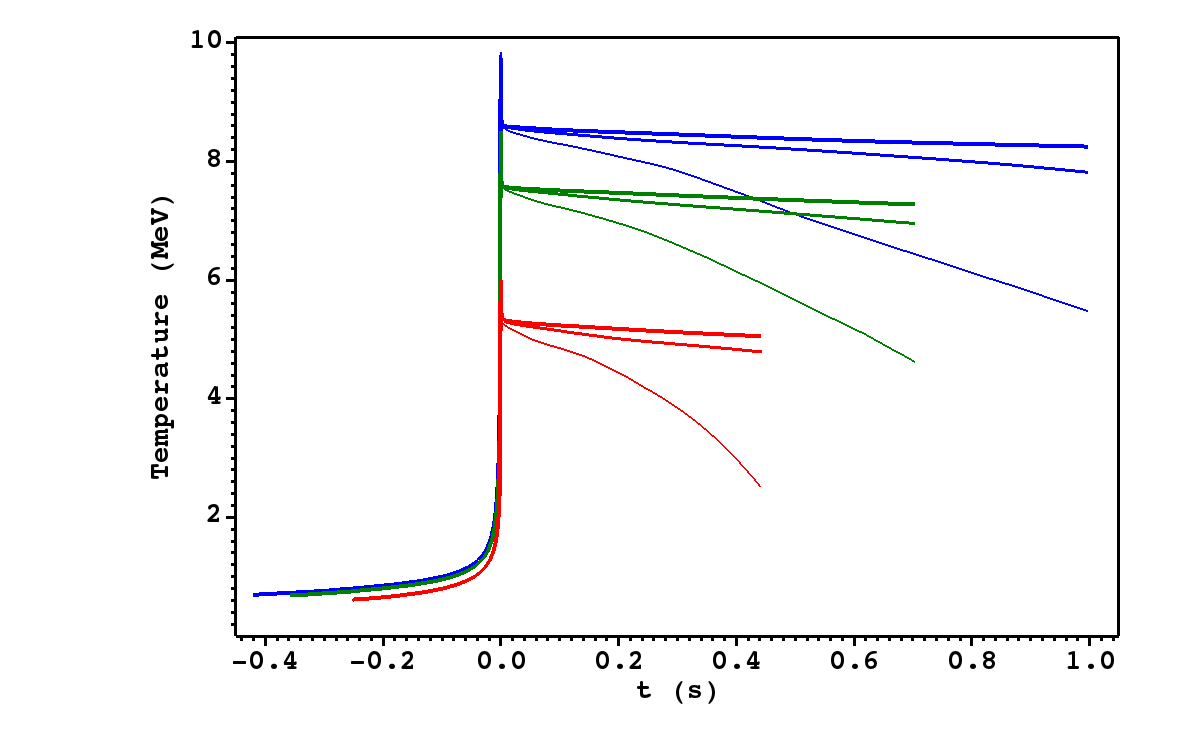}
\includegraphics[width=0.49\textwidth]{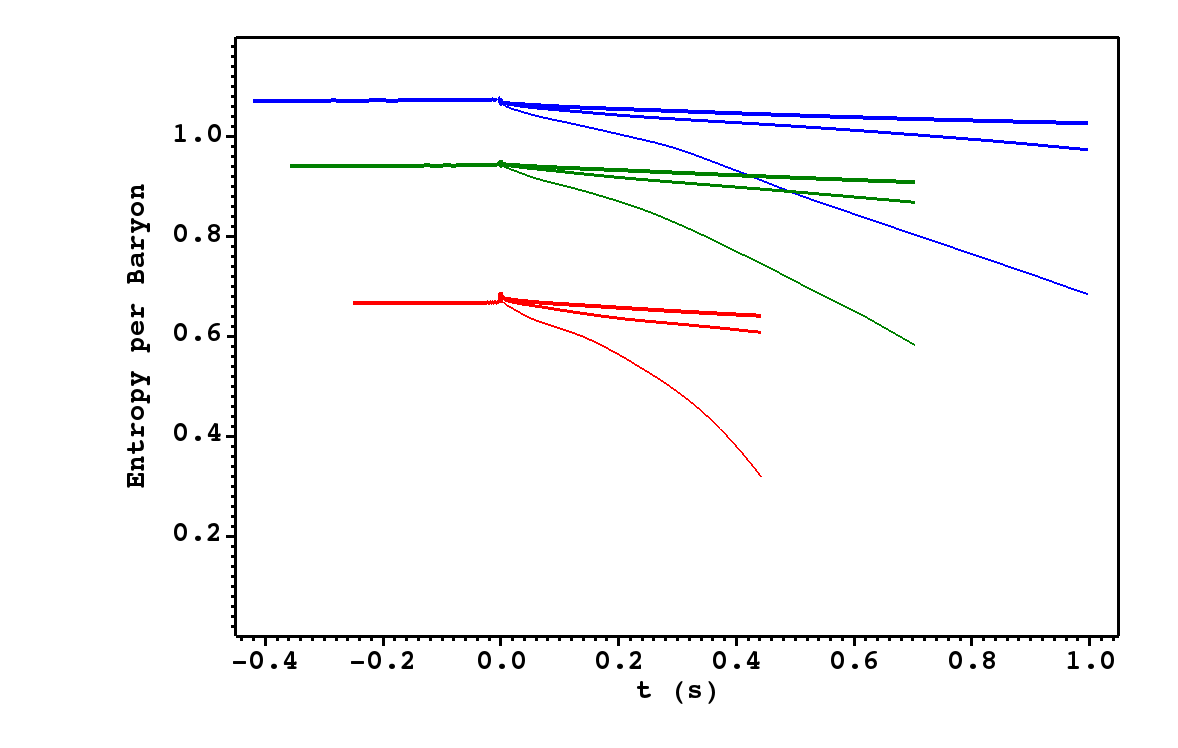}
\caption{Central values of baryon density, electron fraction, temperature, and entropy per baryon as a function of time for adiabatic models S12 (red), S25 (green), and S40 (blue), for three different resolutions ($N_\theta = 128, 192, 256$ and $\Delta r_\mathrm{min} \approx (1.600, 1.067, 0.8000) \times 10^{-1}~\mathrm{km}$, denoted by curves of increasing thickness).
Departures from expected steady values of central temperature and entropy per baryon after bounce improve with increased resolution.
\change{Data are from 1D runs, but results from 2D and 3D runs are essentially the same visually for this spherically symmetric problem.}}
\label{Fig:WH_A_Central}
\end{figure*}

We have numerically integrated this system to obtain the reference solution shown in Figure~\ref{Fig:YL_analytic}.
The enclosed mass and velocity vanish at the origin, and following \citet{Yahil:1983} Table~2 we adopt $D(0)=1.75$ for $\gamma=1.30$. 
When integrating Equation~\eqref{eq:odeYL:D}, extra care must be taken to integrate through the critical point, where $U^{2}=\gamma\,D^{\gamma-1}$, separating the subsonic inner core and the supersonic outer core.  
In particular, we linearize both the numerator and denominator on the right-hand side near the critical point and force the mass density to decrease monotonically with increasing radius.  
%
We performed self-similar collapse simulations with \genasis, with initial state (and reference solutions at checkpoints) obtained from the self-similar solution described above.
The polytropic constant $\kappa$ is specified by setting the initial ($t_i = 0\mathrm{~s}$) central density and pressure to $\rho_i=7\times10^9$~$\mathrm{g~cm^{-3}}$ and $p_i=6\times10^{27}$ $\mathrm{Ba}$, which roughly mimic a pre-supernova progenitor star, and correspond to catastrophe time $t_\infty \approx 6.120 \times 10^{-2}\mathrm{~s}$.  
We evolve to a final central density $\rho_f= 1 \times10^{14}$~$\mathrm{g~cm^{-3}}$, which corresponds to $t_f \approx 6.069 \times 10^{-2}\mathrm{~s}$.
 Figure~\ref{Fig:YahilLattimer_Profile} shows density and velocity profiles at several points during collapse, comparing the results from \genasis\ with the reference solution. 
Convergence of the final configuration with increasing resolution is displayed in Figure~\ref{Fig:YahilLattimer_Convergence}.

\subsection{Adiabatic Collapse, Bounce, and Explosion of Pre-supernova Progenitor Stars}
\label{sec:AdiabaticExplosion}

Using \genasis, we computed the adiabatic (without neutrino interactions) collapse, bounce, and explosion of 11 pre-supernova progenitor stars \citep{Woosley2007Nucleosynthesis}.
The pre-supernova progenitors with which we begin our computations have themselves been evolved from initial conditions of solar metallicity (`S') and initial `zero-age main sequence' masses of $(12,13,14,15,20,23,25,27,30,35,40) M_\odot$ that subsequently have been altered by mass loss due to stellar winds.
Our models are labeled accordingly (e.g. `S12'), and begin with collapse already underway.
Our outer boundary of $\approx 1 \times 10^4~\mathrm{km}$ lies in the oxygen shell.
In 1D and 2D we computed models with $N_\theta = 128, 192, 256$ 
(corresponding to $\Delta r_\mathrm{min} \approx (1.600, 1.067, 0.8000) \times 10^{-1}~\mathrm{km}$ as described at the beginning of \S\ref{sec:Tests}),
but in 3D we did not go beyond $N_\theta = 192$ and only computed models S12, S25, and S40.
These adiabatic models remain spherically symmetric, and results from 1D, 2D, and 3D runs are visually indistinguishable.
For the illustrative purposes of this work we used a publicly available\footnote{\texttt{https://stellarcollapse.org}} LS220 table based on the work of \citet{Lattimer:1991}.

The time evolution of various global energy measures for model S12, obtained from integration over position space, is displayed in Figure~\ref{Fig:WH_A_Energies_S12}.
Plots for the other models are qualitatively similar.
The time axis is adjusted so that `bounce', the point of maximum central density, is at $t = 0$.
The internal+kinetic (overwhelmingly internal) energy and gravitational energy dramatically increase in magnitude just before bounce and nearly offset one another: their sum, the total energy, is about two orders of magnitude smaller than these after bounce, as is the kinetic energy of the outflow (`explosion').
The total energy should remain constant, and indeed the change in total energy remains nearly zero compared with the total and kinetic energies.
Results from computations with three different resolutions are displayed, but differences in these global energy measures are barely perceptible to the naked eye.
The aspect of interest---the kinetic energy, that is, the explosion---is well converged.
\change{A small secular drift in total energy is discernible, and it is possible to address this with more sophisticated formulations and discretizations with self-gravity (e.g. \citet{Muller2010A-New-Multi-dim,Mullen2021An-Extension-of,Zhang2022Energy-conservi}).
But as for instance \citet{Muller2010A-New-Multi-dim} acknowledges, on the time scales of interest here and for purposes of computing the explosion dynamics, more advanced approaches seem not to be necessary.
In the results presented here no such special measures are employed to control global energy errors; but our $\Delta r_\mathrm{min}$ may be smaller than often employed, perhaps resulting in better global energy conservation than sometimes is seen without special measures.}

The impact of resolution is more pronounced in the evolution of the central values of some of the variables displayed in Figure~\ref{Fig:WH_A_Central} for models S12, S25, and S40.
The equations governing the baryon number density and (for these adiabatic models) electron fraction have no source terms, and the time series of their central values show no visible change with increasing resolution. 
As with central density, the central temperature increases dramatically just before bounce; and as with density, it should remain basically constant after bounce in these adiabatic models, but the degree to which this holds visibly depends on resolution.
And in these adiabatic models the central entropy per baryon should, like the central electron fraction, remain constant; but again, achieving this expectation depends visibly on resolution.
The major difference for the temperature and entropy per baryon is that they are much more indirectly related to the balance equations that are actually solved, particularly under degenerate conditions in which the thermal energy makes only a small contribution.
(The kinetic energy, small in global terms as indicated by Figure~\ref{Fig:WH_A_Energies_S12}, is not as sensitive to resolution because the momentum equation provides more direct leverage through its close connection to the velocity.)
While thermal variables may be of less immediate interest than the kinetic energy (i.e. explosion), a plunge to zero temperature causes numerical problems, and for models with neutrinos the interactions depend on temperature. 
Solving a balance equation for the entropy per baryon would more directly address the thermal energy, but would give erroneous results in the presence of a shock, requiring a resort to the internal+kinetic energy balance equation in cells where a shock is detected. 
The `well-balanced' finite volume scheme of \citet{Kappeli-R.2016A-well-balanced} may also be helpful.

\begin{figure*}
\centering
\includegraphics[width=0.49\textwidth]{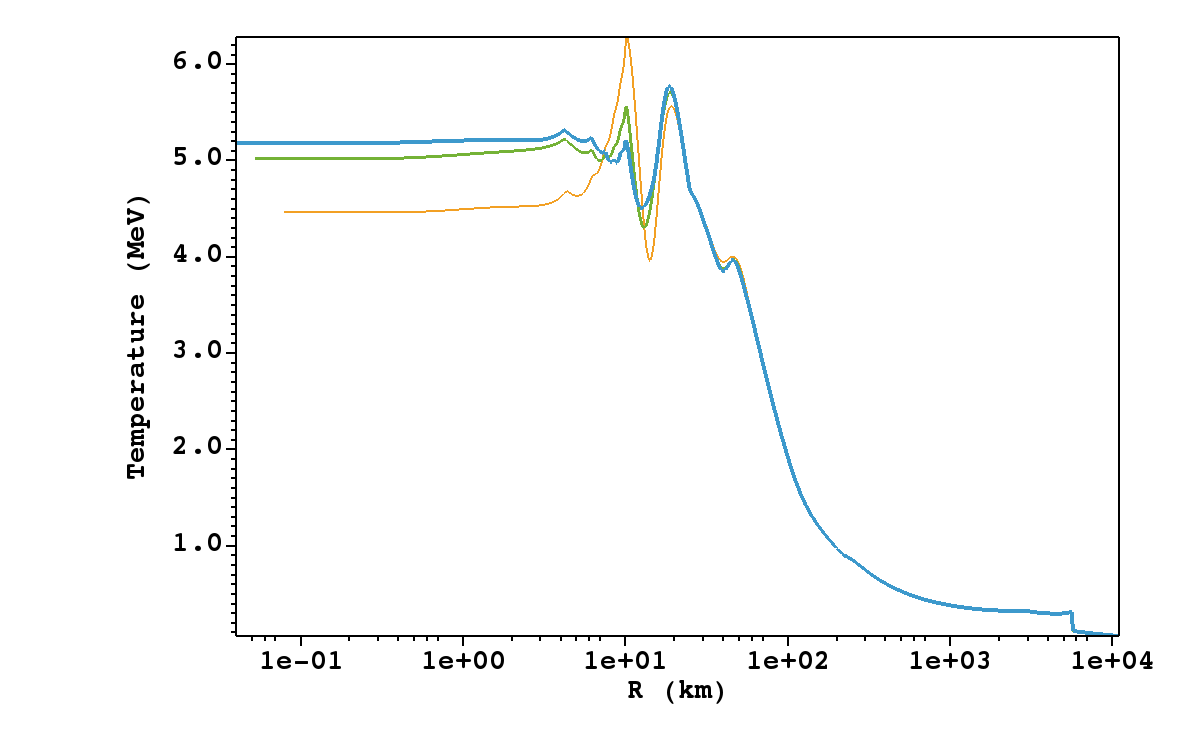}
\includegraphics[width=0.49\textwidth]{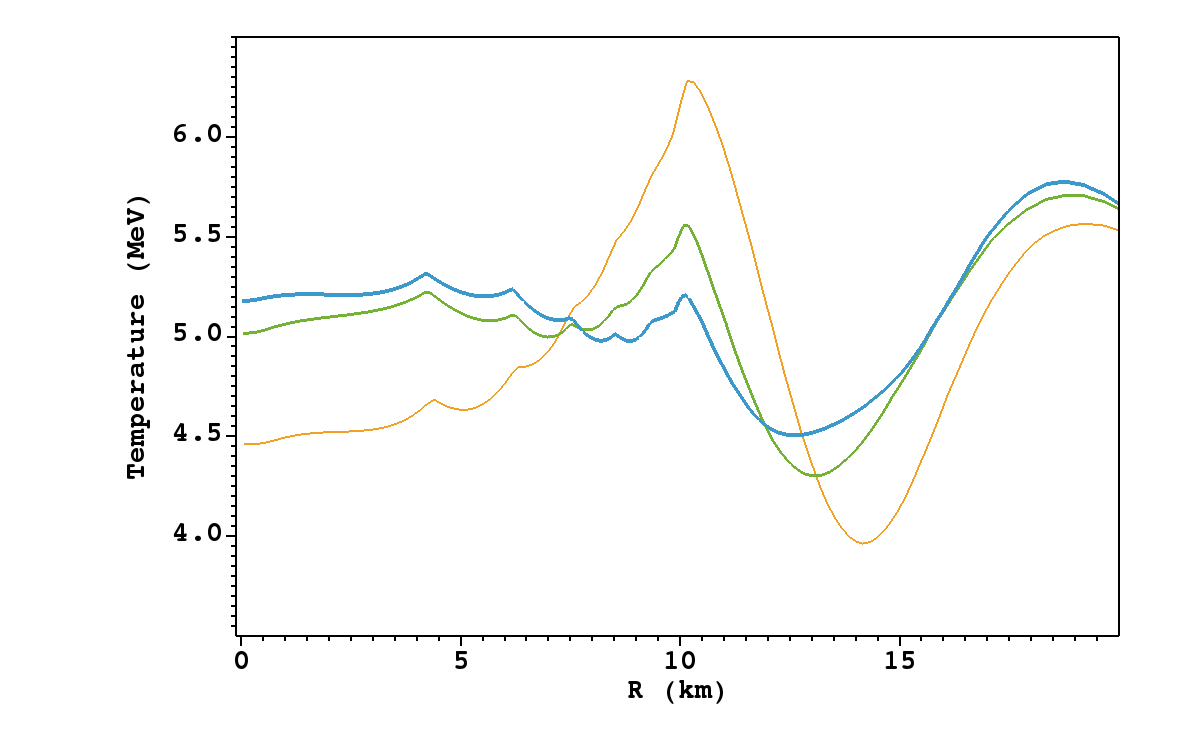}
\includegraphics[width=0.49\textwidth]{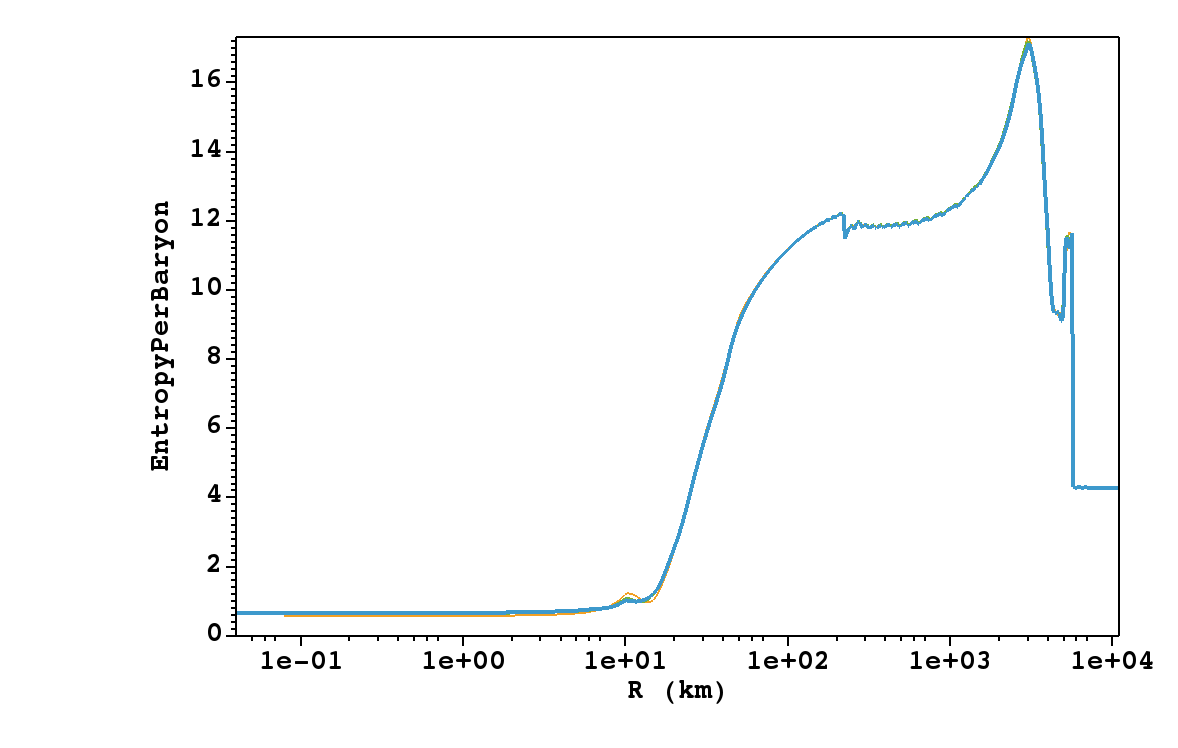}
\includegraphics[width=0.49\textwidth]{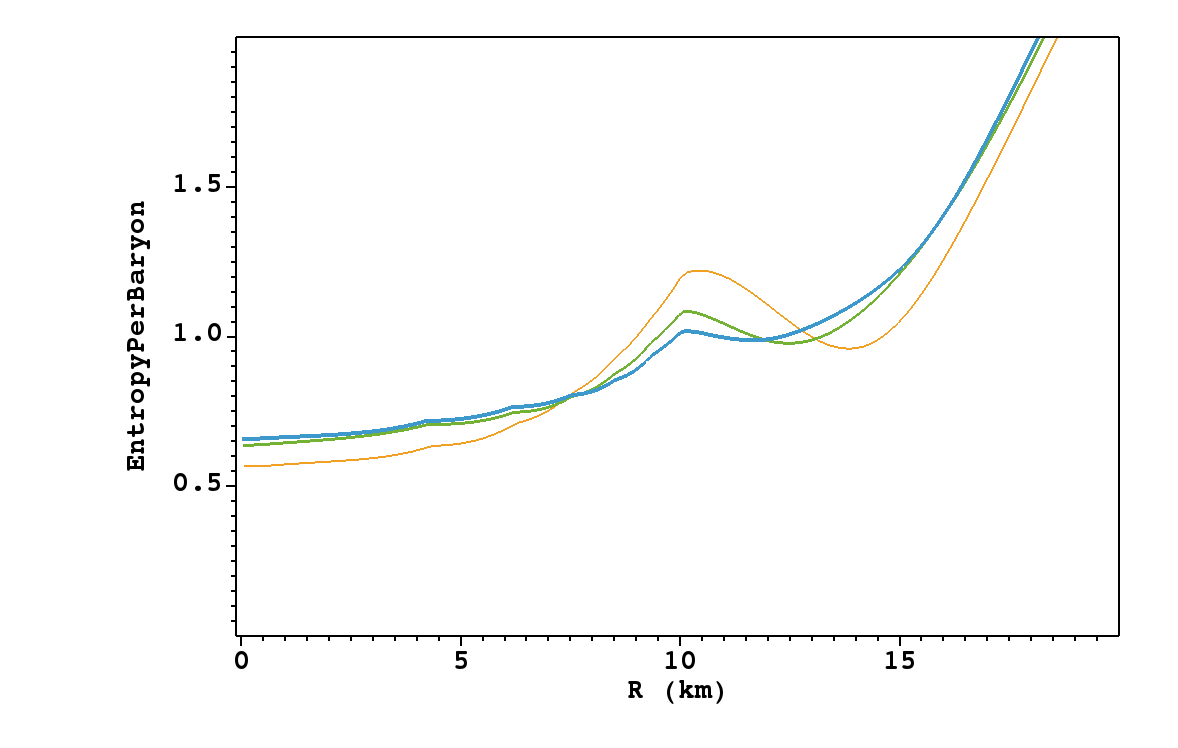}
\includegraphics[width=0.49\textwidth]{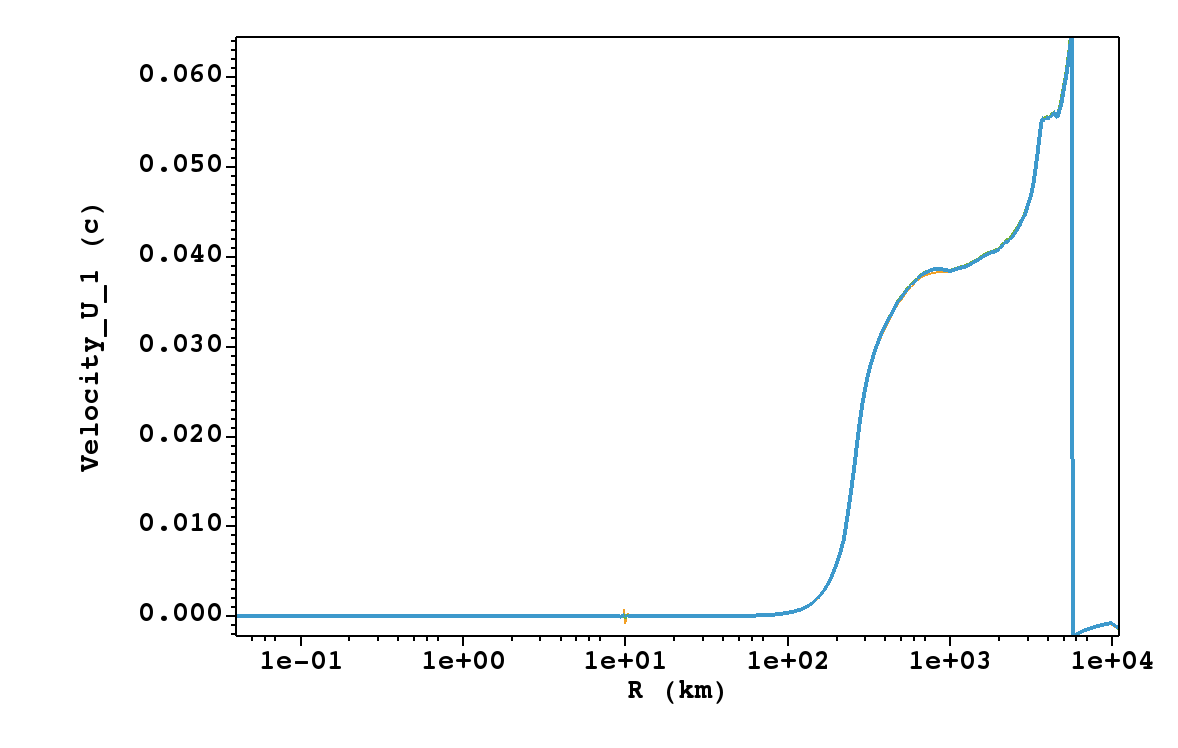}
\includegraphics[width=0.49\textwidth]{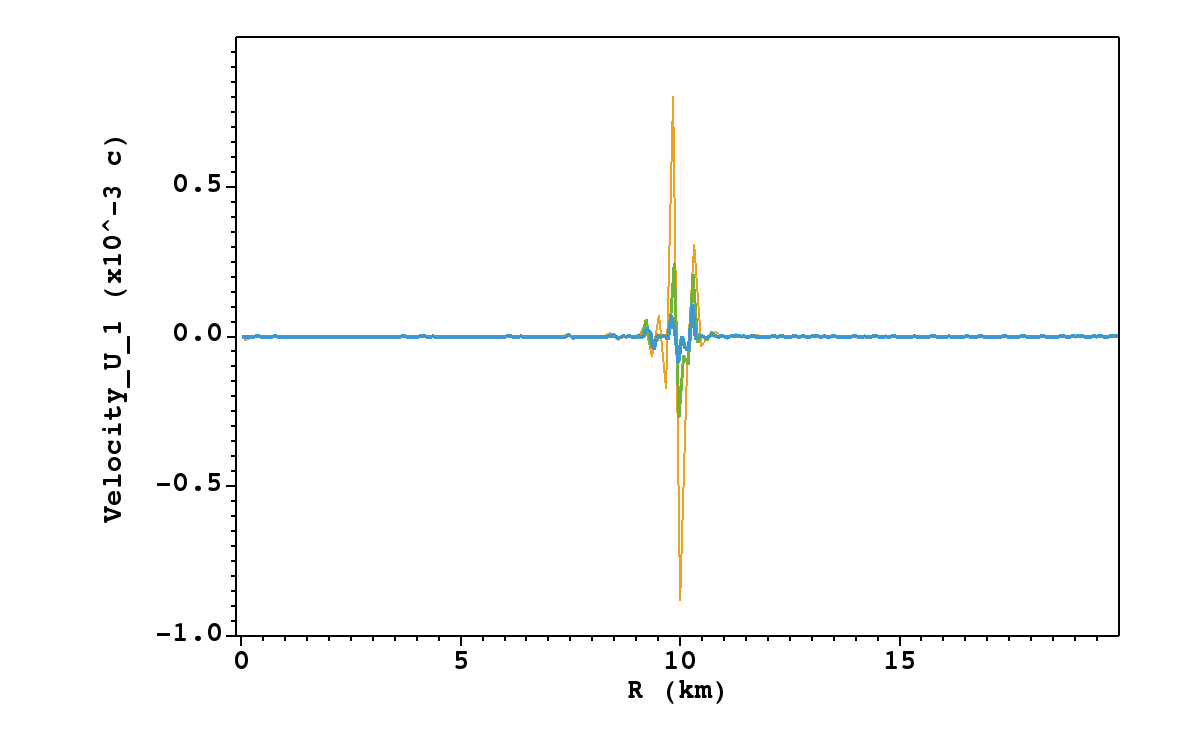}
\caption{Profiles from adiabatic model S12 for three different resolutions ($N_\theta = 128, 192, 256$ and $\Delta r_\mathrm{min} \approx (1.600, 1.067, 0.8000) \times 10^{-1}~\mathrm{km}$, denoted by orange, green, and blue curves of increasing thickness) at about 0.2~s after bounce.
Focusing on the region near $r \approx 10 \text{ km}$ reveals the impact of the nuclear phase transition in the equation of state: a local failure to achieve full hydrostatic equilibrium, resulting in entropy generation by the finite-volume shock-capturing scheme, but ameliorated by increased resolution.
\change{Data are from 1D runs, but results from 2D and 3D runs are essentially the same visually for this spherically symmetric problem.}} 
\label{Fig:WH_A_VelocityGlitch}
\end{figure*}

The drop in temperature and entropy per baryon near the center of the collapsed progenitor is also visible in the profiles from model S12 at about 0.2~s after bounce in Figure~\ref{Fig:WH_A_VelocityGlitch}, along with another infelicity that is also ameliorated by increased resolution.
A positive bump in temperature and entropy per baryon is observed at a radius of about 10~km.
The nuclear phase transition in the equation of state seems to cause a numerical instability that prevents strict local hydrostatic equilibrium from being established.
Confronted with the resulting jaggedness in velocity, the shock-capturing finite-volume scheme generates entropy.
In models with neutrino physics this jaggedness in velocity needs to be kept under control, by adequate resolution if nothing else, to avoid problems with the nonlinear semi-implicit neutrino solver.

\begin{figure*}[t]
\begin{interactive}{animation}{WH\_A\_3D\_nCP192\_0.5X.mp4}
\centering
\subfigure{\includegraphics[width=0.38\textwidth]{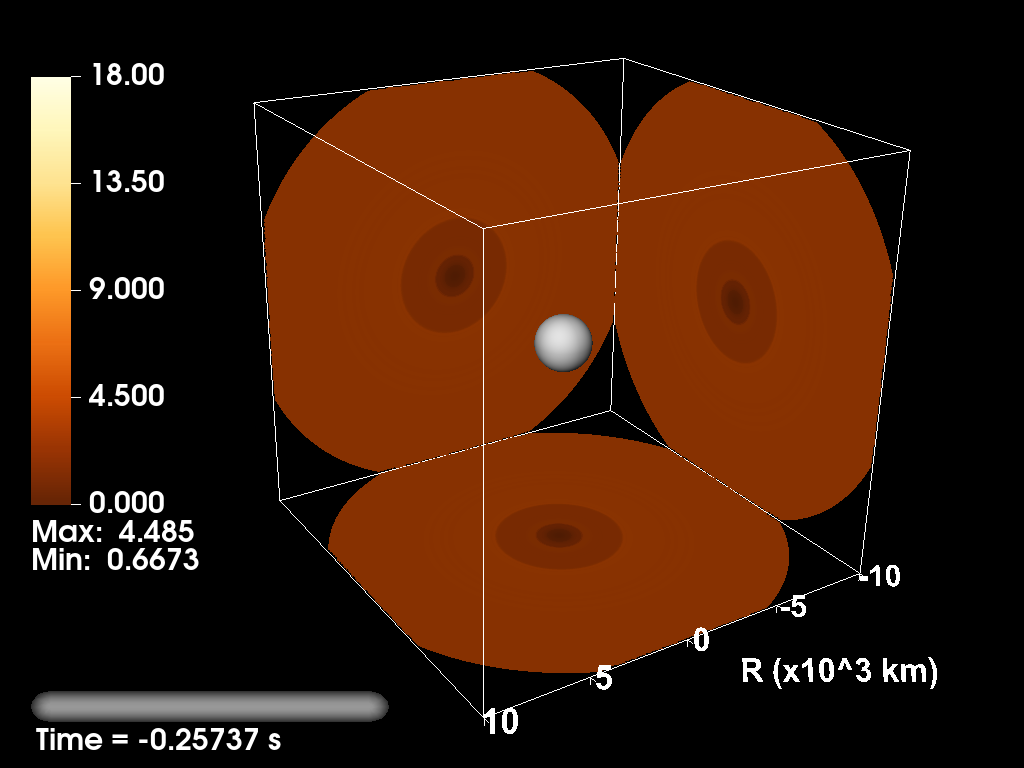}}\quad
\subfigure{\includegraphics[width=0.38\textwidth]{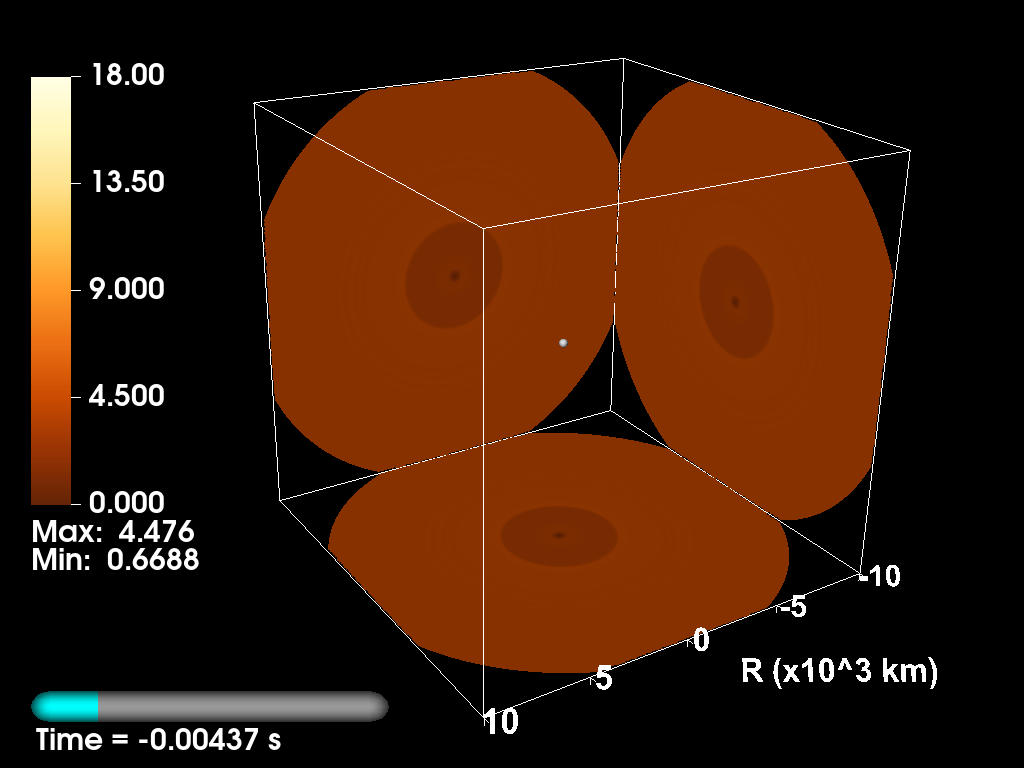}}
\subfigure{\includegraphics[width=0.38\textwidth]{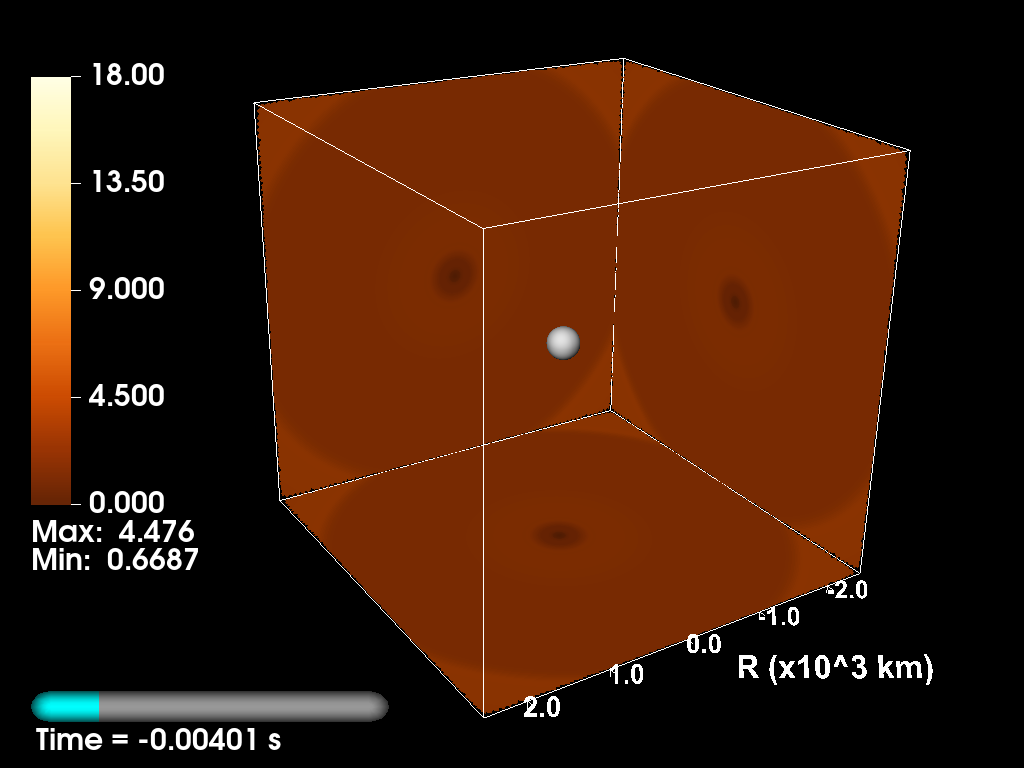}}\quad
\subfigure{\includegraphics[width=0.38\textwidth]{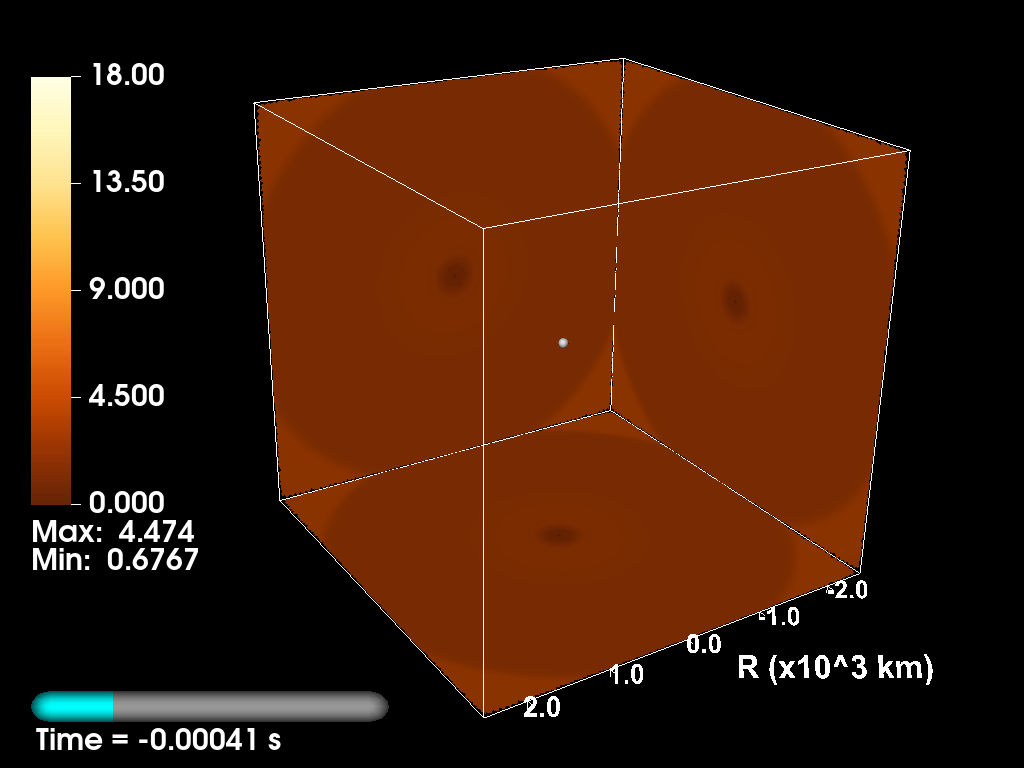}}
\subfigure{\includegraphics[width=0.38\textwidth]{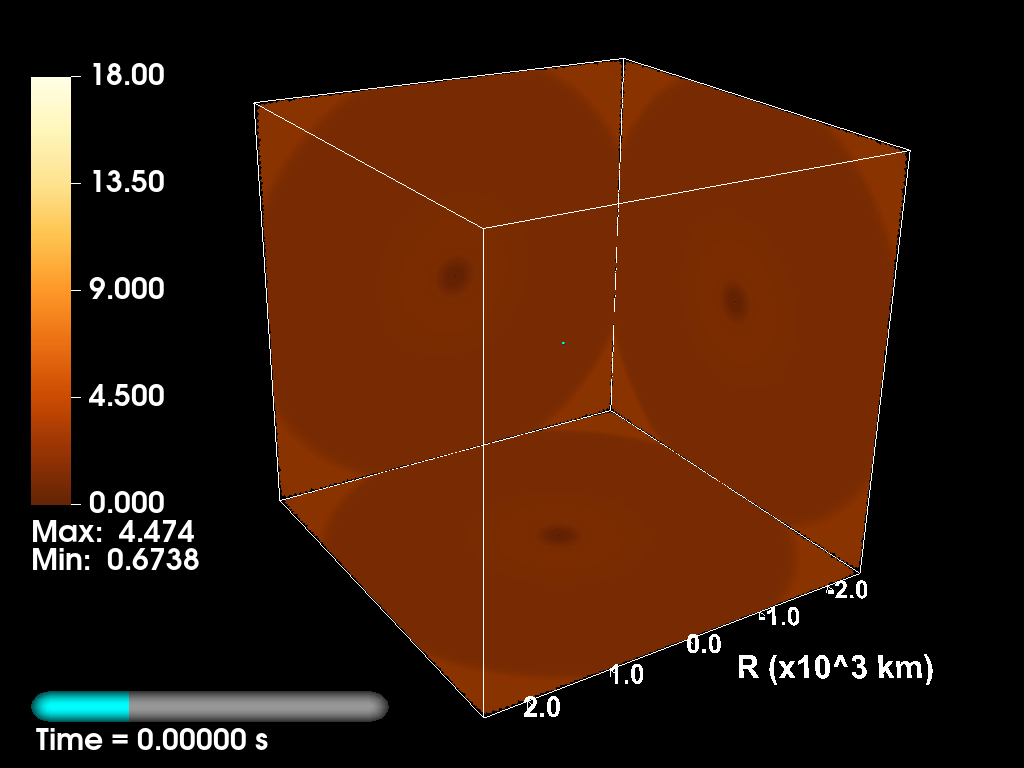}}\quad
\subfigure{\includegraphics[width=0.38\textwidth]{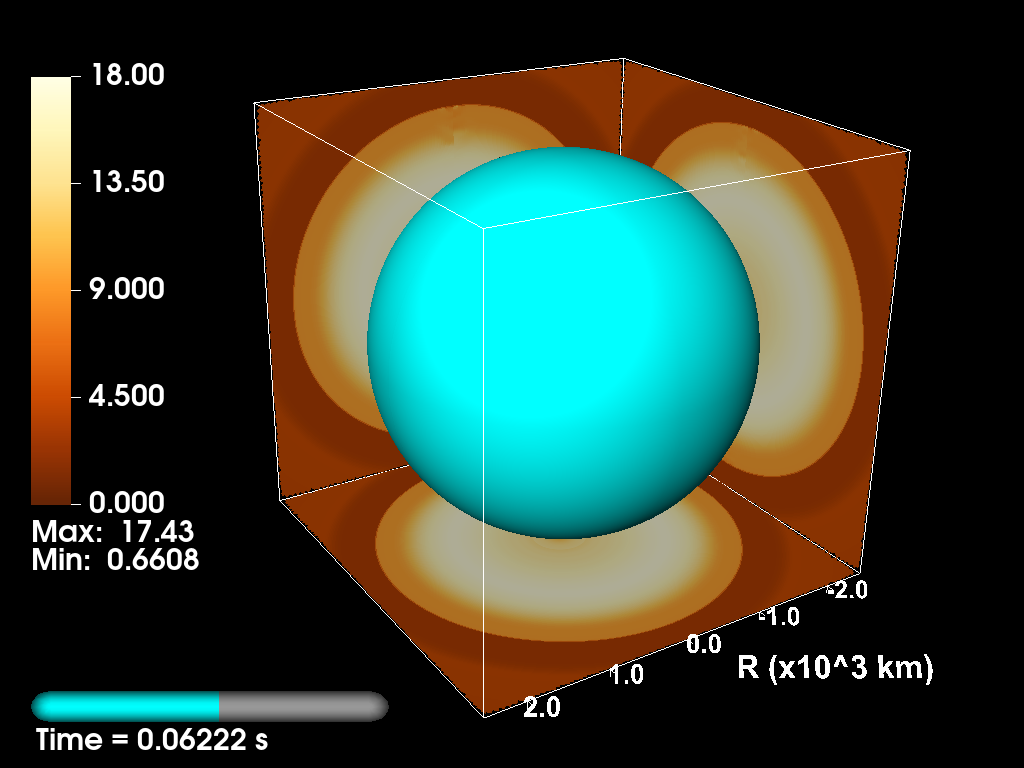}}
\subfigure{\includegraphics[width=0.38\textwidth]{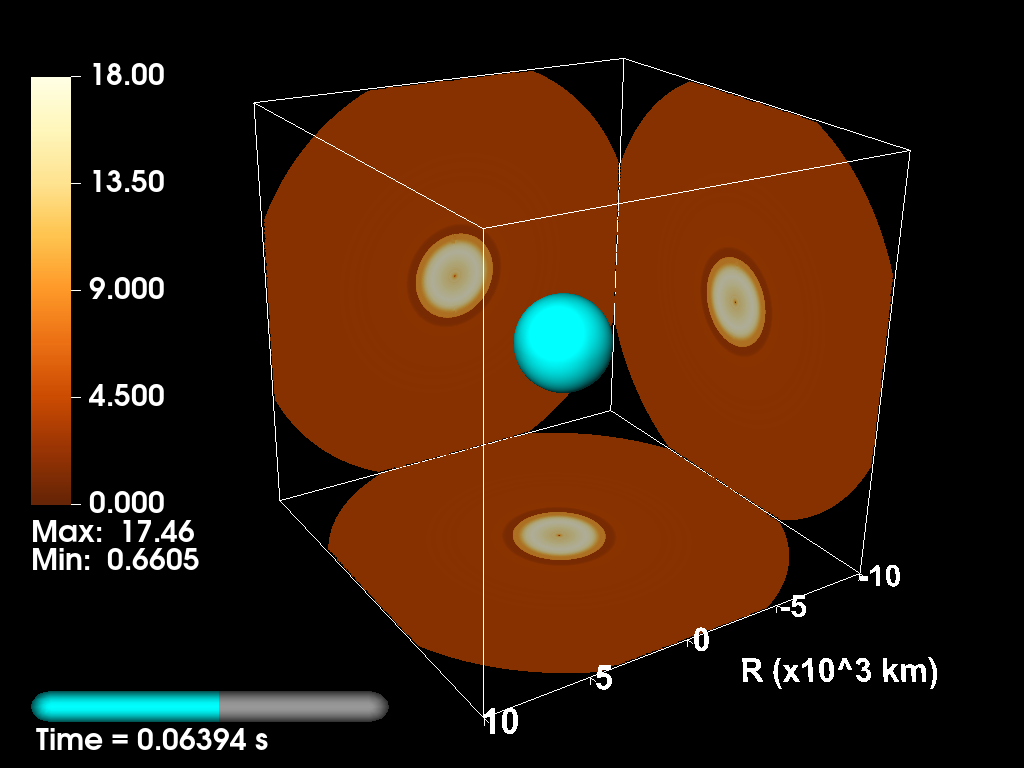}}\quad
\subfigure{\includegraphics[width=0.38\textwidth]{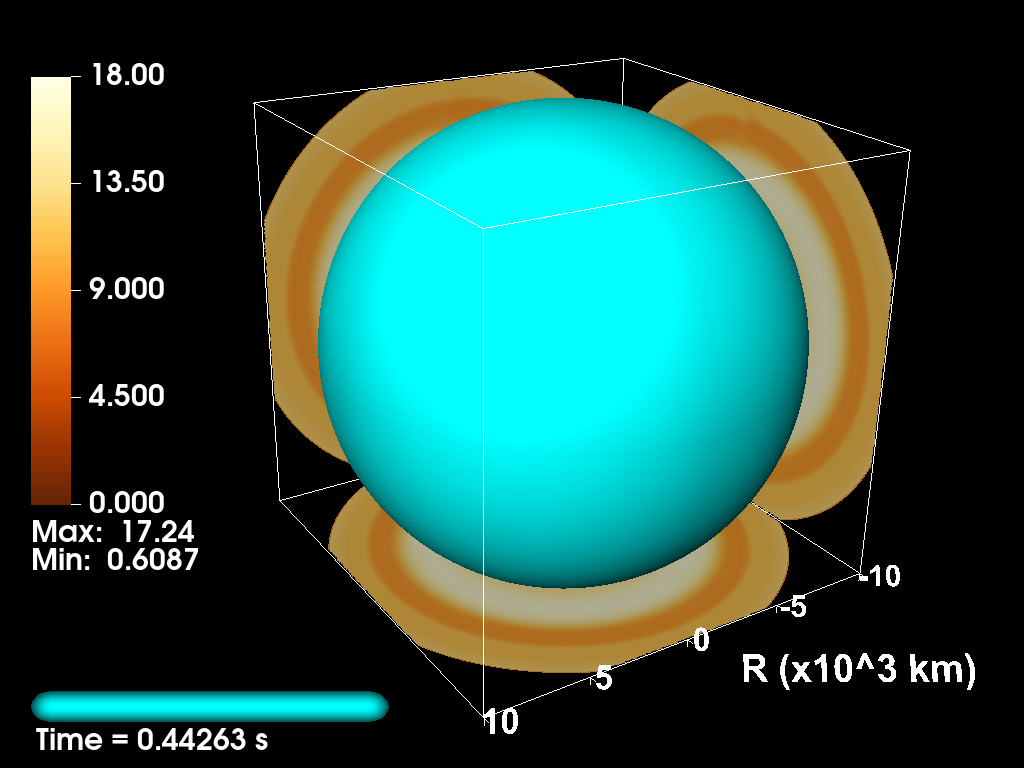}}
\end{interactive}
\caption{
\change{Snapshots of the 3D version of adiabatic model S12 from a 49-second animation available in the online version of the article.
The snapshots depict the animation progression from left to right and top to bottom, as indicated by the time label on the bottom left corner of each panel.}
In order to give a sense of the dynamic range in radius implicated in gravitational collapse, bounce, and explosion, the middle two rows are zoomed in on a closer view.
The grey surface in the first two rows indicates the radius of maximum infall velocity during collapse.
The light blue surface in the second two rows indicates the shock radius during post-bounce expansion.  
Cross sections of the entropy (slices through the origin) are projected onto the rear walls of the visualization box.
}
\label{Fig:WH_A_3D_Snapshot}
\end{figure*}

\begin{figure*}[t]
\centering
\includegraphics[width=1.0\textwidth]{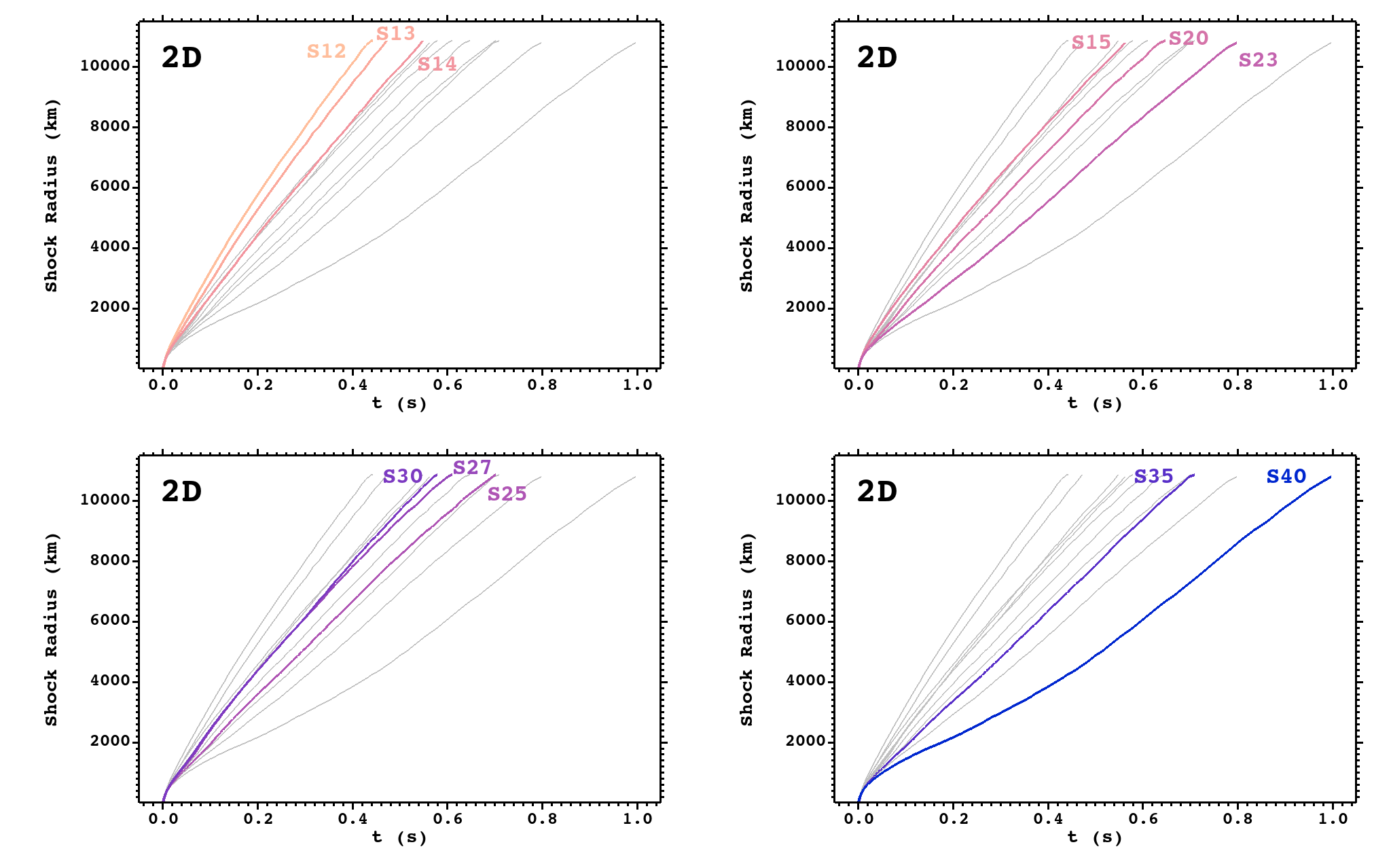}
\caption{
Shock radius in adiabatic models evolved from eleven \citet{Woosley2007Nucleosynthesis} pre-supernova progenitors as a function of post-bounce time.
Shock radius curves for all progenitors are shown in lightweight grey curves in each panel. For clarity, selected curves are highlighted in color in individual panels, revealing that the speed of shock expansion is not monotonic with the zero-age main sequence mass with which the models are labeled.
\change{Data are from 2D runs, but results from 1D and 3D runs are essentially the same visually for this spherically symmetric problem.}}
\label{Fig:WH_A_Shock}
\end{figure*}

\begin{figure*}[t]
\centering
\includegraphics[width=1.0\textwidth]{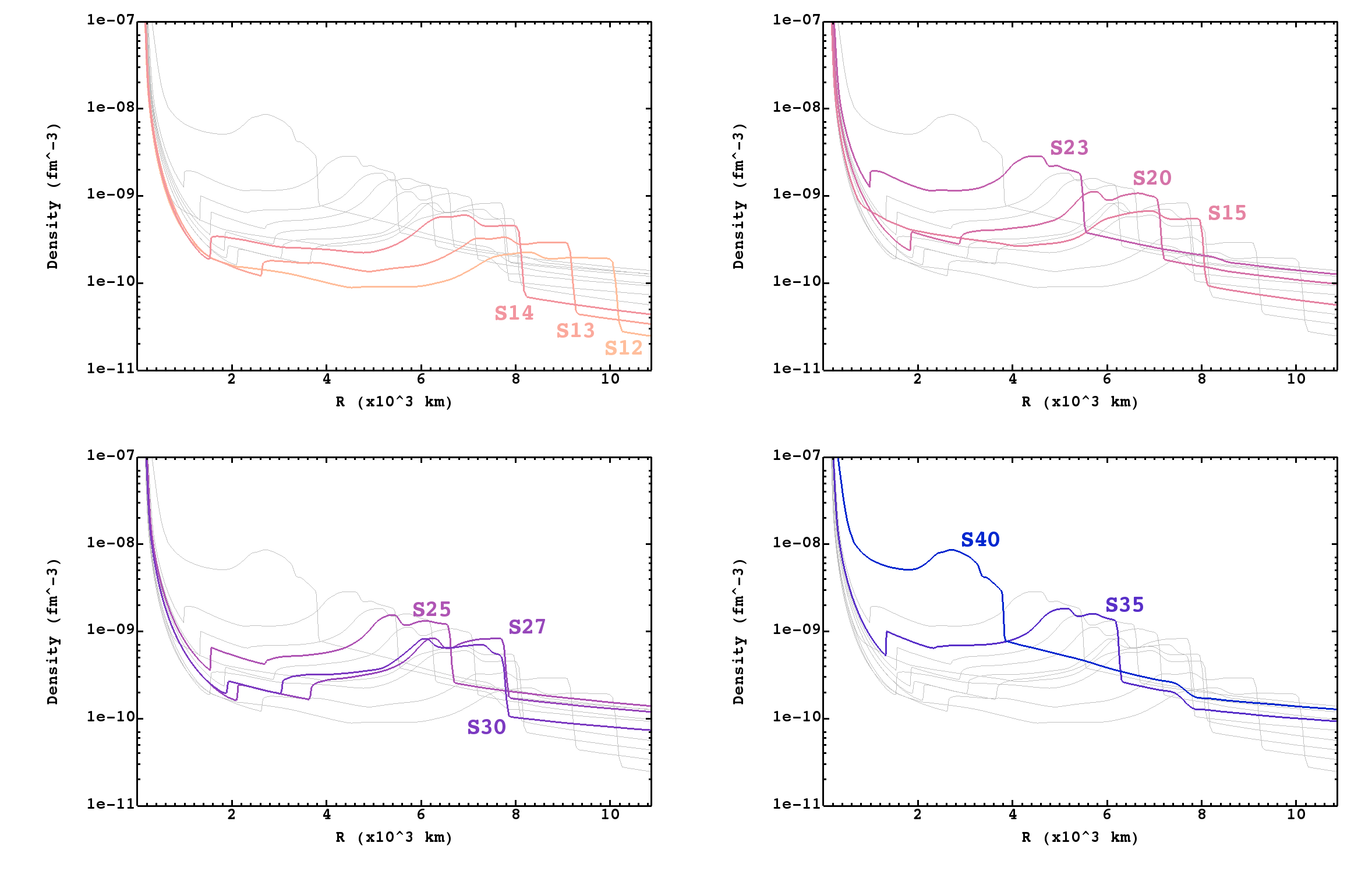}
\caption{
Density profiles in adiabatic models evolved from eleven \citet{Woosley2007Nucleosynthesis} pre-supernova progenitors at approximately 0.4~s after bounce.
Density profiles for all progenitors are shown in lightweight grey curves in each panel. For clarity, selected profiles are highlighted in color in individual panels, revealing that the shock proceeds more slowly when it plows through an envelope of higher density.
\change{Data are from 1D runs, but results from 2D and 3D runs are essentially the same visually for this spherically symmetric problem.}}
\label{Fig:WH_A_DensityEnsemble}
\end{figure*}

Unlike more realistic models that include the essential physics of neutrino interactions, these adiabatic models explode without delay.
The snapshots of the 3D version of model S12 displayed in Figure~\ref{Fig:WH_A_3D_Snapshot} give a sense of the dynamic range in radius implicated in gravitational collapse.
The shock generated at bounce moves promptly and robustly outward as shown in Figure~\ref{Fig:WH_A_Shock}.

While adiabatic explosions are prompt for all eleven progenitors, it is also clear from Figure~\ref{Fig:WH_A_Shock} that the speed of shock expansion is not monotonic with the initial (zero-age main sequence) mass of the pre-supernova progenitors with which the models are labeled. 
Shock expansion is generally slower for larger initial mass for models S12 through S23; but this trend is temporarily reversed for models S25 through S30, before resuming with models S35 and S40. 
Figure~\ref{Fig:WH_A_DensityEnsemble}, which displays density profiles for all eleven \citet{Woosley2007Nucleosynthesis} progenitors at approximately $0.4\,\text{s}$ after bounce, indicates that the shock proceeds more slowly when it plows through an envelope of higher density.
The consequences of higher density implicate both momentum and energy: in addition to moving against a higher ram pressure, in traversing a given range in radius the shock must expend energy in dissociating a larger mass of heavier nuclei into its constituent neutrons and protons.

\begin{figure*}
\centering
\includegraphics[width=0.525\textwidth]{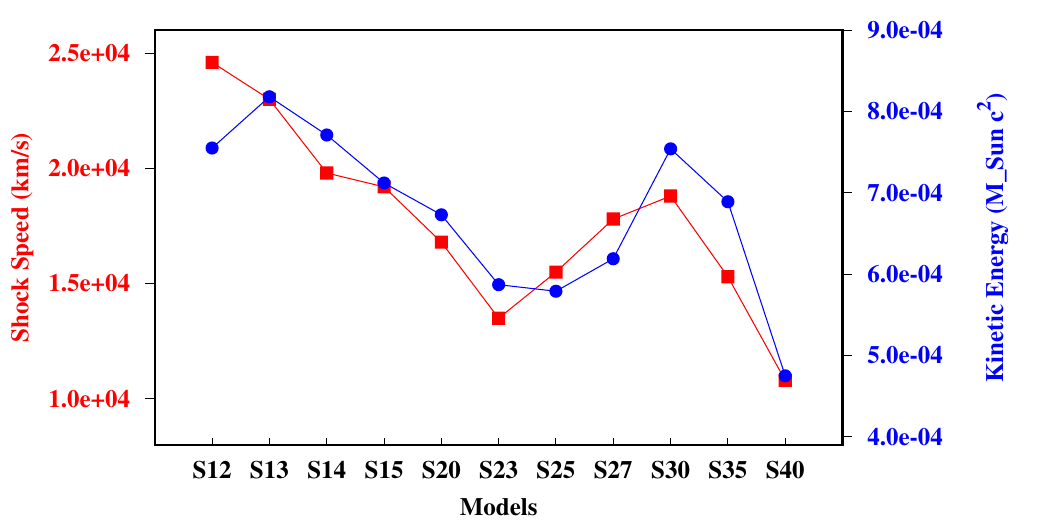}
\includegraphics[width=0.465\textwidth]{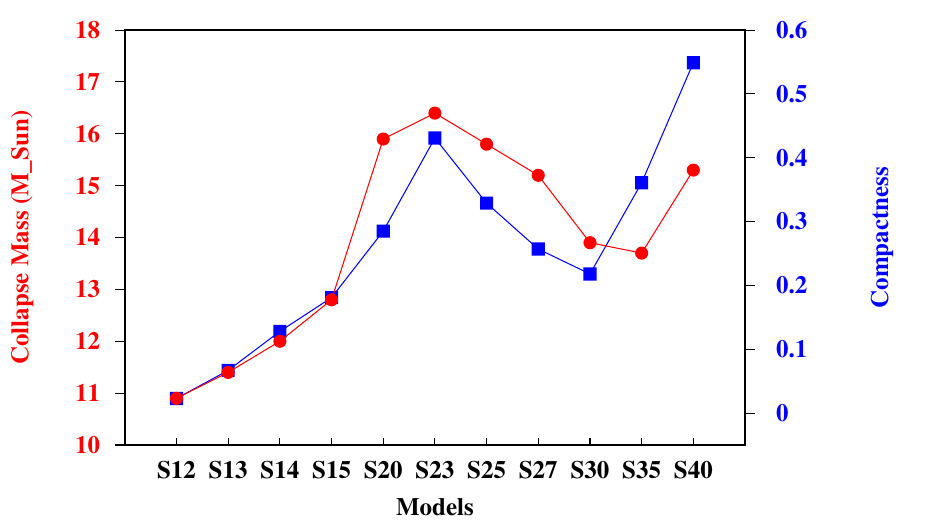}
\caption{The average shock expansion speed and total kinetic energy in adiabatic models are not monotonic as a function of the zero-age main sequence mass with which the models are labeled (left panel).
In mirror image, the stellar mass at the onset of collapse and the compactness parameter are similarly non-monotonic (right panel).}
\label{Fig:WH_Model}
\end{figure*}

\begin{figure*}
\centering
\includegraphics[width=0.49\textwidth]{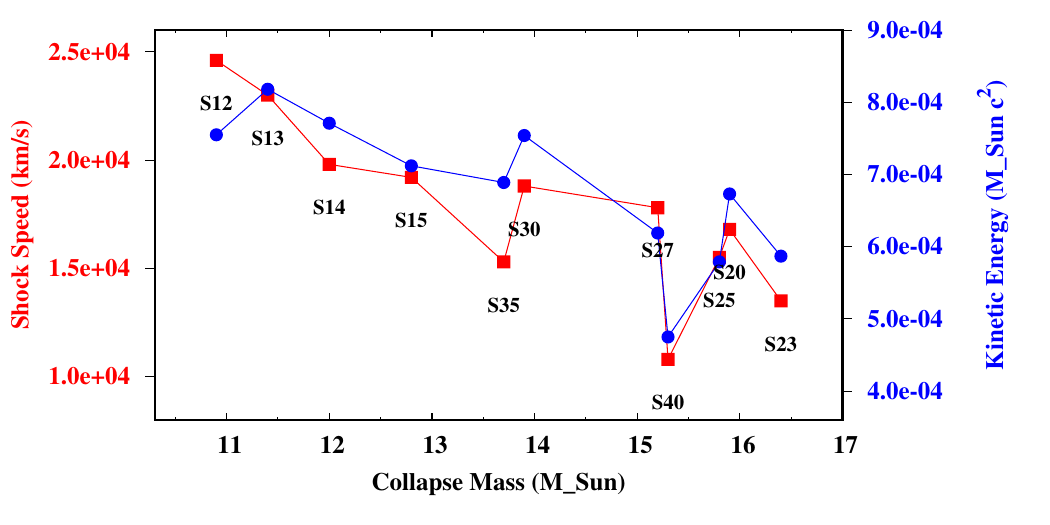}
\includegraphics[width=0.49\textwidth]{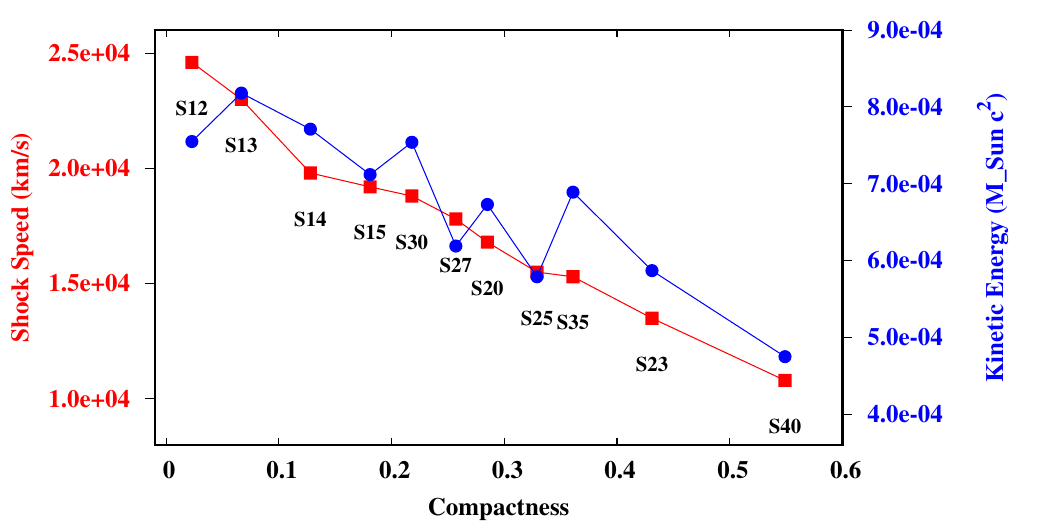}
\caption{The average shock speed and total kinetic energy in adiabatic models are inversely correlated with both the stellar mass at the onset of collapse (left panel) and with the compactness parameter (right panel).
Both are more closely (anti-)correlated with compactness, and the (anti-)correlation of average shock speed with compactness is especially tight.}
\label{Fig:WH_Model2}
\end{figure*}

This non-monotonicity of shock expansion speed with zero-age main sequence mass in these adiabatic models is further explored in Figures~\ref{Fig:WH_Model} and \ref{Fig:WH_Model2}.
The non-monotonicity of average shock expansion speed---and, similarly, total kinetic energy---in the left panel of Figure~\ref{Fig:WH_Model} reflects the pre-supernova progenitors' individual histories of stellar evolution.
(The average shock expansion speed is computed from two points: the position and time when the shock is first detected, and the position and time just before the shock passes the grid boundary.)
For starters, these stars are subject to stellar winds that reduce their total mass.
As a result, the familiar adage ``mass determines destiny'' qualitatively extends to adiabatic explosions: the stellar mass at the onset of collapse shows a similar (but mirror image) non-monotonicity as a function of zero-age main sequence mass in the right panel of Figure~\ref{Fig:WH_Model}.
In more detail, the density structure in the inner layers depends on the particular history of burning stages each star experiences.
Even this can be reasonably captured by a single parameter: the compactness 
\begin{equation}
\xi_{m} = \frac{m/M_\odot}{R(m)/1000\,\text{km}} 
\end{equation}
introduced by \citet{OConnor2011Black-Hole-Form}, evaluated for $m = 2.5\,M_\odot$ at the onset of collapse in the models considered here and also shown in the right panel of Figure~\ref{Fig:WH_Model}, exhibits a similar non-monotonicity.
\change{(The value $m = 2.5\,M_\odot$, roughly the maximum neutron star mass, was chosen by \citet{OConnor2011Black-Hole-Form} to study the post-bounce time of black hole formation.
Because the density change at the passage of the Si/O boundary through the shock can help trigger a neutrino-driven explosion, some authors focused on explodability have chosen a mass coordinate corresponding to this region, $m = 1.75\,M_\odot$, with compactness $\xi_{1.75}$ evaluated at bounce (e.g. \citet{Janka2025Long-Term-Multi}).
But the value $m = 2.5\,M_\odot$ also remains in common use (e.g. \citet{da-Silva-Schneider2020Equation-of-Sta}), and we choose it for three reasons. First, convenience: it is at large enough radius that it changes little between the onset of collapse and core bounce, so that $\xi_{2.5}$ can be evaluated on the pre-collapse progenitor profile.
Second, as a property computed from the progenitor, it is disentangled from the choice of equation of state used in the core-collapse simulation.
Third, by encompassing more of the matter through which the shock must pass, it is more relevant to the global measures with which we correlate it.)}
The anti-correlation evident in Figure~\ref{Fig:WH_Model} between average shock expansion speed and total kinetic energy on the one hand (left panel), and stellar mass at the onset of collapse and compactness on the other (right panel), is shown more directly in Figure~\ref{Fig:WH_Model2}.
The anti-correlation between average shock speed and compactness is especially tight (red curve in right panel of Figure~\ref{Fig:WH_Model2}).
\change{That the average shock speed shows a tighter anti-correlation than the total kinetic energy may be an indication that, for adiabatic explosions, momentum considerations (ram pressure) are somewhat more important and/or more consistent among progenitors than energetic considerations (nuclear dissociation, `$p \, \mathrm{d}V$' work).}

\section{Conclusion}
\label{sec:Conclusion}

Following up on Paper~I, in which the code \genasis\ was introduced with basic fluid dynamics on a centrally refined mesh, this work introduces Newtonian self-gravity and an updated fluid dynamics solver on a single-level spherical coordinate mesh with a coarsening scheme.
The method for a multipole Poisson solver that makes efficient use of GPUs is spelled out with detailed pseudocode. 
This Poisson solver will also be useful for widely-used approximations to general relativity. 
The fluid dynamics solver now also makes efficient use of GPUs, and has been updated with parabolic reconstruction and the ability to use tabulated microphysical equations of state that address composition with a representative heavy nucleus.
The results of test problems show that the Poisson solver functions as expected in 2D and 3D problems, including in the dynamical context of gravitational collapse.
In the adiabatic collapse, bounce, and explosion of pre-supernova progenitors with a tabulated microphysical equation of state we see overall $\sim15$X speedup of GPU over CPU-only  performance in `proportional resource tests'---per OLCF Frontier node, a 64-core AMD EPYC CPU (56 cores reserved for the application) vs. 4 MI250X AMD accelerators (`8 GPUs' as presented to the application).

We propose that adiabatic core collapse, bounce, and prompt explosion become a standard benchmark among groups working on the simulation of core-collapse supernovae.
\change{A dataset from this and the other problems in this paper being made publicly available in \cite{GenASiS_II_Dataset} is a step in this direction.
The dataset includes selected outputs from 1D, 2D, and 3D simulations for several resolutions ($N_\theta$) discussed in \S \ref{sec:OppenheimerSnyder}, \S\ref{sec:LinMestelShu}, \S\ref{sec:YahilLattimer}, and \S\ref{sec:AdiabaticExplosion}.}

While the absence of physically crucial neutrinos renders this a test problem not directly relevant to comparisons with astronomical observations, persistent and systematic differences in the results obtained with different core-collapse supernova codes \citep{Janka2025Long-Term-Multi} suggest the wisdom of stepping back to a simplified problem in order to detect and diagnose some of the potential sources of those differences.
Prior to the complications and costs of neutrino radiation fluid dynamics, modeling of the adiabatic collapse, bounce, and explosion of pre-supernova progenitors probes such issues as the handling of wide ranges of spatial and temporal scales; treatments of self-gravity; the nuclear phase transition and evolution of nuclear composition; and code performance, including the use of hardware accelerators (typically GPUs) on architectures now characteristic of capability supercomputers.
Works reporting on energy conservation in the adiabatic explosion of a single progenitor include \citet{Skinner2019FORNAX:-A-Flexi} and \citet{Bruenn2020Chimera:-A-Mass}; a more comprehensive exposition of a single adiabatic explosion is presented by \citet{Pochik2021thornado-hydro:}.

Study of adiabatic explosions also sets baseline expectations for `explodability' studies with more complete physics \citep{Janka2025Long-Term-Multi}.
In our own study of a small ensemble of the spherically symmetric (but evolved in 1D/2D/3D) adiabatic explosions of eleven pre-supernova progenitors we find that the average shock speed and total kinetic energy are not monotonic with the mass with which a star is born (zero-age main sequence mass).
These measures of adiabatic explosion speed and strength depend on the density of the infalling layers through which the shock proceeds.
By the time of collapse the star's mass is reduced due to stellar winds, and the shock speed and total kinetic energy are more closely correlated with this---and even more closely with the compactness parameter at collapse.
The extent to which these patterns hold up in the face of stochastic multidimensional dynamics---resulting from convective burning in 3D progenitors, and the stationary accretion shock instability and neutrino-driven convection after bounce---is an interesting subject of ongoing and future study.

\acknowledgments
%

We thank David Pochik for assistance with the reference solution for self-similar gravitational collapse of a polytropic sphere.
This material is based upon work supported by the U.S. Department of Energy, Office of Science, Office of Nuclear Physics under contract number DE-AC05-00OR22725 and the National Science Foundation under Grant No. 1535130. 
This research used resources of the Oak Ridge Leadership Computing Facility, which is a DOE Office of Science User Facility supported under Contract DE-AC05-00OR22725.

\change{

\appendix

Following the initial submission and review of this manuscript, we implemented a suggestion in the main text to solve the problem of falling entropy and temperature in the nascent neutron star after bounce depicted in Figures~\ref{Fig:WH_A_Central} and \ref{Fig:WH_A_VelocityGlitch}, with excellent results.
Specifically, we added a balance equation for fluid entropy to the balance equations indicated by Eq.~(\ref{eq:BalancedVariablesGravitational}).
The balanced entropy $S$ (not to be confused with momentum density $\mathbf{S}$) is associated with flux $S  \mathbf{v}$ and, as with the electron number density in these adiabatic models, a vanishing source term.
The associated primitive variable is the entropy per baryon $s = S / n_b$.
Both the energy and entropy balance equations are evolved everywhere, but only one of these updates is used, on a cell-by-cell basis, in the inversion to primitive variables.
The energy update is used where $n_b < 10^{11} \, \text{g} \, \text{cm}^{-3} / m_u$ (that is, safely outside the nascent neutron star), and also where a shock is detected at any density.
The criterion for shock detection is a fractional change in pressure across a cell interface of absolute value larger than $0.2$ coupled with compression (indicated by a negative velocity difference across the interface, value at larger coordinate minus value at smaller coordinate).
Figures~\ref{Fig:WH_A_Central_New} and \ref{Fig:WH_A_VelocityGlitch_New} show that entropy and temperature at high densities now hold steady as expected for adiabatic runs with all indicated resolutions, indicating that this particular problem is no longer an issue necessitating higher resolution.
Shock trajectories are essentially the same visually as those in Fig.~\ref{Fig:WH_A_Shock}.
This success also depends on use of the HLLC Riemann solver described in Paper~I; the adiabatic models presented in the main text had been computed with the HLL solver, because use of the HLLC solver without entropy evolution had exacerbated the velocity jaggedness at the phase transition.
A dataset from these new results is made publicly available in \cite{GenASiS_II_Dataset_Appendix}.

\begin{figure*}
\centering
\includegraphics[width=0.49\textwidth]{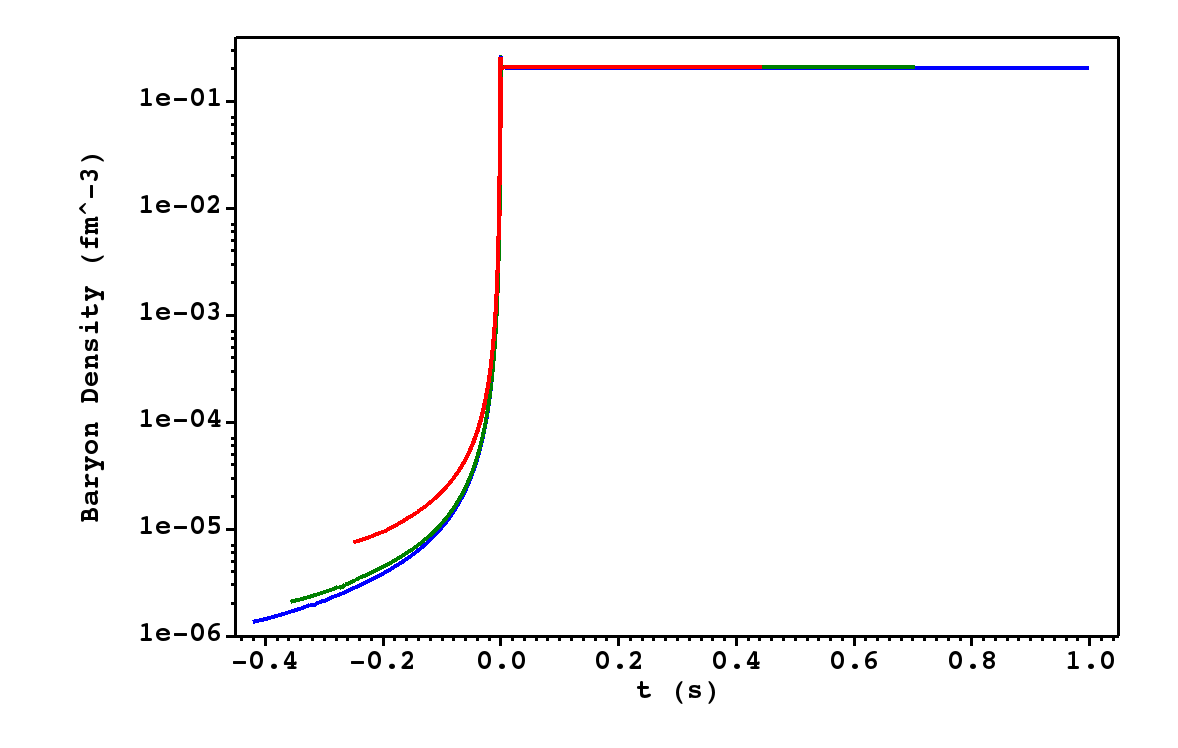}
\includegraphics[width=0.49\textwidth]{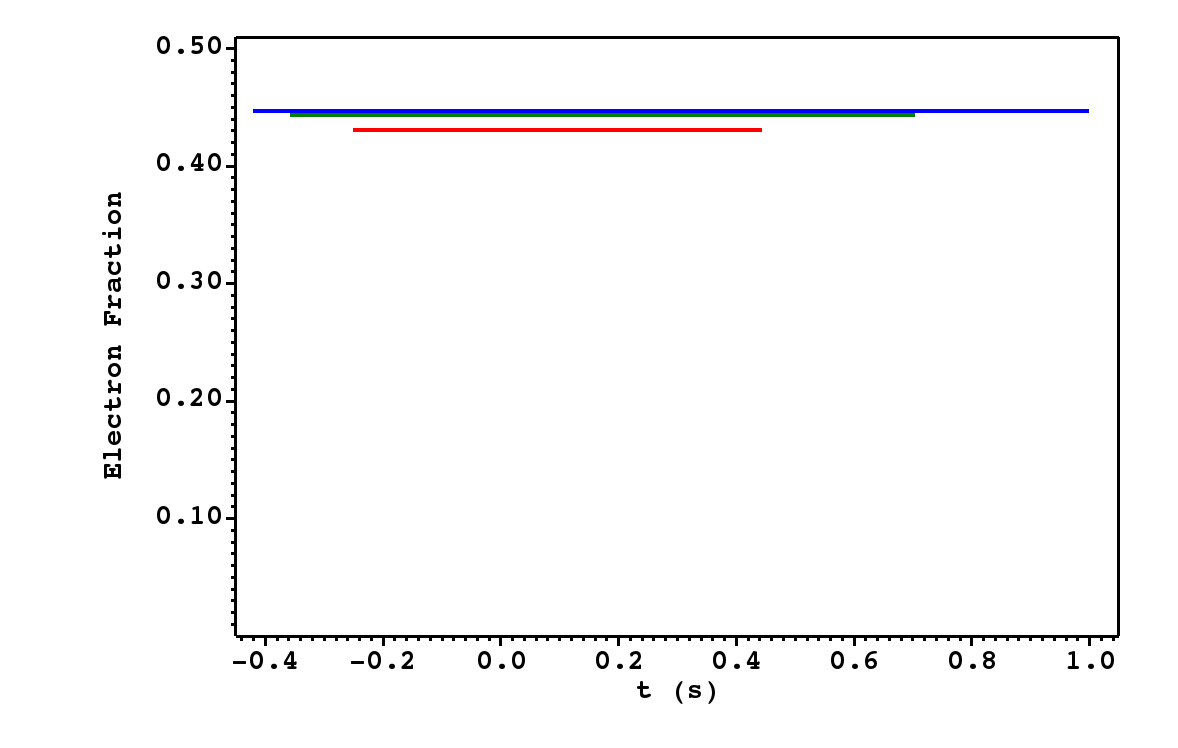}
\includegraphics[width=0.49\textwidth]{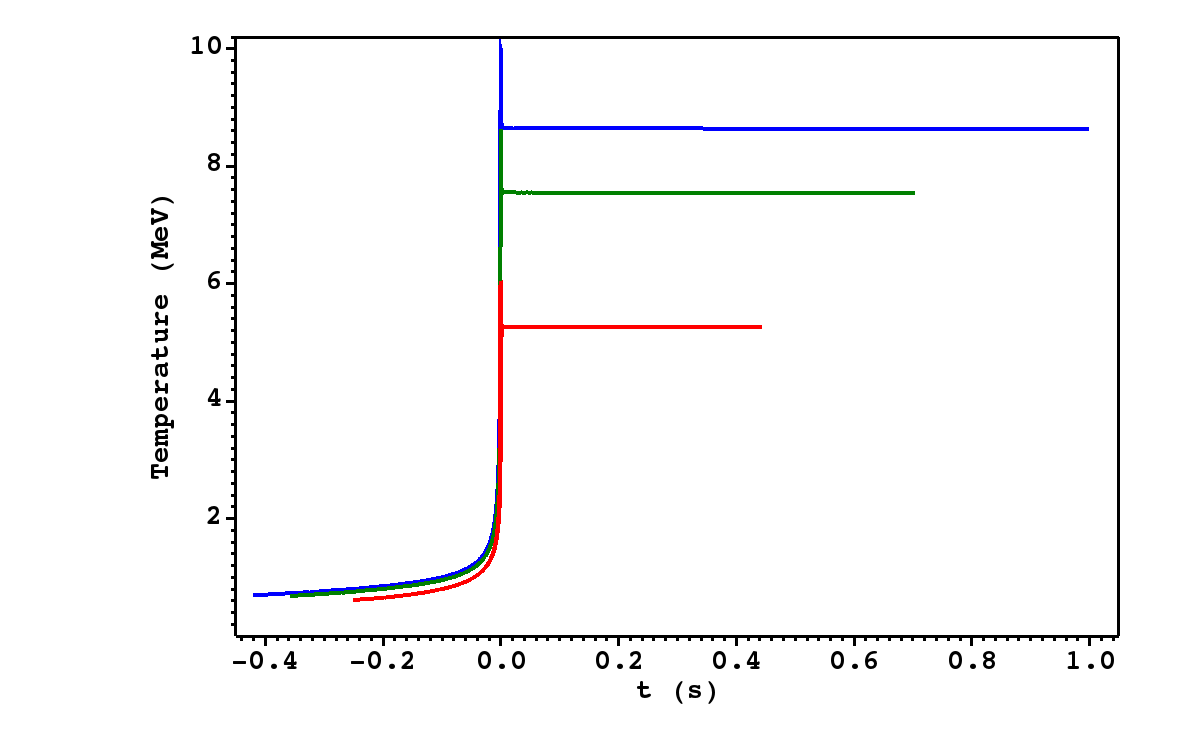}
\includegraphics[width=0.49\textwidth]{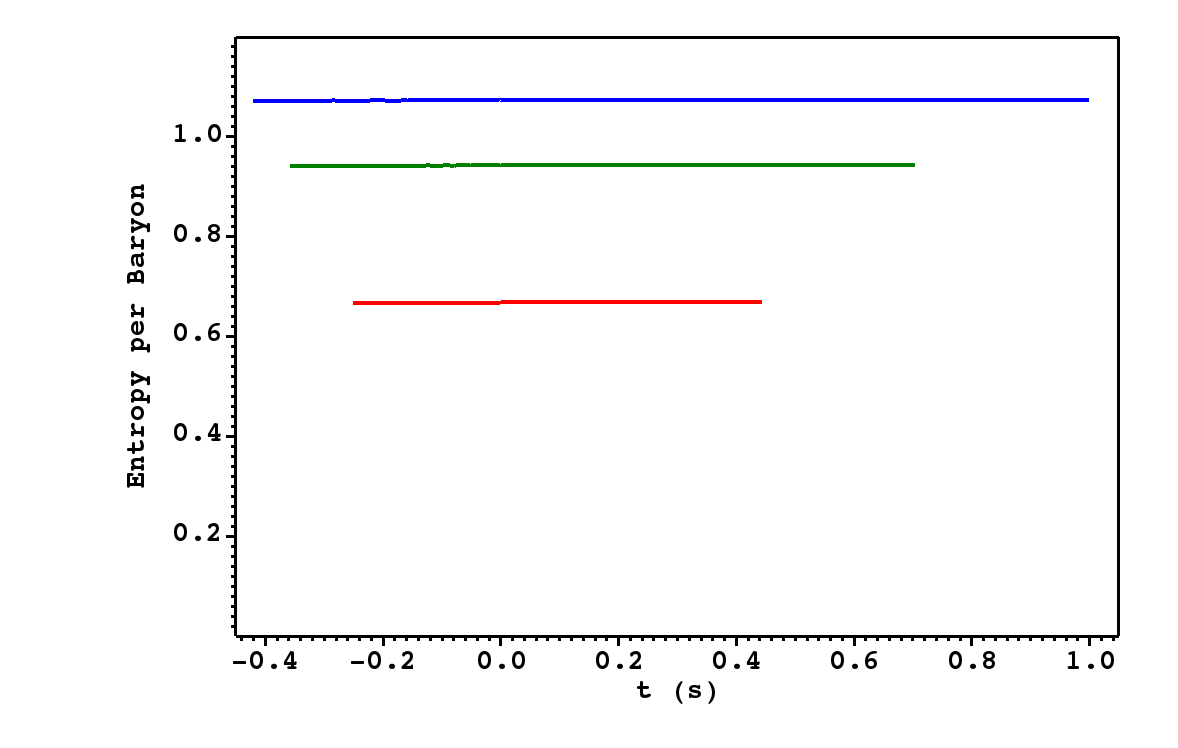}
\caption{\change{Central values of baryon density, electron fraction, temperature, and entropy per baryon as a function of time for adiabatic models S12 (red), S25 (green), and S40 (blue), for three different resolutions ($N_\theta = 128, 192, 256$, denoted by curves of increasing thickness), now using entropy evolution at high density and in the absence of a shock.
This prevents the departures from expected steady values of central temperature and entropy per baryon evident in Figure~\ref{Fig:WH_A_Central}.
Data are from 1D runs, but results from 2D and 3D runs are essentially the same visually for this spherically symmetric problem.}}
\label{Fig:WH_A_Central_New}
\end{figure*}
\begin{figure*}
\centering
\includegraphics[width=0.49\textwidth]{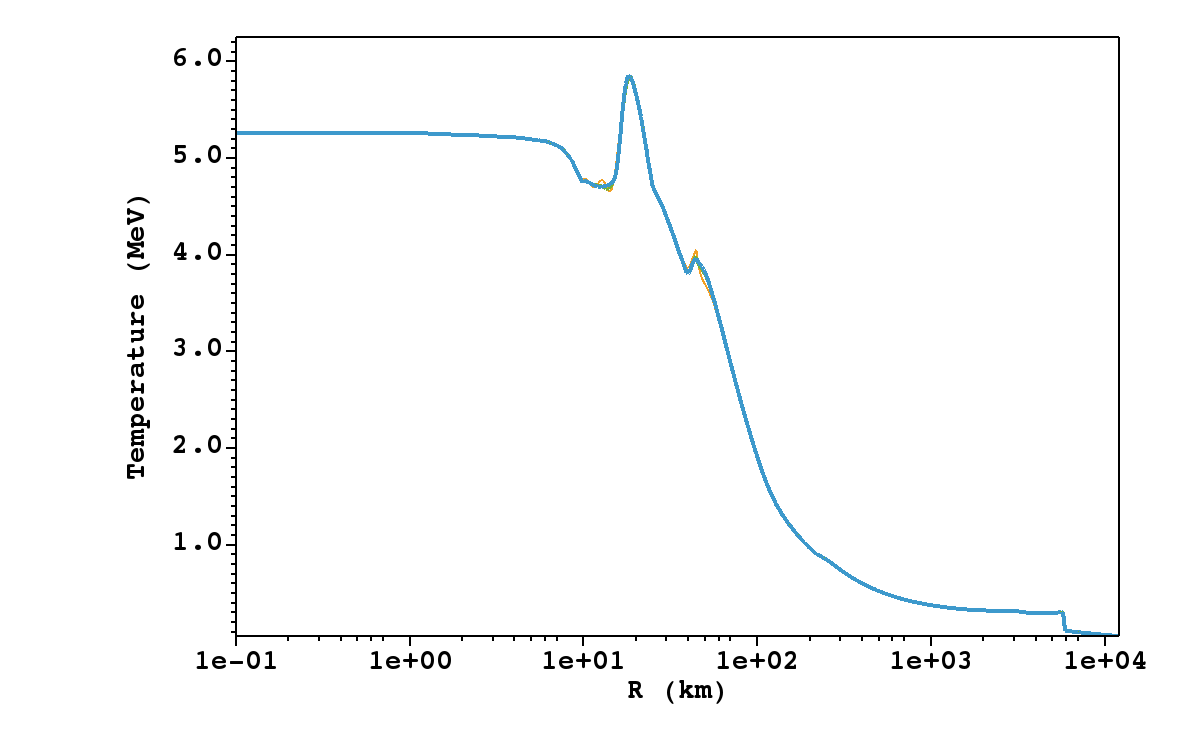}
\includegraphics[width=0.49\textwidth]{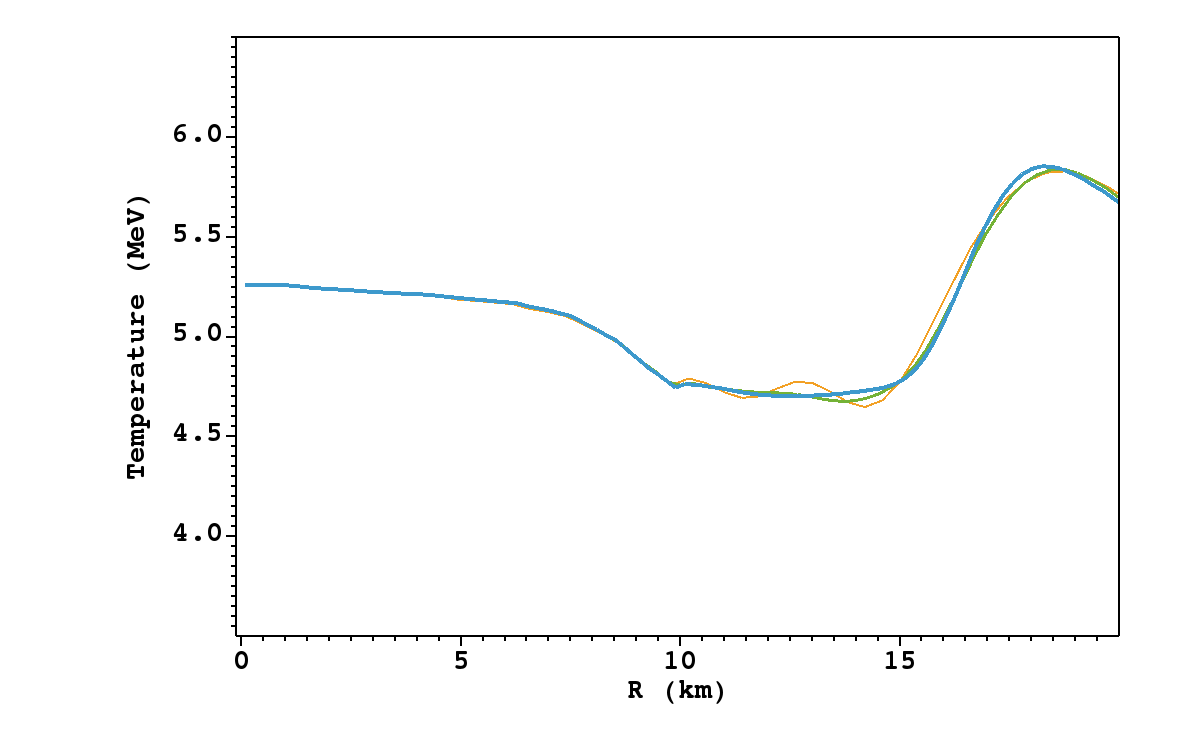}
\includegraphics[width=0.49\textwidth]{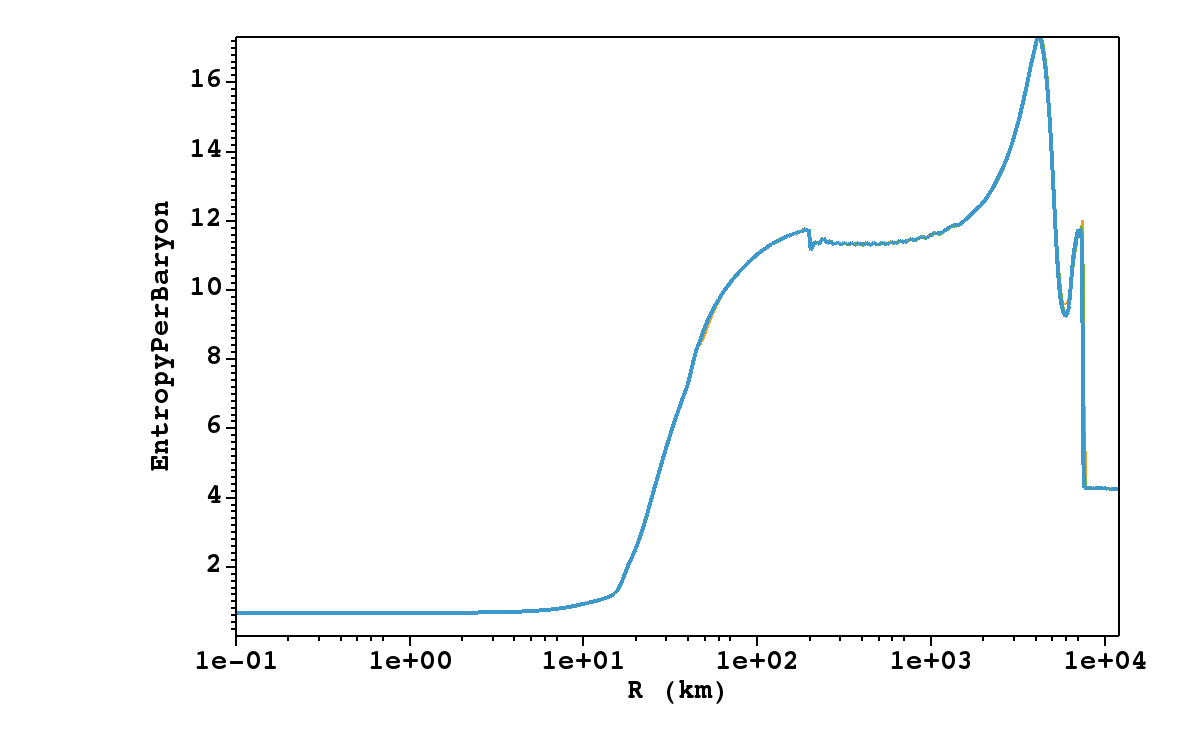}
\includegraphics[width=0.49\textwidth]{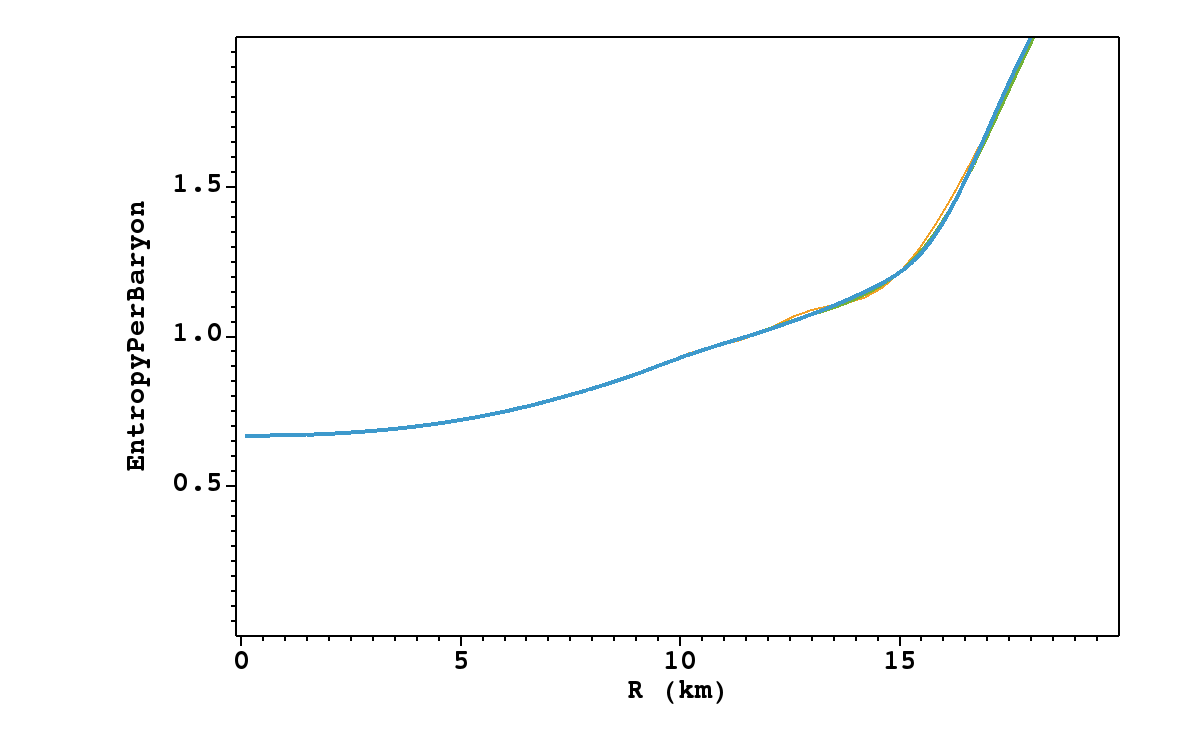}
\includegraphics[width=0.49\textwidth]{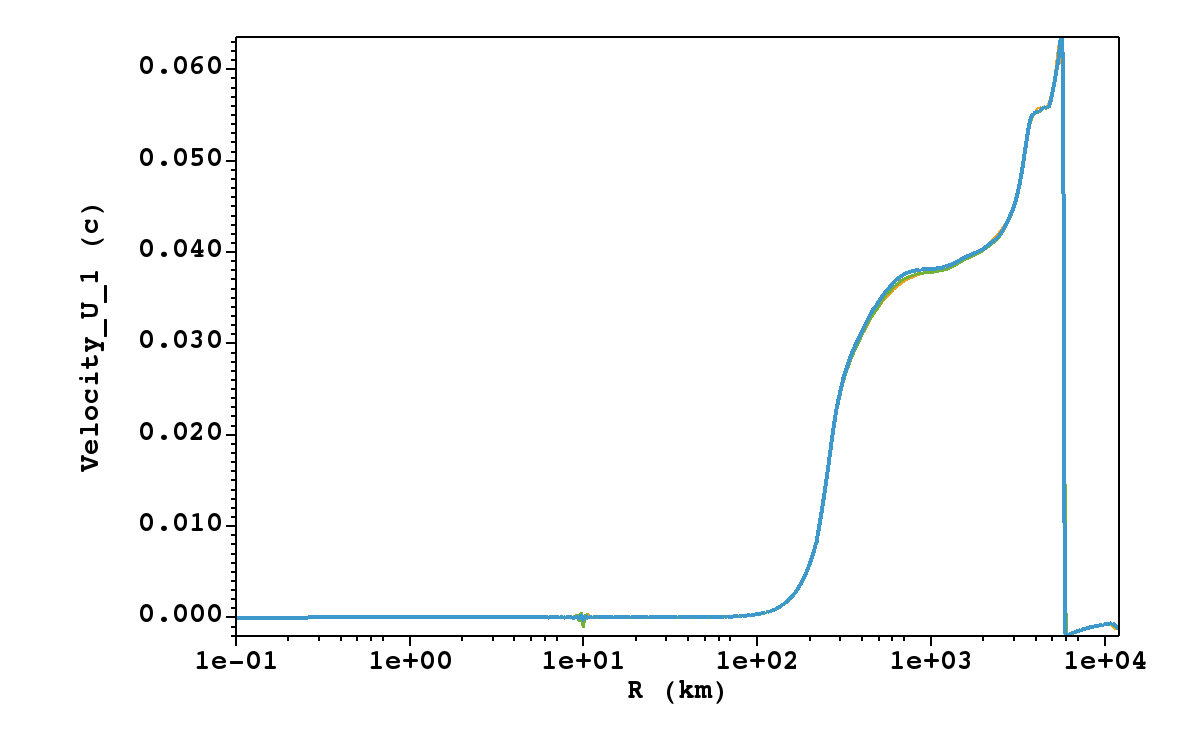}
\includegraphics[width=0.49\textwidth]{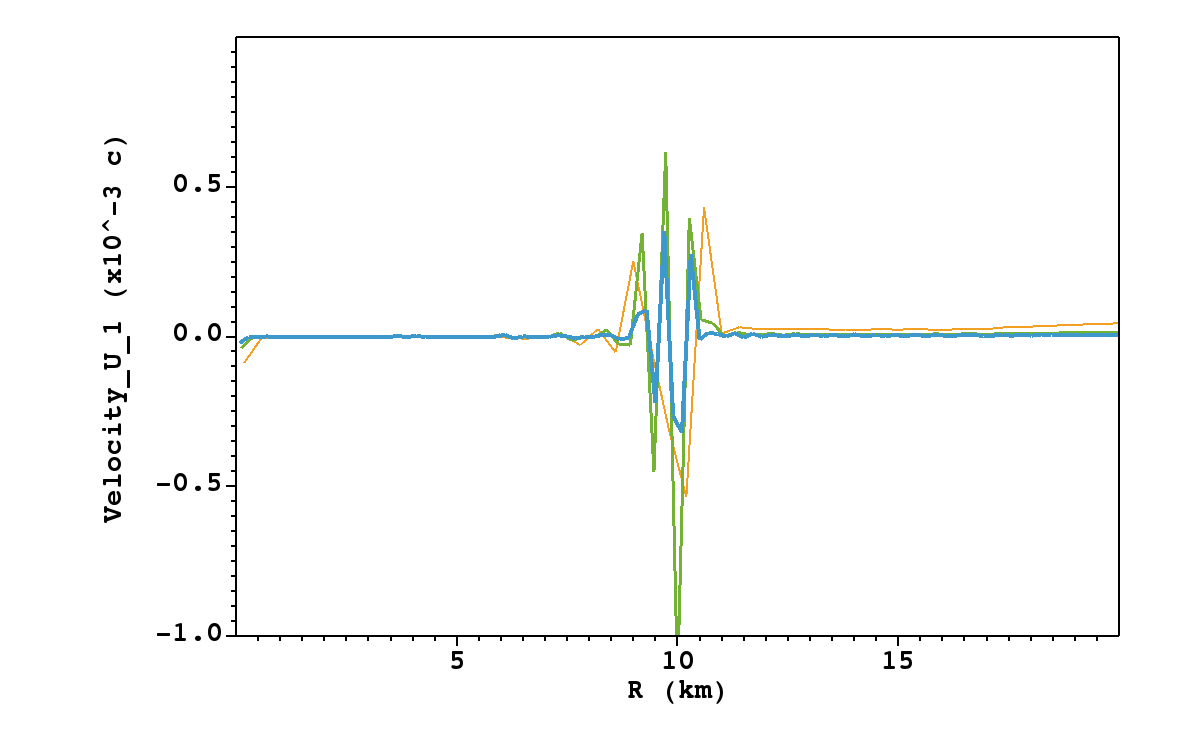}
\caption{\change{Profiles from adiabatic model S12 for three different resolutions ($N_\theta = 128, 192, 256$, denoted by orange, green, and blue curves of increasing thickness) at about 0.2~s after bounce, now using entropy evolution at high density and in the absence of a shock.
This prevents the declining entropy and temperature at high density and entropy generation at the nuclear phase transition evident in Figure~\ref{Fig:WH_A_VelocityGlitch}.
Data are from 1D runs, but results from 2D and 3D runs are essentially the same visually for this spherically symmetric problem.}} 
\label{Fig:WH_A_VelocityGlitch_New}
\end{figure*}
}

\def\cpc{Comp. Phys. Commun.}

\bibliographystyle{aasjournal}

\bibliography{Bib}

\end{document}